\documentclass[fleqn,usenatbib]{mnras}

\usepackage{newtxtext,newtxmath}

\usepackage[T1]{fontenc}

\DeclareRobustCommand{\VAN}[3]{#2}
\let\VANthebibliography\thebibliography
\def\thebibliography{\DeclareRobustCommand{\VAN}[3]{##3}\VANthebibliography}

\usepackage{graphicx}	
\usepackage{deluxetable}
\usepackage{subcaption}
\usepackage{gensymb}
\usepackage{ulem}
\usepackage{academicons}
\usepackage{xcolor}
\usepackage{scalerel}
\usepackage{tikz}
\usetikzlibrary{svg.path}

\definecolor{orcidlogocol}{HTML}{A6CE39}
\tikzset{
  orcidlogo/.pic={
    \fill[orcidlogocol] svg{M256,128c0,70.7-57.3,128-128,128C57.3,256,0,198.7,0,128C0,57.3,57.3,0,128,0C198.7,0,256,57.3,256,128z};
    \fill[white] svg{M86.3,186.2H70.9V79.1h15.4v48.4V186.2z}
                 svg{M108.9,79.1h41.6c39.6,0,57,28.3,57,53.6c0,27.5-21.5,53.6-56.8,53.6h-41.8V79.1z M124.3,172.4h24.5c34.9,0,42.9-26.5,42.9-39.7c0-21.5-13.7-39.7-43.7-39.7h-23.7V172.4z}
                 svg{M88.7,56.8c0,5.5-4.5,10.1-10.1,10.1c-5.6,0-10.1-4.6-10.1-10.1c0-5.6,4.5-10.1,10.1-10.1C84.2,46.7,88.7,51.3,88.7,56.8z};
  }
}

\newcommand\orcidicon[1]{\href{https://orcid.org/#1}{\mbox{\scalerel*{
\begin{tikzpicture}[yscale=-1,transform shape]
\pic{orcidlogo};
\end{tikzpicture}
}{|}}}}

\usepackage{hyperref} 

\title[Line Verifications and a Spectroscopic Atlas]{A 1.46--2.48~$\mu$m Spectroscopic Atlas of a T6 Dwarf (1060~K) Atmosphere with IGRINS: First Detections of H$_2$S and H$_2$, and Verification of H$_2$O, CH$_4$, and NH$_3$ Line Lists}

\author[Tannock et al.]{Megan E. Tannock,$^{1}$\thanks{E-mail: mtannock@uwo.ca}\orcidicon{0000-0002-9445-2870}, 
Stanimir Metchev$^{2,3}$\thanks{E-mail: smetchev@uwo.ca}\orcidicon{0000-0003-3050-8203}, 
Callie E. Hood$^{4}$\orcidicon{0000-0003-1150-7889}, 
Gregory N. Mace$^{5}$\orcidicon{0000-0001-7875-6391}, 
\newauthor
Jonathan J. Fortney$^{4}$\orcidicon{0000-0002-9843-4354}, 
Caroline V. Morley$^{5}$\orcidicon{0000-0002-4404-0456}, 
Daniel T. Jaffe,$^{5}$\orcidicon{0000-0003-3577-3540}, 
Roxana Lupu$^{6}$\orcidicon{0000-0003-3444-5908}
\\
$^{1}$Department of Physics and Astronomy, The University of Western Ontario, 1151 Richmond St, London, Ontario, N6A~3K7, Canada\\
$^{2}$Department of Physics and Astronomy, Institute for Earth and Space Exploration, The University of Western Ontario, 1151 Richmond St, London,\\ Ontario, N6A~3K7, Canada\\
$^{3}$Department of Astrophysics, American Museum of Natural History, 200 Central Park West, New York, New York, 10024--5102, USA\\
$^{4}$Department of Astronomy \& Astrophysics, University of California, Santa Cruz, CA 95064, USA\\
$^{5}$Department of Astronomy, The University of Texas, Austin, TX 78712, USA\\
$^{6}$Eureka Scientific Inc, Oakland, CA 94602, USA
}

\date{Accepted 2022 May 18. Received 2022 May 17; in original form 2022 January 15}

\pubyear{2022}

\begin{document}
\label{firstpage}
\pagerange{\pageref{firstpage}--\pageref{lastpage}}
\maketitle

\begin{abstract}

We present Gemini South/IGRINS observations of the 1060~K T6 dwarf 2MASS~J08173001$-$6155158 with unprecedented resolution ($R\equiv\lambda/\Delta\lambda=45\,000$) and signal-to-noise ratio (SNR$>$200) for a late-type T dwarf. We use this benchmark observation to test the reliability of molecular line lists used up-to-date atmospheric models. We determine which spectroscopic regions should be used to estimate the parameters of cold brown dwarfs and, by extension, exoplanets. We present a detailed spectroscopic atlas with molecular identifications across the $H$ and $K$ bands of the near-infrared. We find that water (H$_2$O) line lists are overall reliable. We find the most discrepancies amongst older methane (CH$_4$) line lists, and that the most up-to-date CH$_4$ line lists correct many of these issues. We identify individual ammonia (NH$_3$) lines, a hydrogen sulfide (H$_2$S) feature at 1.5900~$\mu$m, and a molecular hydrogen (H$_2$) feature at 2.1218~$\mu$m. These are the first unambiguous detections of H$_2$S and H$_2$ in an extra-solar atmosphere. 
With the H$_2$ detection, we place an upper limit on the atmospheric dust concentration of this T6 dwarf: at least 500 times less than the interstellar value, implying that the atmosphere is effectively dust-free.
We additionally identify several features that do not appear in the model spectra. 
Our assessment of the line lists is valuable for atmospheric model applications to high-dispersion, low-SNR, high-background spectra, such as an exoplanet around a star. We demonstrate a significant enhancement in the detection of the CH$_4$ absorption signal in this T6 dwarf with the most up-to-date line lists.



\end{abstract}




\section{Introduction}
\label{sec:introduction}

Reliable determinations of the effective temperatures, radii, and masses of isolated, self-luminous brown dwarfs and giant exoplanets are dependent on accurate modelling of their spectra. However, it is known that the laboratory-based experimental line lists used to generate model spectra are inconsistent with each other and are even missing lines for some molecular species (e.g., \citealt{Saumon2012, Canty2015}). Even the most up-to date spectral models do not completely reproduce observed spectral features in cold brown dwarfs, limiting our ability to constrain their basic properties. 

Methane and ammonia are of particular interest for T dwarfs. At the time of their discovery, the distinction between L and T dwarfs was based on whether methane lines were present in low-resolution ($R\equiv\lambda/\Delta\lambda\sim100$) spectra \citep{Oppenheimer1995,Geballe1996}. Similarly, ammonia was used to mark the end of the T-sequence and is the distinguishing opacity source of Y dwarfs \citep{Cushing2011}. \citet{Noll2000} have since shown that the onset of methane absorption occurs as early as L5 in $R\approx600$ near-infrared spectra, and \citet{Cushing2006} report the appearance of ammonia bands at spectral type T2 in the mid-infrared. In the latest T dwarfs (T8, T9), ammonia becomes a major opacity source \citep{Cushing2006}. 

Previous spectroscopic studies of late-T and Y dwarfs with broad wavelength coverage have been limited to $R = 6000$ or less, making the identification of specific molecular absorption features difficult. Additionally, older generations of photospheric models have not been able to fit the available data well (e.g., \citealt{Bochanski2011,Leggett2012,Leggett2019,Beichman2014,Canty2015,Schneider2015,Luhman2016,Miles2020,Tannock2021}).

A current hurdle in characterizing cold brown dwarfs and giant exoplanets are systematic uncertainties in the wavelengths and strengths of absorption lines in theoretical photospheres. Missing lines or inaccurate line lists make detections of molecules and determinations of radial velocities and projected rotation velocities difficult or impossible, especially in low signal-to-noise observations of exoplanet atmospheres. It is therefore necessary to confirm the accuracy of line lists by comparing to high signal-to-noise observations. Isolated brown dwarfs, free from the overwhelming light of a companion star, have atmospheres containing some of the key opacity sources in exoplanets, making them suitable laboratories for testing the accuracy of line lists. Improvements in the atmospheric opacity estimates for cold substellar atmospheres would also be invaluable for the characterization of potentially habitable exoplanets. Methane and ammonia have been suggested as biosignature gases in exoplanet atmospheres (e.g., \citealt{Leger1996,Seager2013}). Water, while not a biosignature gas, is also an important signature of habitability and is a major opacity source in brown dwarfs. 

We present a high signal-to-noise (SNR $>$ 200) spectrum of a T6 dwarf with unprecedented $R= 45\,000$ resolution and 1.45--2.48~$\mu$m coverage, observed with the Immersion GRating INfrared Spectrometer (IGRINS; \citealt{Yuk2010,Park2014,Mace2016,Mace2018}) on Gemini South. We perform a detailed study of absorption features due to water, methane, ammonia, carbon monoxide, and hydrogen sulfide. Our target, 2MASS~J08173001$-$6155158 (also known as DENIS J081730.0-615520; hereafter 2M0817) was discovered by \citet{Artigau2010} through a photometric cross match between the Two Micron All Sky Survey (2MASS) and the DEep Near-Infrared Survey of the Southern sky (DENIS) point-source catalogues, and spectroscopically identified as a T6 dwarf. It is at a heliocentric distance of only 5.2127 $\pm$ 0.0113 pc \citep{Gaia2016,Gaia2021}, and is one of the brightest late-type T dwarfs ($K$-band magnitude 13.52; \citealt{Skrutskie2006}). \citet{Radigan2014} find a rotation period of 2.8 $\pm$ 0.2 h for 2M0817 from ground-based $J$-band observations spanning four hours.


\section{Spectroscopy with IGRINS on Gemini South}
\label{sec:IGRINS}

We observed 2M0817 with IGRINS on Gemini South under Gemini program ID GS-2018A-Q-304 (PI: M. Tannock). IGRINS is a high-resolution ($R= 45\,000$), cross-dispersed spectrograph that simultaneously covers the $H$ and $K$ bands from 1.45 to 2.48~$\mu$m.

Observations took place over four nights in April and May 2018 while IGRINS was on Gemini South. The slit was oriented at the default position angle of 90 degrees (east-west) for IGRINS, and exposures were taken along an ABBA dither pattern. We observed an A0 V star before or after each observation of the target at a similar airmass, with the same telescope and instrument configuration. We summarize these observations in Table~\ref{table:observations}.

\begin{table*}
	\centering
	\caption{Gemini South/IGRINS Spectroscopic Observations of 2MASS~J08173001$-$6155158 under Gemini program ID GS-2018A-Q-304 (PI: M. Tannock). The given SNR values are for the final, combined spectra.
	The FWHM of the trace includes both atmospheric seeing and effects from telescope and instrument optics. Typical atmospheric seeing at Gemini South is 0.5\arcsec.}
	\label{table:observations}
	\begin{tabular}{lcccccccc} 
		\hline
		Date & Exposure & Exposure & Target & Telluric & Telluric & $H$-band & $K$-band & FWHM of the \\
		Observed & Time & Sequence & Airmass & A0 V & Standard & SNR (at & SNR (at & Trace in the \\
		 & (s) & & & Standard & Airmass & 1.589 $\mu$m) & 2.101 $\mu$m) & $H$ Band (\arcsec ) \\
		\hline
		2018 Apr 5 & 1200 & AB & 1.18--1.20 & HIP 40621 & 1.14 & 85 & 44 & 0.9 \\ 
        2018 Apr 5 & 600 & ABBA & 1.20--1.25 & HIP 35393 & 1.28 & 63 & 34 & 0.9 \\ 
        2018 May 7 & 600 & ABBA & 1.22--1.28 & HIP 36489 & 1.31 & 56 & 33 & 0.8 \\ 
        2018 May 22 & 518 & ABBAAB & 1.27--1.37 & HIP 40621 & 1.36 & 174 & 112 & 0.6 \\ 
        2018 May 22 & 518 & ABBAAB & 1.40--1.58 & HIP 40621 & 1.58 & 181 & 113 & 0.6 \\ 
        2018 May 23 & 518 & ABBAAB & 1.26--1.37 & HIP 40621 & 1.35 & 184 & 119 & 0.6 \\ 
		\hline
	\end{tabular}
\end{table*}


\subsection{Data Reduction} 
\label{sec:reductions}

The data were reduced with Version 2 of the IGRINS Pipeline Package (PLP; \citealt{plp}), at each epoch individually. The PLP performs sky subtraction, flat-fielding, bad-pixel correction, aperture extraction, wavelength calibration, and telluric correction. The PLP outputs wavelength-calibrated, telluric-corrected fluxes and the signal-to-noise ratio (SNR) for each point in the spectrum. For each observing epoch, the wavelength solution was derived from a combination of OH emission lines and telluric absorption lines. OH emission lines in the observed spectrum were removed through A–B pair subtraction, and telluric absorption lines were removed by dividing the target spectrum by an A0 V standard spectrum. The target spectrum was also multiplied by a standard Vega model to remove any features from the A0 V standard itself.

We found that the strongest telluric features were not completely removed in our data reduction, and left residuals that affected the chi square statistic ($\chi^2$) when comparing to models (Section~\ref{sec:modelscomparison}). 
To identify and mask strong telluric features, we generated transmission spectra for the Earth's atmosphere with the Planetary Spectrum Generator (PSG; \citealt{PSG})\footnote{\url{https://psg.gsfc.nasa.gov/}}. 
We used the Earth's Transmittance template with the longitude, latitude, and altitude of the Gemini South Observatory. We found that masking atmospheric lines with $>$ 65 per cent absorption strengths in the PSG Earth transmittance spectrum, along with strong OH emission lines (identified from the atlas of \citealt{Rousselot2000}) significantly improved the quality of our model photosphere fits. 
The PSG spectra and the 65 per cent absorption threshold beyond which we masked features are shown in the Figures of the Appendix.

We used a custom IDL code to combine the individual spectra. We first corrected for the barycentric velocity at each epoch. We then processed the $H$ and $K$ bands separately: we normalized the flux to peak at unity in each of the $H$ and $K$ bands, and then resampled the data to identical wavelength values. We computed the weights from the SNR values computed by the PLP ($w_i = (SNR_i / f_i)^2$), where $f_i$ is the flux at each epoch) and computed the weighted average ($\bar{f} = \sum_{i=1}^{N} (f_i w_i / w_t)$) where $w_t$ is the sum of the weights for $N$ epochs) and uncertainties ($\sigma = w_t^{1/2}$) across all epochs.

\begin{figure*}
	\includegraphics[width=\textwidth]{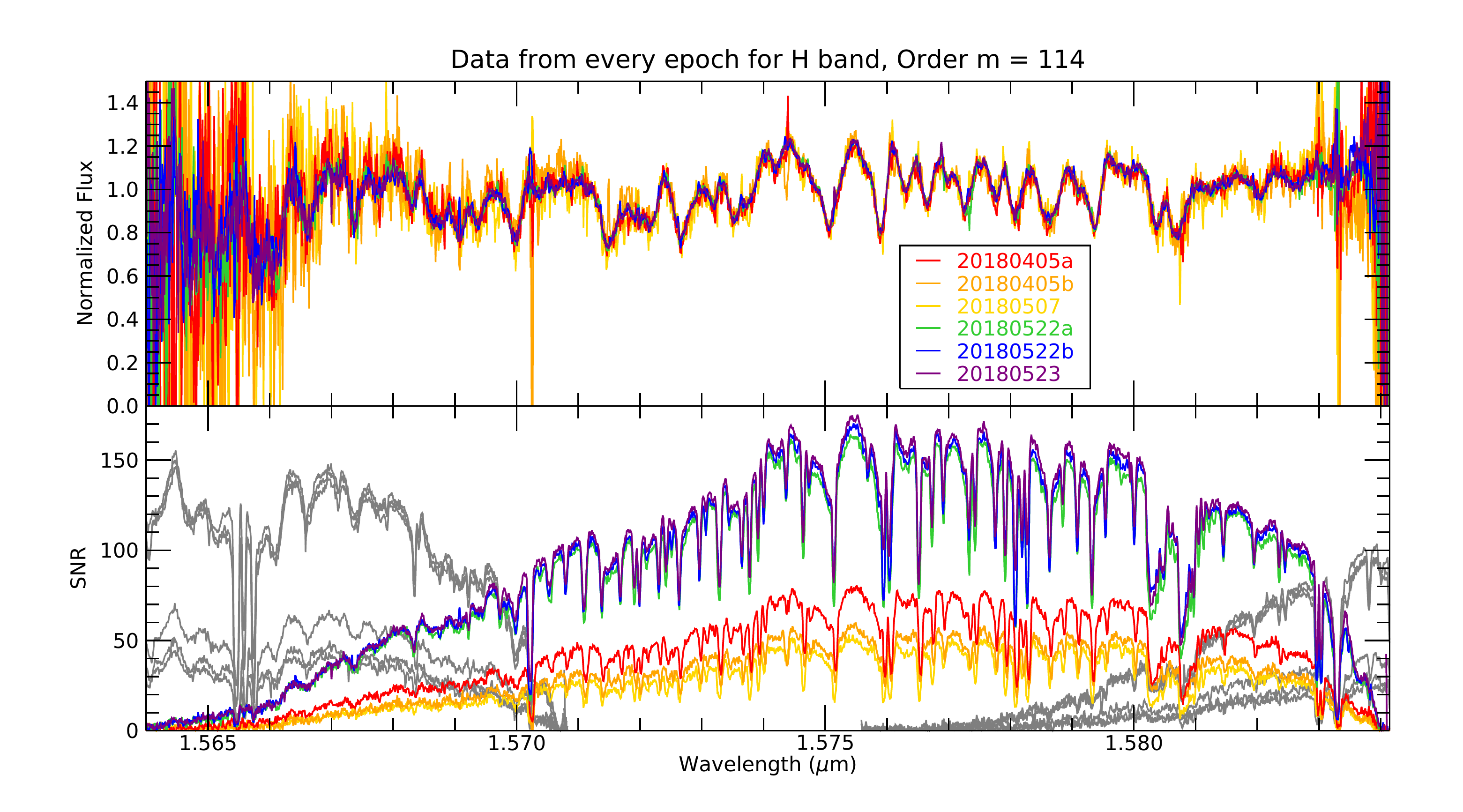}
    \caption{ A sample order near the centre of the $H$ band, showing spectra from each of the six observing epochs. The normalized flux is shown in the top panel, and the deep absorption features in this order are due to H$_2$O. The SNR is shown in the bottom panel. The SNR of the neighbouring orders are also shown in grey, to show that IGRINS has good SNR coverage at all wavelengths. The IGRINS instrument transmission profile (blaze) is imprinted on the SNR spectrum, and is the reason for the fall-off in SNR at the edges of the order. The three highest-SNR spectra, obtained on 2018 May 22 and 23, were combined to create the final spectrum shown in Fig.~\ref{fig:HKstitched}. }
    \label{fig:allepochs}
\end{figure*}

\begin{figure*}
	\includegraphics[width=\textwidth]{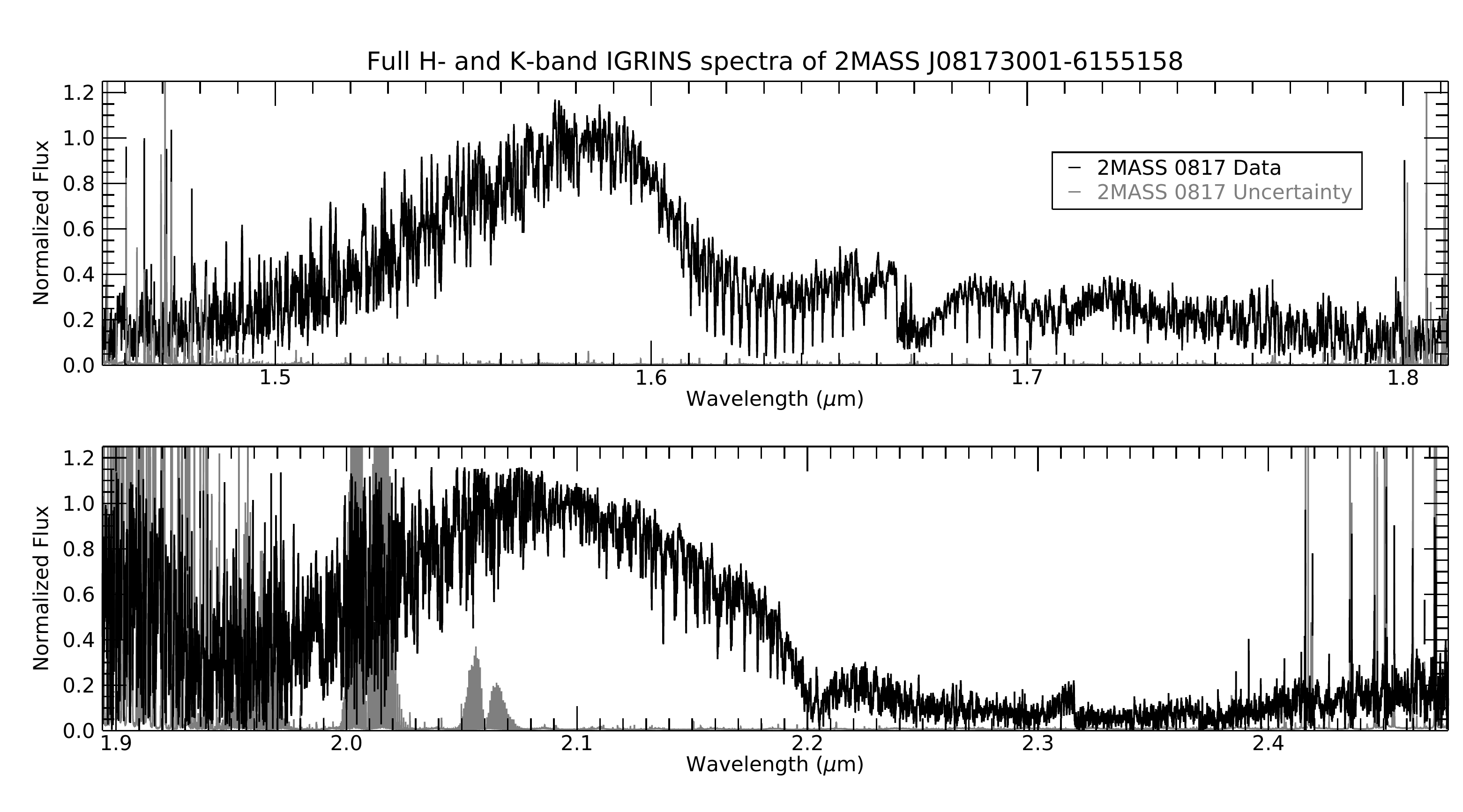}
    \caption{The full $H$- and $K$-band IGRINS spectra of 2MASS~J08173001$-$6155158 with epochs combined and the orders stitched together. This figure does not include the quadratic correction described in Section~\ref{sec:fittingmodels}. These data appear noisy, but in fact have SNR $\simeq$ 300 at the peak of the $H$-band spectrum and SNR $\simeq$ 200 at the $K$-band peak. The apparent noise spikes are all absorption features, and can be seen in detail in the full set of figures in the Appendix.
    }
    \label{fig:HKstitched}
\end{figure*}

We found that in some cases, the IGRINS PLP produced fluxes of $\sim$0, but with disproportionately high SNR values, resulting in large weights.
This produced large downward spikes in the weighted average spectrum. We obtained the highest SNR combined spectrum free of such spikes when we combined the three highest SNR epochs: 2018 May 22 (both sequences) and 2018 May 23. In Fig.~\ref{fig:allepochs} we show the data from each epoch in an order at the centre of the $H$ band. Three of the nights stand out with their higher SNR. We performed the remainder of our analysis with the weighted average of these three epochs. Our final combined spectrum (Fig.~\ref{fig:HKstitched}) had a signal-to-noise of $\sim$300 at the peak of the $H$ band and $\sim$200 at the peak of the $K$ band.

There is some overlap between the orders in the spectrum (see Table~\ref{table:orderswavelengths} for a list of the orders and their wavelength coverage). For our analysis, we analysed each order individually. The instrument blaze profile results in the short-wavelength ends of the order having lower SNR than the long-wavelength ends (see the bottom panel of Fig.~\ref{fig:allepochs}). We show the complete spectra with the orders stitched together in Fig.~\ref{fig:HKstitched}. For this stitched spectrum, in each region of overlap, we averaged the fluxes from the two orders.


\subsection{Confirmation of Wavelength Calibration}
\label{sec:wavelengthcal}

We verified our wavelength calibration by comparing the telluric lines in the spectra of our A0 V standard stars to the Earth's transmittance spectrum from the PSG, generated over the wavelength coverage of IGRINS at 1.5 times the resolution of IGRINS. 
In each IGRINS order, between five and ten lines spread over the order were selected (very deep lines and blended lines were avoided), and we measured the wavelength at the minimum flux for each of these lines. We found an average offset of less than half an IGRINS pixel (0.110~\AA~at the centre of the $H$ band), confirming our wavelength calibration.

\section{Model Fitting and Parameter Determination}
\label{sec:modelscomparison}

We compared our observed spectra to the models of \citet[hereafter, BT-Settl]{Allard2012,Allard2014}, \citet[hereafter, Morley]{Morley2012}, \citet[hereafter, Sonora Bobcat]{Marley2021}, and an alternative version of the Sonora Bobcat models with updated molecular line lists from Hood et al. (in preparation, hereafter, Bobcat Alternative A). The BT-Settl models are based on the PHOENIX code \citep{Allard1995,Hauschildt1999}. The latter three model sets are all based on the same 1D radiative-convective equilibrium model atmosphere code (e.g., \citealt{Marley1996,Fortney2008,Marley2015}). 
The Morley models include the effect of clouds that may be relevant for T dwarf atmospheres by applying the \citet{Ackerman2001} cloud model. In contrast, the Sonora Bobcat models assume a cloud-free atmosphere. The Sonora Bobcat models include post-2012 updates to the gas opacity database, described in \citet{Freedman2014}, \citet{Lupu2014}, and \citet{Marley2021}. 
The Bobcat Alternative A models are thermal emission spectra generated from the Sonora Bobcat atmospheric structures with the code described in the Appendix of \citet{Morley2015}. Only a selection of opacities, that dominate at near infrared wavelengths, are included: H$_2$O, CH$_4$, CO, NH$_3$, H$_2$S, and collision-induced opacity of H$_2$-H$_2$, H$_2$-He, and H$_2$-CH$_4$. The opacity data for these sources are the same as the Sonora Bobcat models, with the notable exceptions of updated H$_2$O \citep{Polyansky2018}, CH$_4$ \citep{Hargreaves2020}, and NH$_3$ \citep{Coles2019} line lists.

\subsection{Fitting of Photospheric Models}
\label{sec:fittingmodels}

The models are provided on fixed grids of effective temperature ($T_{\rm eff}$) and surface gravity ($\log g$, with $g$ in units of cm s$^{-2}$), and we do not interpolate between models to intermediate values. We allowed our model fitting to explore $T_{\rm eff}$ grids between 700~K and 1300~K (the expected range of 900--1100~K in $T_{\rm eff}$ for a T6 dwarf, $\pm$200~K; \citealt{Filippazzo2015}), in steps of 50~K or 100~K, depending on the model family. For $\log g$ we explored grids between $\log g=4.0$ and 5.5, in steps of 0.5~dex for all model families except the Sonora Bobcat models, which are in steps of 0.25~dex. The Morley models also have a sedimentation efficiency ($f_{\rm sed}$) parameter on a grid from 2 to 5 in integer steps.

We explored a radial velocity (RV) grid by applying a Doppler shift to the wavelength of the models. We also expect that our observed spectrum will have significant rotational broadening from its known axial rotation. We explored a grid of projected rotation velocities ($v \sin i$), by simulating rotational broadening in the model spectra. We convolved the model spectra with the standard rotation kernel from \citet{Gray1992}, as described in \citet{Tannock2021}. For both RV and $v \sin i$ we first explored coarse grids with steps of 2 km\,s$^{-1}$ over a broad range of values, then narrowed our grid and repeated the fitting with finer steps of 0.1 km\,s$^{-1}$. After shifting and broadening the model spectra, we resampled the model spectra to the wavelengths of the observed spectrum using IDL's \texttt{interpol} function.\footnote{We have not broadened the model spectra by the $\approx$6.5~km~s$^{-1}$ line width of the instrument profile. The potential effect on the $v\sin i=22.5\pm0.9$~km~s$^{-1}$ that we ultimately find (Table~\ref{table:parameters}) would be to decrease it by $\approx$0.6~km~s$^{-1}$.}

We also observed an instrumental effect resulting in an upward curving in the residuals when compared to models at the ends of the orders. To minimize this effect and analyse the highest SNR regions of the data, we removed the ends of each order, leaving $\sim$1--2~nm overlap between orders. 
We additionally divided out a quadratic function that minimized the $\chi^2$ statistic between the data and the model to further remove this instrumental effect. With these corrections we obtained the $\chi^2$ values as:
\begin{equation}
\label{eq:chisq}
\chi^2 = \sum_{i=1}^N \left( \frac{O_i/(a\lambda^2 + b\lambda + c) - M_i}{\sqrt{\sigma_i^2 + \sigma_M^2}}\right)^2 ,
\end{equation} 
\noindent where $O_i$ is the observed flux, $M_i$ is the flux of the model, $\sigma_i$ is the uncertainty of the data, $\sigma_M$ is a systematic uncertainty assigned to the model, and $\lambda_i$ is the wavelength of the corresponding data point. The coefficients of the quadratic correction are $a$, $b$, and $c$. Following a similar process to the one described in \citet{Suarez2021}, we set the partial derivatives of Equation~\ref{eq:chisq} to zero and solved the resulting system of equations to find the values of $a$, $b$, and $c$. We determined the coefficients of the quadratic for every model on the model grid individually.

We identified the best-fitting order of the entire spectrum (order $m=85$ of the $K$ band for the Bobcat Alternative A model) and determined the value of $\sigma_M$ that produced a reduced $\chi^2$ statistic of 1.0. The adopted value of $\sigma_M$ was approximately half of the average uncertainty of the data. This systematic uncertainty was added in quadrature to the observational uncertainties in every order. 
As a final measure of the goodness of fit for each order, in our figures we report the $\Delta\chi^2_{\rm reduced}$ with respect to the minimum $\chi^2_{\rm reduced}=1$ value for the best-fitting order. 
That is, for the best-fitting ($m=85$) order the goodness of fit is $\Delta\chi^2_{\rm reduced}=0$, while for orders with poorer fits, the goodness of fit is $\Delta\chi^2_{\rm reduced}=\chi^2_{\rm reduced}-1$.

The total uncertainty, including the systematic uncertainty added in quadrature, is shown in grey in Fig.~\ref{fig:Hcomparemodels} and in all following figures, including in the Appendix. The total uncertainty is still very small in most orders, and appears indistinguishable from the wavelength axis in most figures, as our estimated S/N ratios can be well over 100. Nevertheless, we believe the overall uncertainties to be this small based on the above $\chi^2$ analysis.

\begin{table*}
	\centering
	\caption{The wavelengths of the IGRINS orders and the major molecular absorbers in each order. Diffraction order numbers, $m$, were extrapolated from \citet{Stahl2021}. 
	}
	\label{table:orderswavelengths}
	\begin{tabular}{cccl@{\hskip 0.75in}cccl} 
		\hline
		Band & Order & Wavelength & Major & Band & Order & Wavelength & Major \\
		& ($m$) & Coverage ($\mu$m) & Absorbers & & ($m$) & Coverage ($\mu$m) & Absorbers \\
		\hline
        $H$ & 124	&	1.454--1.460 &	H$_2$O	                & $K$ & 94	&	1.894--1.910	&	H$_2$O	\\          
        $H$ & 123	&	1.459--1.470 &	H$_2$O	                & $K$ & 93	&	1.909--1.930	&	H$_2$O	\\          
        $H$ & 122	&	1.469--1.483 &	H$_2$O	                & $K$ & 92	&	1.929--1.950	&	H$_2$O	\\          
        $H$ & 121	&	1.482--1.494 &	H$_2$O	                & $K$ & 91	&	1.949--1.972	&	H$_2$O, NH$_3$	\\  
        $H$ & 120	&	1.493--1.506 &	H$_2$O	                & $K$ & 90	&	1.971--1.993	&	H$_2$O, NH$_3$	\\  
        $H$ & 119	&	1.504--1.519 &	H$_2$O, NH$_3$	        & $K$ & 89	&	1.992--2.015	&	H$_2$O, NH$_3$	\\  
        $H$ & 118	&	1.517--1.531 &	H$_2$O	                & $K$ & 88	&	2.014--2.038	&	H$_2$O, NH$_3$	\\  
        $H$ & 117	&	1.529--1.543 &	H$_2$O	                & $K$ & 87	&	2.037--2.061	&	H$_2$O, NH$_3$	\\  
        $H$ & 116	&	1.541--1.556 & 	H$_2$O	                & $K$ & 86	&	2.060--2.085	&	H$_2$O, CH$_4$, NH$_3$	\\ 
        $H$ & 115	&	1.554--1.569 &	H$_2$O, CO              & $K$ & 85  &   2.084--2.109	&	H$_2$O, CH$_4$	\\  
        $H$ & 114	&	1.567--1.583 &	H$_2$O                  & $K$ & 84	&	2.108--2.134	&	H$_2$O, CH$_4$, H$_2$	\\ 
        $H$ & 113	&	1.581--1.596 &	H$_2$O, CH$_4$, H$_2$S  & $K$ & 83  &   2.133--2.159	&	H$_2$O, CH$_4$	\\  
        $H$ & 112	&	1.594--1.610 &	H$_2$O, CH$_4$          & $K$ & 82  &.  2.158--2.185	&	CH$_4$	\\          
        $H$ & 111	&	1.608--1.624 &	H$_2$O, CH$_4$	        & $K$ & 81	&	2.184--2.212	&	CH$_4$, NH$_3$	\\  
        $H$ & 110	&	1.622--1.639 &	H$_2$O, CH$_4$	        & $K$ & 80	&	2.211--2.239	&	CH$_4$	\\          
        $H$ & 109	&	1.637--1.653 &	H$_2$O, CH$_4$	        & $K$ & 79	&	2.238--2.267	&	CH$_4$	\\          
        $H$ & 108	&	1.651--1.668 &	H$_2$O, CH$_4$	        & $K$ & 78	&	2.266--2.295	&	H$_2$O, CH$_4$	\\  
        $H$ & 107	&	1.666--1.683 &	H$_2$O, CH$_4$	        & $K$ & 77	&	2.294--2.326	&	H$_2$O, CH$_4$, CO	\\ 
        $H$ & 106	&	1.681--1.699 &	H$_2$O, CH$_4$	        & $K$ & 76	&	2.325--2.355	&	H$_2$O, CH$_4$, CO	\\ 
        $H$ & 105	&	1.697--1.715 &	H$_2$O, CH$_4$	        & $K$ & 75	&	2.354--2.383	&	H$_2$O, CH$_4$, CO	\\ 
        $H$ & 104	&	1.713--1.730 &	H$_2$O, CH$_4$	        & $K$ & 74	&	2.389--2.414	&	H$_2$O, CH$_4$, CO	\\ 
        $H$ & 103	&	1.728--1.747 &	H$_2$O, CH$_4$	        & $K$ & 73	&	2.420--2.445	&	H$_2$O, CH$_4$, CO	\\ 
        $H$ & 102	&	1.745--1.764 &	H$_2$O, CH$_4$          & $K$ & 72	&   2.452--2.478	&	H$_2$O, CH$_4$ \\   
        $H$ & 101	&	1.762--1.781 &	H$_2$O, CH$_4$	\\ 
        $H$ & 100   &	1.779--1.798 &	H$_2$O, CH$_4$	\\ 
        $H$ & 99	&	1.797--1.812 &	H$_2$O, CH$_4$	\\ 
		\hline
	\end{tabular}
\end{table*}


\subsection{Determination of Physical Parameters}
\label{sec:parameters}

\begin{figure*}
	\includegraphics[width=\textwidth]{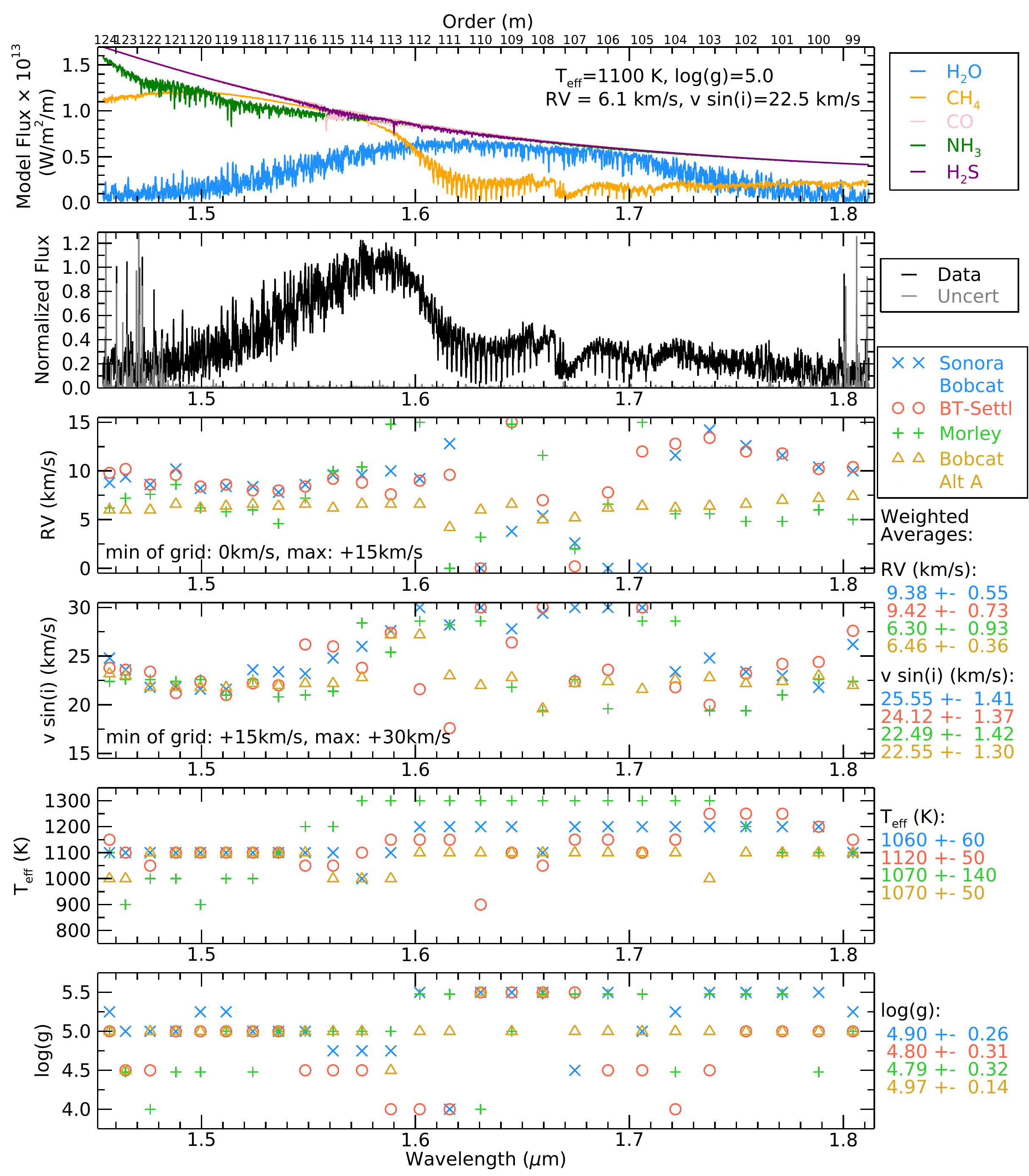}
    \caption{Results of the model fitting for the $H$ band. 
    Top panel: The Bobcat Alternative A model spectra of each major molecule with collision-induced absorption from molecular hydrogen and helium included. The Model Flux (the y-axis of the top panel) is what would be measured at the surface of the object. The models shown in this panel have $T_{\rm eff}=1100$ K, $\log{g}=5.0$ (with $g$ in cm~s$^{-2}$), $v \sin{i}=22.5$ km\,s$^{-1}$, and RV = 6.1 km\,s$^{-1}$, and are also matched to the resolution of the IGRINS data. The IGRINS order numbers ($m$) are given along the top horizontal axis. Second panel from the top: The full $H$-band IGRINS spectrum, with the orders stitched together. 
    Bottom four panels: The parameters of the best-fitting model for each order, from each family of models. From top to bottom the parameters are: RV, $v \sin{i}$, $T_{\rm eff}$, and $\log{g}$. 
    The weighted average of each parameter is given on the right side of the figure. In some cases the best-fitting models are at the maximum and minimum values of the allowed grid, which indicates that these models produce inadequate fits in the particular order.
    These values are still included in the weighted mean, but have very little weight assigned to them due to their large $\chi^2$ statistics. 
    }
    \label{fig:Hcomparemodels}
\end{figure*}

\begin{figure*}
	\includegraphics[width=\textwidth]{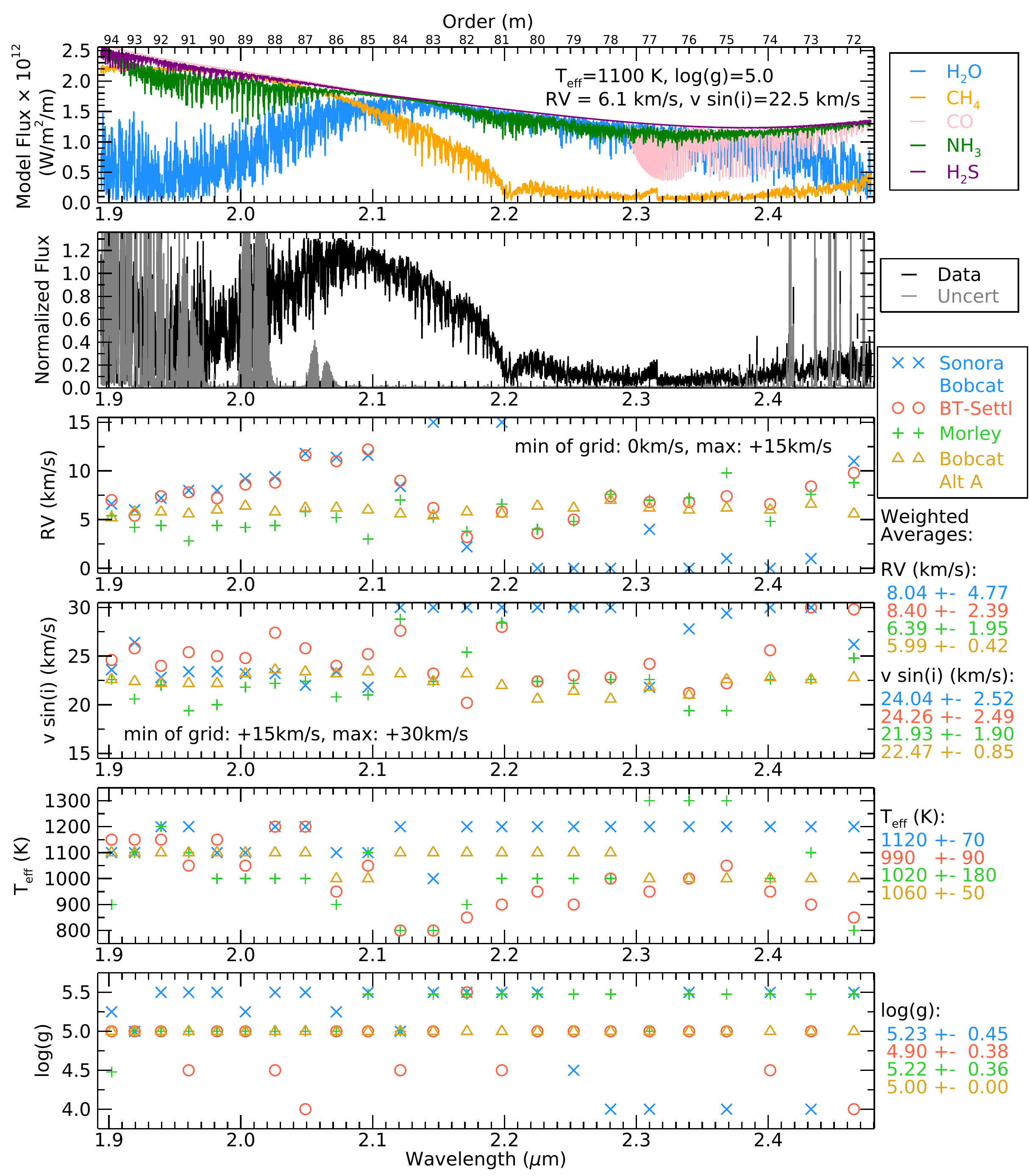}
    \caption{The same layout as Fig.~\ref{fig:Hcomparemodels}, but for the $K$ band. The $\log{g}$ value is extremely consistent for the Bobcat Alternative A models, with $\log{g}$ = 5.0 in every order of the $K$ band. The standard deviation on this weighted average is therefore zero (see Section~\ref{sec:parameters} for details on this calculation). In Table~\ref{table:parameters} we compute the weighted average and standard deviation based on both the $H$ and $K$ bands, so the standard deviation is non-zero for the final adopted value.
    }
    \label{fig:Kcomparemodels}
\end{figure*}

We show the results of the model fitting across all orders for all model families in Fig.~\ref{fig:Hcomparemodels} and \ref{fig:Kcomparemodels}. In the top panel, a Bobcat Alternative A model is used to separate the contribution of each molecular species, in order to identify the dominant molecule or molecules in each order. These `single-molecule models' include a single molecule (e.g., water, methane), plus collision-induced absorption from molecular hydrogen and helium. To help identify particular features and molecules, a panel like this is included at the top of almost all of our figures.

We find that the Bobcat Alternative A models with the updated line lists provide the best fits to the data. 
We adopt the values given by the Bobcat Alternative A models, and present the weighted average of each parameter across all $H$ and $K$ band orders in Table~\ref{table:parameters}. As described in \citet{Tannock2021}, we compute the weighted average and the unbiased weighted sample standard deviation, where the weight is $e^{-\chi^2_{\rm reduced}}$, so that the better fits and more reliable orders are more heavily weighted. The parameter values given in Fig.~\ref{fig:Hcomparemodels} and \ref{fig:Kcomparemodels} are computed in the same way, but for each order separately. 
The Bobcat Alternative A models are the most consistent across all orders, and give the smallest uncertainties on the measured parameters. Overall, all models do fairly well in regions dominated by water, while fits are poor in regions dominated by methane.

For the remainder of our analysis, we will focus on the results of the Bobcat Alternative A models, unless otherwise stated. We show the best fitting Bobcat Alternative A models for all orders of the $H$ and $K$ bands in Fig.~\ref{fig:Horders} and \ref{fig:Korders}, and in the following sections we highlight a few notable orders and regions.

\begin{table}
	\centering
	\caption{Properties of 2MASS~J08173001$-$6155158. Parameters estimated from the spectra presented in this paper are based on all $H$ and $K$ band orders.}
	\label{table:parameters}
	\begin{tabular}{lc} 
		\hline
		Property & Value \\
		\hline
		Spectral Type \tablenotemark{a} & T6  \\ 
        Effective temperature ($T_{\rm eff}$) \tablenotemark{b} & 1060 $\pm$ 50 K  \\
        Surface gravity ($\log{g}$) \tablenotemark{b,c} & 5.0 $\pm$ 0.1 \\
        Projected rotation velocity ($v \sin{i}$) \tablenotemark{b} & 22.5 $\pm$ 0.9 km s$^{-1}$ \\
        Radial velocity (RV) \tablenotemark{b} & 6.1 $\pm$ 0.5 km s$^{-1}$ \\
		\hline
	\end{tabular}
	\tablenotetext{a}{\citet{Artigau2010}.}
	\tablenotetext{b}{This work.}
	\tablenotetext{c}{The units of $g$ are cm s$^{-2}$.}
\end{table}

\section{Molecule-by-Molecule Analysis of the Model Spectra}
\label{sec:detections}

In this section we assess the quality of the fits from each family of models. We examine the parameters determined for each region of the spectrum and what the dominant absorbers are in each region. Water (H$_2$O) and methane (CH$_4$) are the most abundant absorbers in late-type T dwarf spectra \citep{Burgasser2006}. Carbon monoxide (CO) and ammonia (NH$_3$) also play a major role, and hydrogen sulfide (H$_2$S) is the next most abundant absorber. The references for the line lists of the major molecules used in each family of models are listed in Table~\ref{table:linelists}. As 2M0817 is a fairly rapid rotator ($v \sin{i} = 22.5 \pm 0.9$ km s$^{-1}$; Table~\ref{table:parameters}), we see that most lines are in fact blends of the dominant absorbers, most often H$_2$O and CH$_4$. 

\begin{figure*}
	\includegraphics[trim={0cm 4.4cm 0cm 0cm}, clip,width=\textwidth]{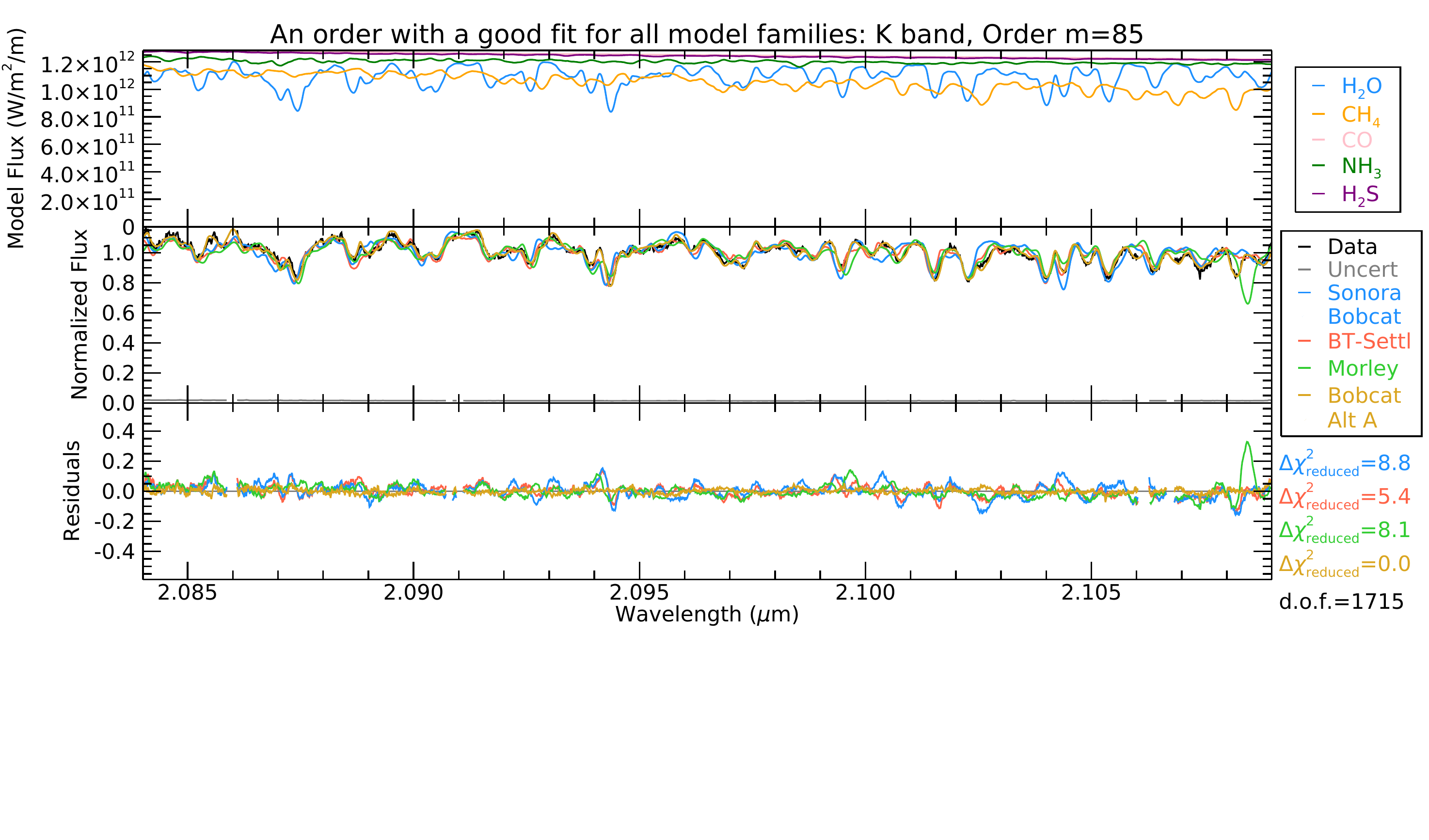} 
    \caption{In order $m=85$ of the $K$ band, 
    all models are very well matched to the data. The dominant absorbers in this region are H$_2$O and CH$_4$. The top panel shows the Bobcat Alternative A model spectra including opacity from one major molecule at a time, in addition to H$_2$/He collision-induced absorption. The middle panel shows the IGRINS data (black; uncertainty shown in grey) with the best fitting models from each model family. The Bobcat Alternative A model spectra in the top panel have the same $T_{\rm eff}$ and $\log{g}$ values as the best fitting Bobcat Alternative A model, are broadened to the same $v \sin{i}$, and have the same RV shift applied. The bottom panel shows the residuals (data - model) on the same vertical scale as the middle panel, with the same colour scheme. The data and residuals contain gaps in the plot where strong telluric lines have been masked out. The $\Delta\chi^2_{\rm reduced}$ statistic is the difference between the $\chi^2_{\rm reduced}$ of the model for the current order and the $\chi^2_{\rm reduced}=1.0$ of the best-fitting model (Bobcat Alternative A model) for order $m=85$. The degrees of freedom (d.o.f.) for the $\chi^2_{\rm reduced}$ for each model are also shown.
    } 
    \label{fig:goodfit}
\end{figure*}

\begin{figure*}
	\includegraphics[trim={0cm 4.4cm 0cm 0cm}, clip,width=\textwidth]{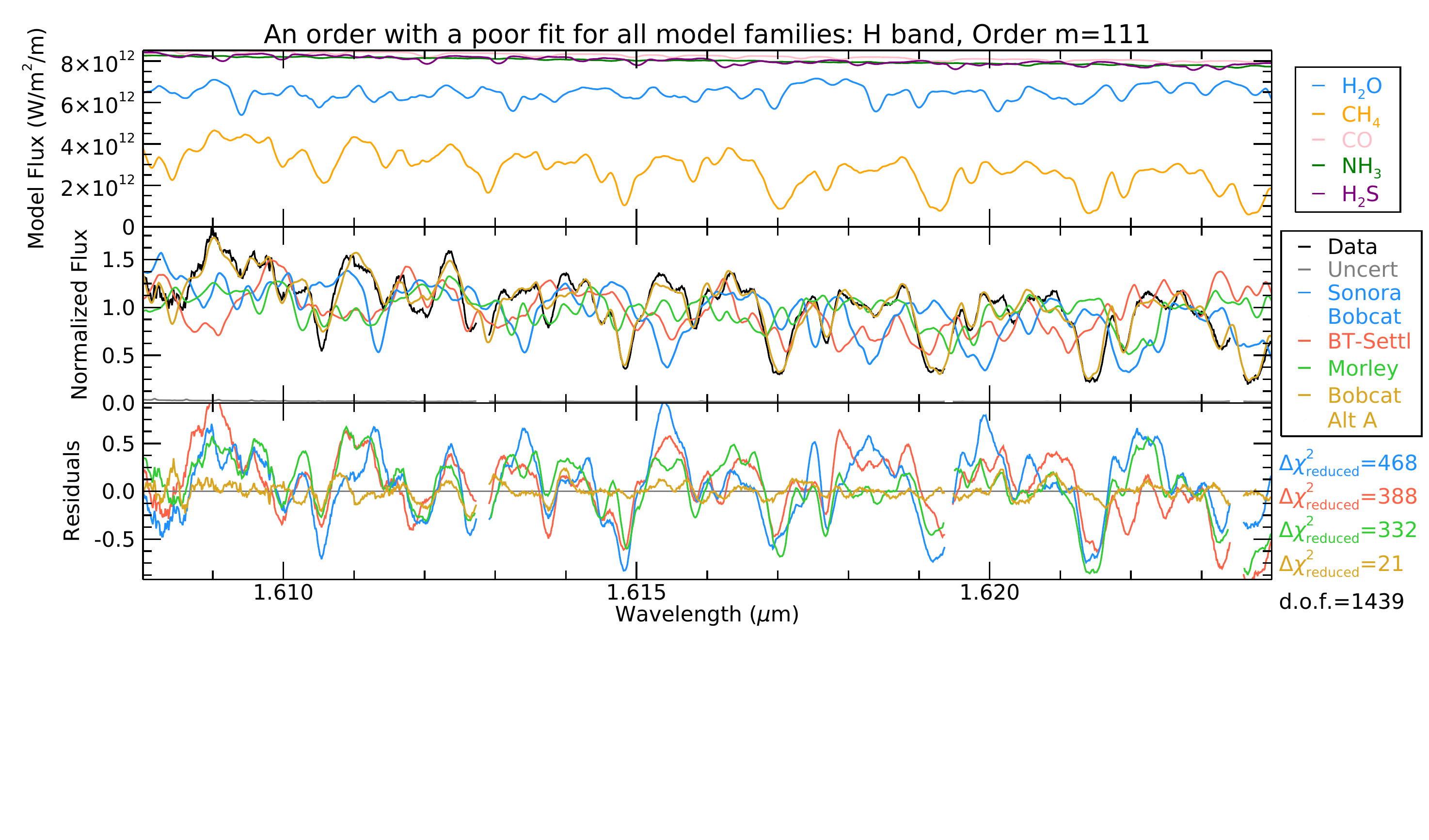} 
    \caption{The same layout as Fig.~\ref{fig:goodfit}, but now showing order $m=111$ of the $H$ band, 
    an order with a poor fit. The dominant absorber in this order is CH$_4$. The most up-to-date line lists for CH$_4$ (\citealt{Hargreaves2020}, used in the Bobcat Alternative A model) provide accurate wavelengths for the deepest lines, but the weaker features in the continuum (likely CH$_4$ blended with H$_2$O) are poorly-fit.}
    \label{fig:badfit}
\end{figure*}

In Fig.~\ref{fig:goodfit} we show order $m=85$ 
of the $K$ band: the order where the models most accurately represent the data. The dominant absorbers in this order are H$_2$O, and CH$_4$. The Bobcat Alternative A model provides the best fit, and the residuals for this model are very flat. The other models also do a fair job in matching the major features. For comparison, in Fig.~\ref{fig:badfit}, we show order $m=111$ 
of the $H$ band: one of the orders where all models provide poor fits. The major absorber in this order is CH$_4$. We see that the locations of the strongest CH$_4$ features are matched in the Bobcat Alternative A model, which has the most up-to-date CH$_4$ line list (Table~\ref{table:linelists}). 
In the following sections, we discuss each molecular absorber separately.

\begin{table*}
	\centering
	\caption{Literature references for the line lists for each of the model photosphere families. For information about specific isotopologues, line widths, and how these sources are combined for each family of models please see the original works listed in the column headers.}
	\label{table:linelists}
	\begin{tabular}{lp{0.2\linewidth}p{0.2\linewidth}p{0.2\linewidth}p{0.2\linewidth}} 
		\hline
	    Molecule & Bobcat Alternative A (Hood et al. in preparation) & Sonora Bobcat \citep{Marley2021} & Morley \citep{Morley2012} & BT-Settl \citep{Allard2012,Allard2014} \\
     \hline
        H$_2$O & ExoMol/POKAZATEL \citep{Polyansky2018}; BT2 \citep{Barber2006} & \citet{Tennyson2018}; BT2 \citep{Barber2006} & \citet{Partridge1997}; HITRAN'08 \citep{Rothman2009} & BT2 \citep{Barber2006} \\
        CH$_4$ & HITEMP \citep{Hargreaves2020} & \citet{Yurchenko2013}; Exomol/10to10 \citep{Yurchenko2014}; Spherical Top Data System \citep{Wenger1998} & Spherical Top Data System \citep{Wenger1998}; HITRAN'08 \citep{Rothman2009}; \citet{Strong1993} & Spherical Top Data System \citep{Wenger1998} \\
        CO & HITEMP 2010 \citep{Rothman2010}; \citet{Li2015} & HITEMP 2010 \citep{Rothman2010}; \citet{Li2015} & \citet{Goorvitch1994}; R. Tipping (1993, private communication); HITRAN'08 \citep{Rothman2009} & \citet{Goorvitch1994} \\
        NH$_3$ & ExoMol/CoYuTe \citep{Coles2019} & BYTe \citep{Yurchenko2011} & BYTe \citep{Yurchenko2011} & \citet{Sharp2007} \\
        H$_2$S & ExoMol \citep{Tennyson2012}; \citet{Azzam2015}; HITRAN 2012 \citep{Rothman2013} & ExoMol \citep{Tennyson2012}; \citet{Azzam2015}; HITRAN 2012 \citep{Rothman2013} & R. Wattson (1996, private communication); HITRAN'08 \citep{Rothman2009} & HITRAN 2004 \citep{Rothman2005} \\
		\hline
	\end{tabular}
\end{table*}

\subsection{Water}
\label{sec:water}

The water-dominated regions of the spectrum provide the most consistent results from fitting models to spectra across all model families (Fig.~\ref{fig:Hcomparemodels} and \ref{fig:Kcomparemodels}). The short-wavelength end of the $H$ band (1.454--1.580~$\mu$m) gives consistent results for each model family, and across the various families. The long-wavelength end of the $H$ band (1.750--1.812~$\mu$m) and the short-wavelength end of the $K$ band (1.894--2.100~$\mu$m) give consistent results within each family of models, but not necessarily across the various model families. 

We note that the Sonora Bobcat and BT-Settl models give higher estimates of the RV, and there is a trend in RV where the RV increases with wavelength (the models are increasingly blue-shifted) in the short-wavelength end of the $K$ band (1.894--2.060~$\mu$m; Fig.~\ref{fig:Kcomparemodels}) for these two models. 
NH$_3$ is also an important absorber in this region but is likely not responsible for this trend in RV because Sonora Bobcat shares the same line lists for ammonia as the Morley models \citep[BYTe]{Yurchenko2011}, and the Morley models do not show this trend. 

The behaviour for the BT-Settl models indicates that the BT2 \citep{Barber2006} H$_2$O line lists, when used alone, are unreliable for RV determinations in this wavelength region. The similar behaviour from Sonora Bobcat indicates that \citet{Tennyson2018}, supplemented with isotopologues from BT2, is also unreliable. The Bobcat Alternative A models use ExoMol/POKAZATEL \citep{Polyansky2018} as the main H$_2$O line list, and also use isotopologue data from BT2. However for this model, we obtain very consistent RV measurements in this wavelength region. The improved accuracy of ExoMol/POKAZATEL line lists appear to make up for any discrepancies in BT2. The HITRAN'08 \citep{Rothman2009} and \citet{Partridge1997} line lists used in the Morley models also give more self-consistent estimates of RV in this region.

Overall we consider water, specifically for the line list used in the Bobcat Alternative A models (ExoMol/POKAZATEL), to be the most reliable molecule for determining the physical parameters of cold brown dwarfs, producing values that we trust.


\subsection{Methane}
\label{sec:methane}

As seen in Fig.~\ref{fig:Hcomparemodels} and \ref{fig:Kcomparemodels}, there is much greater variation in the parameters estimated in the methane-dominated regions (1.60--1.73 $\mu$m in the $H$ band and 2.11--2.40 $\mu$m in the $K$ band) compared to the water-dominated regions, and the $v \sin{i}$ values are particularly discrepant. 
Each family of models uses a different set of line lists for CH$_4$, though there is some overlap between the Sonora Bobcat, Morley, and BT-Settl models which use multiple sources for their CH$_4$ line lists (Table~\ref{table:linelists}). 
Uncertainty has been reported for theoretical CH$_4$ band positions previously: \citet{Canty2015} report offsets between the absorption features in their observed data and the peaks of CH$_4$ opacity from the Exomol/10to10 line list \citep{Yurchenko2014} between 1.615 and 1.710 $\mu$m. 

\begin{figure*}
	\includegraphics[trim={0cm 4.1cm 0cm 0cm}, clip,width=\textwidth]{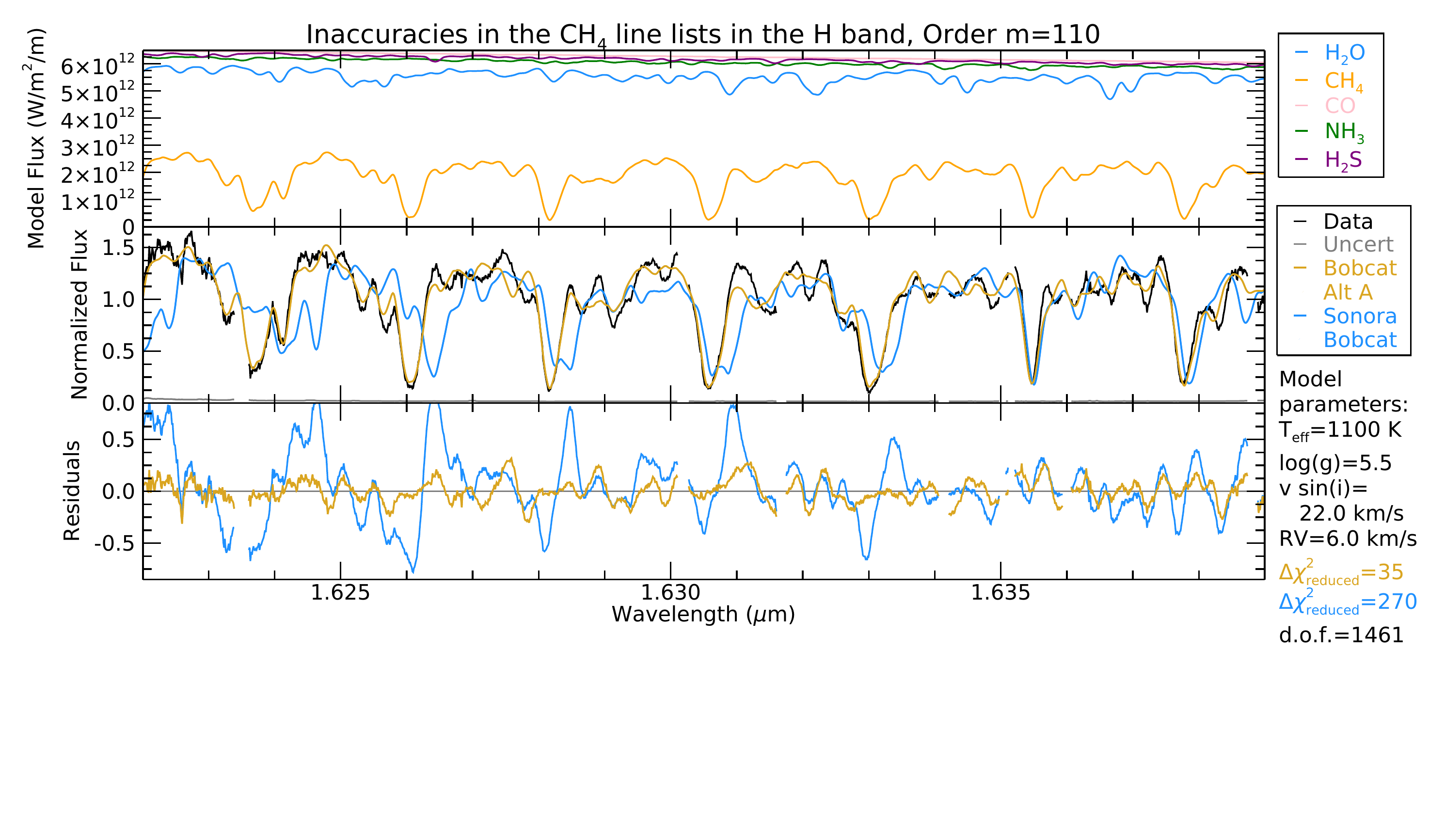}
    \caption{The improvement made with the newer CH$_4$ line lists is most apparent in 
    order $m=110$ of the $H$ band. 
    The top panel is the same as Fig.~\ref{fig:goodfit}: the Bobcat Alternative A model spectra of each major molecule with H$_2$/He collision-induced absorption are shown. 
    Here the middle panel shows the IGRINS data (black), the Sonora Bobcat model (light blue), and the Bobcat Alternative A model (gold). The models have identical physical parameters, rotational broadening ($v \sin{i}$), and RV shift. The bottom panel shows the residuals (data - model) on the same vertical scale, with the same colour scheme. The deepest features are CH$_4$, and the weaker features in the continuum are mainly CH$_4$ or CH$_4$ blended with H$_2$O. The Bobcat Alternative A model shows excellent agreement with the data in the major features, while the Sonora Bobcat model appears to have a stretch causing misalignment in the major features when compared to the IGRINS data.}
    \label{fig:CH4stretch}
\end{figure*}

In Fig.~\ref{fig:CH4stretch} we show a Sonora Bobcat model and a Bobcat Alternative A model with identical physical parameters for an order in the methane region of the $H$ band (order $m=111$ of the $H$ band, 
1.608--1.624 $\mu$m). The CH$_4$ lines used in the Sonora Bobcat models (the same as examined by \citealt{Canty2015}; Table~\ref{table:linelists}) do not match the data well, and appear to have a stretch across this order. Both models poorly fit the weaker lines and the continuum in this region. Radial velocities estimated by the Sonora Bobcat models are discrepant in the methane-dominated regions, due to these inaccurate line positions.
We find significant improvement from the line lists used in the Bobcat Alternative A models (HITEMP, \citealt{Hargreaves2020}) over older models in regions dominated by CH$_4$, in particular in the $H$ band. However, the regions dominated by CH$_4$, even in the Bobcat Alternative A models, still have the most variation in the estimates of the physical parameters. We summarize these regions in Table~\ref{table:mysterylines}, noted as `CH$_4$ regions.' Models using older CH$_4$ line lists should therefore be used with caution. Unaccounted for disequilibrium chemistry may impact these weaker features. We do not explore disequilibrium chemistry for CH$_4$ or H$_2$O in this work, but see Section~\ref{sec:carbonmonoxide} for details on CO disequilibrium chemistry.

Recent theoretical line lists are far more complete than the previously-used laboratory-measured line lists, which are designed to have very accurate line positions but capture fewer lines due to the limits on resolution in laboratory experiments. 
Therefore, theoretical line lists should improve accuracy in regions of the spectrum with weaker bands present, if those bands were unresolved in the laboratory lists. 
A recent improvement in the available line lists has been the combinations of theoretical line lists with laboratory measurements (e.g., HITEMP, \citealt{Hargreaves2020}). Such combination lists provide the best of both worlds, as we show here, where we find a dramatic improvement to high resolution spectroscopic fits.

\subsection{Carbon Monoxide}
\label{sec:carbonmonoxide}

For effective temperatures $\lesssim$ 1300~K (near the L/T transition), the dominant carbon-bearing molecule in the visible part of atmospheres of brown dwarfs switches from CO to CH$_4$ \citep{Fegley1996,Burrows1997}. There are still signatures of CO in the spectra of cold brown dwarfs, and carbon exists abundantly as CO deeper in the atmosphere, where temperatures are higher.

\begin{figure*}
	\includegraphics[trim={0cm 4.4cm 0cm 0cm}, clip,width=\textwidth]{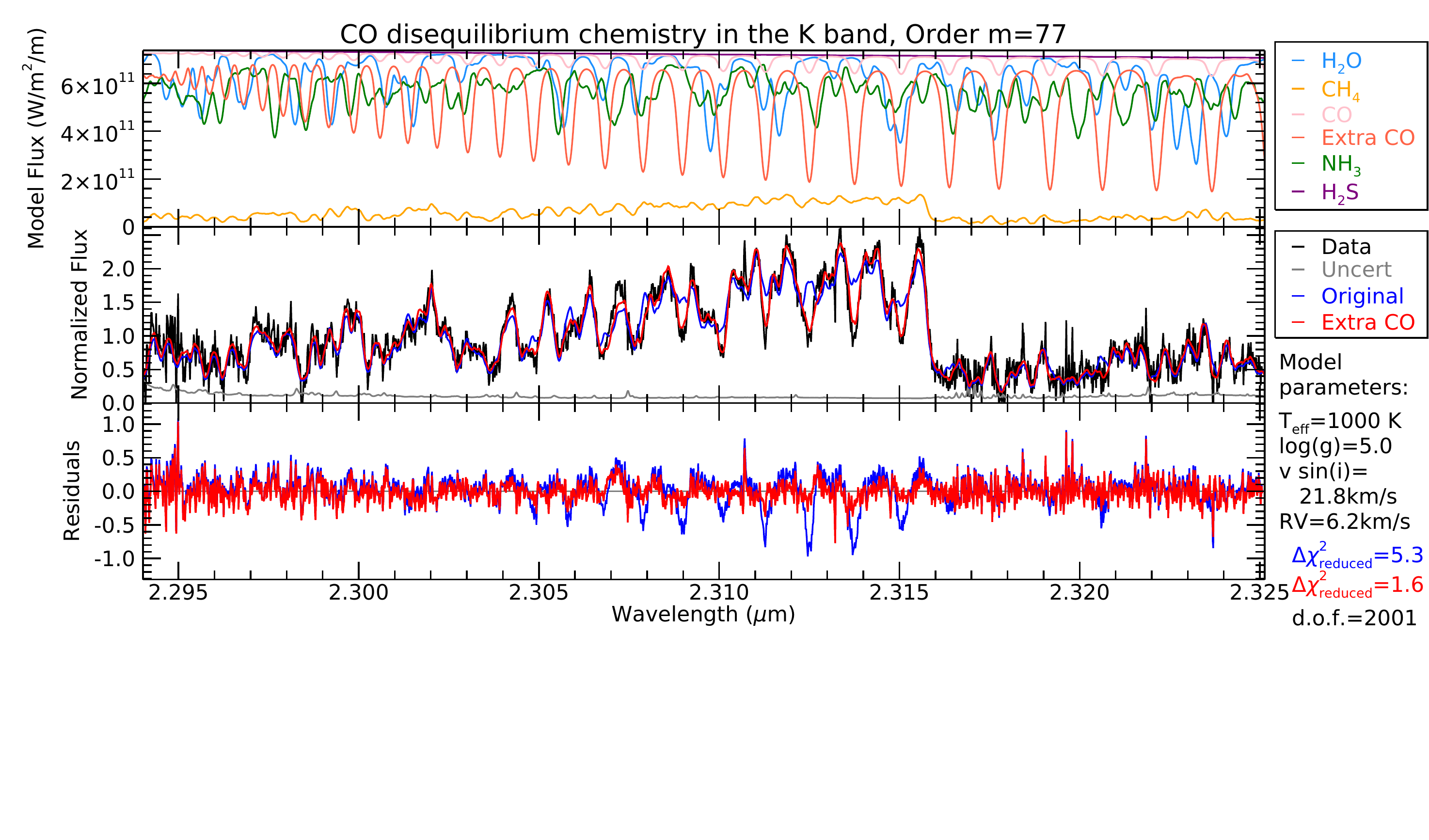} 
    \caption{ Order $m=77$ of the $K$ band 
    is dominated by CH$_4$, but CO and H$_2$O are also important absorbers. Here the top panel is similar to Fig.~\ref{fig:goodfit}, with an extra line for a model with an increased CO abundance (CO volume mixing ratio of $3\times10^{-4}$, labelled `Extra CO'). The middle panel shows the observed spectrum with two versions of a Bobcat Alternative A model, one with the CO as estimated by chemical equilibrium (labelled `Original') and one with an increased abundance of CO (labelled `Extra CO'). The bottom panel shows the residuals for the two models. The CO strength in particular is important to improve the accuracy of the models in the long end of the $K$ band. It is clear that the depth of the features in the `Original' model is too weak at the positions of the CO features, implying that vertical mixing must be taking place in this atmosphere.}
    \label{fig:COstrength}
\end{figure*}

We found that at the CO bands our model fitting selected higher effective temperatures ($T_{\rm eff}\sim$ 1200~K) compared to other orders. Accordingly, we observed several notable features in the residuals of 
orders $m=77$ through $m=73$ of the $K$ band 
(2.294--2.445 $\mu$m), as well as in 
order $m=115$ of the $H$ band 
(1.554--1.569 $\mu$m), where a CO band head is present. The features in the residuals aligned with CO absorption features. We show an example of this in Fig.~\ref{fig:COstrength}, along with a model with increased CO abundance, providing an improved fit. The model with increased CO abundance is also shown in Fig. \ref{fig:Hcomparemodels}, \ref{fig:Kcomparemodels}, and the Appendix figures, and is described in detail below. 

This increased CO abundance implies disequilibrium chemistry, which can occur when vertical mixing occurs in the atmosphere \citep{Lodders2002,Saumon2003}. If CO is being brought from deeper, hotter layers to the upper atmosphere faster than the chemical reaction that converts CO to CH$_4$, there will be more CO in the upper layers of the atmosphere than predicted from chemical equilibrium. 

The Sonora Bobcat and Bobcat Alternative A models use the same cloudless, rainout chemical equilibrium structure models \citep{Marley2021}. These structure models assume chemical equilibrium and give the pressure, temperature, and chemical abundances throughout the atmosphere. To improve our fitting, we generated a small grid of Bobcat Alternative A models with varied amounts of CO, deviating from the chemical equilibrium assumptions used in the Sonora Bobcat structure models. We took a simple approach where we fixed the volume mixing ratio (VMR) for CO to values of $10^{-6}$, $3\times10^{-5}$, $10^{-5}$, $3\times10^{-4}$, $10^{-4}$, $3\times10^{-3}$, and $10^{-3}$. This is a zeroth-order approximation, as 1) the CO VMR is not constant throughout the entire atmosphere, 2) other abundances like CH$_4$ and H$_2$O will also be affected by disequilibrium chemistry, and 3) we are using the temperature-pressure profile from the chemical equilibrium Sonora Bobcat models, but a much higher CO abundance could affect the temperature-pressure profile.

We found a CO VMR of $3\times10^{-4}$ provided the best fits to our data. Fig.~\ref{fig:COstrength} shows a comparison of the original equilibrium chemistry model to the model with this fixed CO VMR value (Fig.~\ref{fig:Hcomparemodels}, \ref{fig:Kcomparemodels}, and the Appendix figures also show models with this fixed CO VMR). In equilibrium models, the CO VMR ranges from $10^{-7}$ to $2.5\times10^{-4}$ for pressures probed by the $K$ band. 
The CO VMR value of the best fitting model is slightly higher than the range in values expected for equilibrium, and explains why our initial fitting selected models with higher effective temperatures, as the CO abundance would be higher in the hotter models. 

Following the method in Section 6.1 of \citet{Miles2020}, we use the quench pressure to estimate the eddy diffusion coefficient ($\log{K_{zz}}$, where $K_{zz}$ has units of cm$^2$ s$^{-1}$). We find $\log{K_{zz}=6.4}$ for 2M0817. This value is in line with the range of relatively low inferred $K_{zz}$ values ($\sim$100 $\times$ lower than expected from mixing length theory of convection) from \citet{Miles2020} for colder ($T_{\rm eff} \leq 750$~K) brown dwarfs. Miles et al. attribute the low $K_{zz}$ values to quenching in radiative regions of the atmosphere, where mixing is likely more sluggish than in convective regions. Interestingly, we find the same behaviour here at $\sim$ 300~K hotter $T_{\rm eff}$. The Sonora model pressure-temperature profile is completely radiative down past the quench pressure of 20 bars, to a pressure of 30 bars, where the atmosphere transitions to the deep convective region.

Disequilibrium chemistry for CO has been observed spectroscopically and inferred photometrically in many other late-type T dwarfs and Y dwarfs (e.g.,~\citealt{Noll1997,Oppenheimer1998,Golimowski2004,Geballe2009,Leggett2012,Sorahana2012,Miles2020}), and has been known in Jupiter for decades \citep{Prinn1977, Noll1988}. The growing number of T and Y dwarfs with evidence for CO disequilibrium chemistry indicates that vertical mixing is an important factor in accurately modelling brown dwarf spectra even at cold temperatures.

\subsection{Ammonia}
\label{sec:ammonia}

Water and methane are the dominant absorbers in the spectra of late-type T dwarfs, but ammonia is important too, especially at T$<$700~K (the coldest T dwarfs and Y dwarfs), where it becomes the dominant nitrogen-bearing molecule \citep{Lodders2002}. Ammonia is of special significance as it is the defining species in the spectra of Y dwarfs \citep{Cushing2011}. 

The choice of an ammonia line list (among published lists) does not appear to significantly impact the physical parameters derived by comparing to models, but ammonia lines are clearly present in the observed spectrum and are important to include in the models. We are able to detect ammonia clearly in several regions of our spectrum.
 
This T6 dwarf joins the handful of T dwarfs with confirmed NH$_3$ detections in the near-infrared. 
\citet{Saumon2000} find evidence for NH$_3$ in the $H$- and $K$-band spectra of Gliese 229B (spectral type T6.5p, $T_{\rm eff} \sim 950$~K) and \citet{Canty2015} report the detection of several NH$_3$ absorption features in the $H$ and $K$ bands in a T8 and T9 dwarf. \citet{Bochanski2011} additionally report detections of NH$_3$ in a T9 dwarf, however, \citet{Saumon2012} question whether some of those detections are indeed attributable to NH$_3$. \citet{Saumon2012} do confirm the stronger NH$_3$ features at $\sim$2 $\mu$m in the spectrum of \citet{Bochanski2011}. We re-confirm the strongest NH$_3$ identified in these works, but some of the weaker lines identified in these later spectral types do not appear in our warmer T6 dwarf.

\begin{figure*}
	\includegraphics[trim={0cm 4.4cm 0cm 0cm}, clip,width=\textwidth]{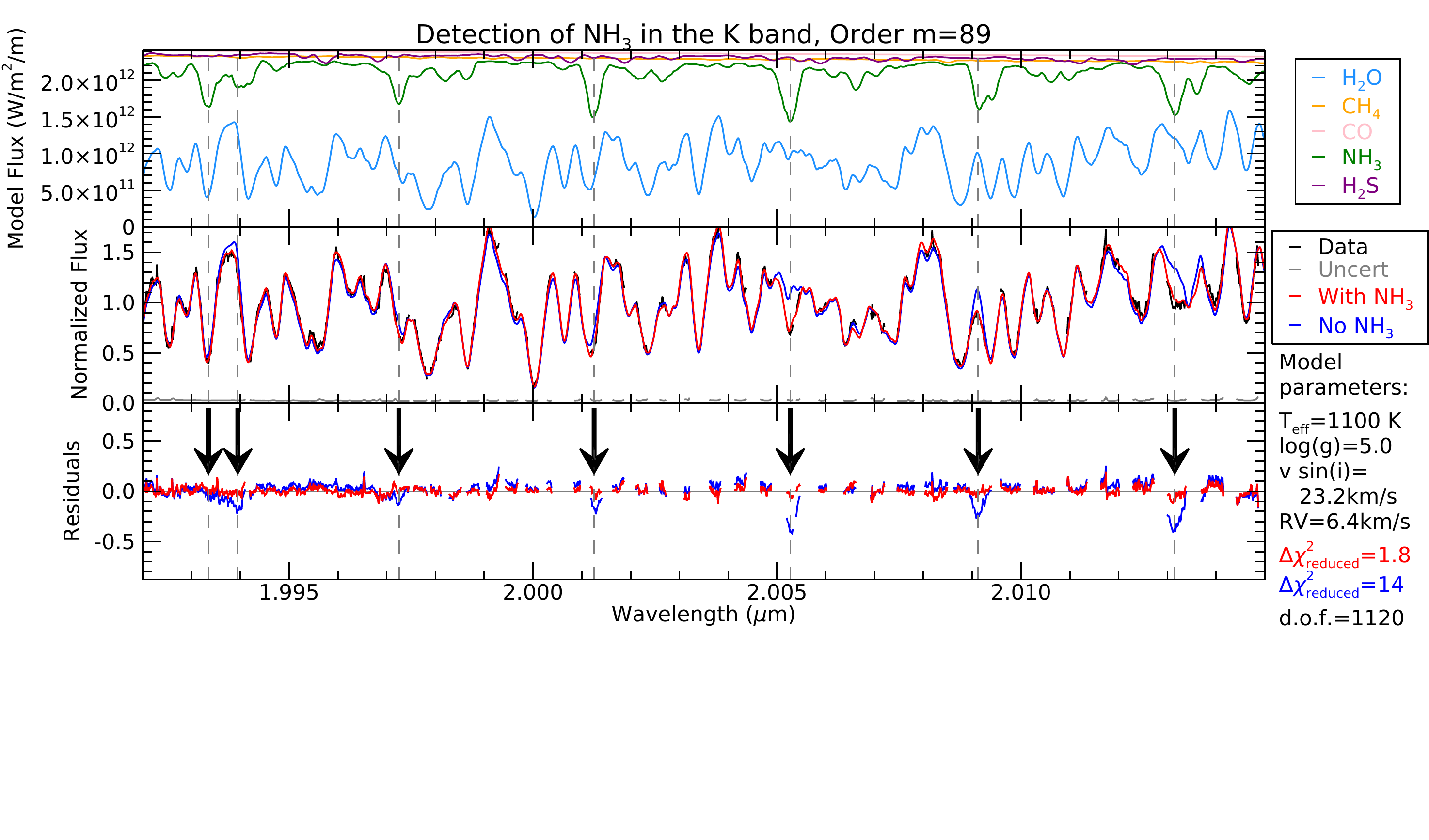}
    \caption{The same layout as Fig.~\ref{fig:COstrength}, but showing Bobcat Alternative A models with and without NH$_3$. Arrows indicate where the model without NH$_3$ deviates from the data. The NH$_3$ lines are blended with stronger H$_2$O lines, but we see significant improvement in the $\chi^2_{\rm reduced}$ values when NH$_3$ is included in the model. 
    Order $m=89$ of the $K$ band 
    has many strong telluric lines, but is still well fit by the models. It is difficult to discern the data from the model containing NH$_3$, and the quality of the fit is reflect in the flat residuals and low $\chi^2_{\rm reduced}$ value.
    }
    \label{fig:NH3detection}
\end{figure*}

\citet{Cushing2021} indicate NH$_3$ features should be present in the infrared at 1.03, 1.21, 1.31, 1.51, 1.66, 1.98, and 2.26 $\mu$m, but would be blended with stronger H$_2$O and CH$_4$ lines making them difficult to detect. While the features at 1.03, 1.21, and 1.31 $\mu$m are outside of our wavelength coverage, we do have clear detections of NH$_3$ at 1.51, 1.98, and 2.26 $\mu$m using the Bobcat Alternative A models. The ammonia lines in our observed spectra are indeed blended with stronger H$_2$O lines, but we are able to detect them nonetheless. We compared Bobcat Alternative A models, with and without NH$_3$, and the presence of the NH$_3$ is clear in the comb-like residuals of Fig.~\ref{fig:NH3detection}. We also see significant improvement in the reduced $\chi^2$ statistic when NH$_3$ is included in the model. We find that the NH$_3$ at 1.66 $\mu$m is far too weak to detect amongst the much stronger H$_2$O and CH$_4$ features in this region for an object of this temperature. 
While NH$_3$ has been detected in early T dwarfs in the mid-infrared \citep{Roellig2004,Cushing2006}, 2M0817 is the warmest brown dwarf with individual NH$_3$ lines detected in the near-infrared.

More recently, \citet{Line2015,Line2017} and \citet{Zalesky2019} constrained the NH$_3$ abundance for multiple cold brown dwarfs (spectral types T7 and later, including several Y dwarfs) with low-resolution ($R<300$ with IRTF/SpeX and HST/WFC3) retrievals. These studies are sensitive to how NH$_3$ opacities influence the spectroscopic appearance of cold brown dwarfs, but the low-resolution of the observations prevents identification of individual NH$_3$ lines in the spectra.

\subsection{Hydrogen Sulfide}
\label{sec:hydrogensulfide}

\begin{figure*}
	\includegraphics[trim={0cm 4.4cm 0cm 0cm}, clip,width=\textwidth]{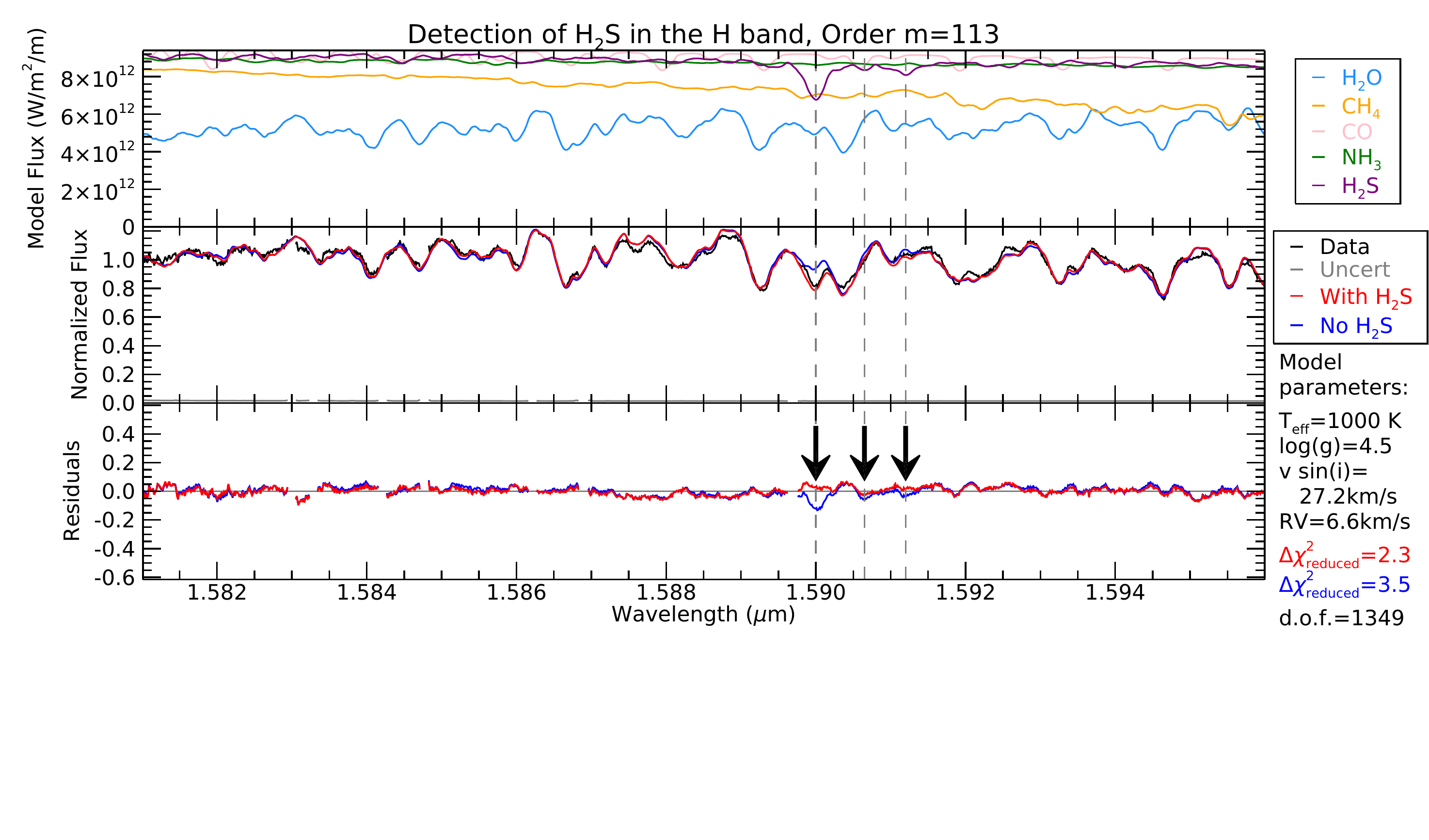}
    \caption{The same layout as Fig.~\ref{fig:NH3detection}, but showing Bobcat Alternative A models with and without H$_2$S. This order shows a clear H$_2$S detection at 1.5900 $\mu$m. 
    Order $m=113$ of the $H$ band 
    is well fit by models. The H$_2$S line of interest is blended with an H$_2$O line, but we see the impact of H$_2$S in the residuals, and improvement in the $\chi^2_{\rm reduced}$ value for the model including H$_2$S.
    }
    \label{fig:H2Sdetection}
\end{figure*}

We present clear, unambiguous detections of H$_2$S in 2M0817. Our most notable detection is a feature at 1.5900~$\mu$m. This feature is blended with a weak H$_2$O line at the same position, so we show our data compared to Bobcat Alternative A models with and without H$_2$S in Fig.~\ref{fig:H2Sdetection}. We see the clear signature of this H$_2$S line in the residuals, as well as the presence of other weaker H$_2$S lines nearby at 1.5906~$\mu$m and 1.5912~$\mu$m.

There is only one other report of a possible H$_2$S detection in a brown dwarf in the literature. \citet{Saumon2000} note an H$_2$S absorption feature at 2.1084 $\mu$m in the spectrum of Gliese 229B (spectral type T6.5p). However, we do not confirm this line in our data, nor do our updated models predict any H$_2$S lines at this position. 

H$_2$S has been identified in the giant planets of our Solar System: \citet{Irwin2018} detect H$_2$S in the atmosphere of Uranus and \citet{Irwin2019} present a tentative detection of H$_2$S in Neptune, both in the 1.57--1.59 $\mu$m region, the same region in which we have our clearest detection. Detections of H$_2$S in Jupiter have also been debated \citep{Noll1995,Niemann1998}. Our spectrum of 2M0817 exhibits the only convincing detection of H$_2$S in an extra-solar atmosphere to date.

We estimated the column density of H$_2$S ($N_{\rm H_2S}$) in the atmosphere of 2M0817 to compare against the \citet{Irwin2018,Irwin2019} estimates for Uranus and Neptune.
We first determined the pressure ($P$) corresponding to the brightness temperature ($T_b$) at the centre of the strongest H$_2$S line in the Bobcat Alternative A model spectrum ($P=14$ bars and $T_b=1420$ K). Then, for each layer in the model atmosphere, we calculated the local number density of all gas molecules using the ideal gas law. We obtained the local number density of H$_2$S specifically by multiplying with the H$_2$S VMR (H$_2$S has an approximately constant equilibrium VMR of  of $2.5\times10^{-5}$ throughout the Bobcat Alternative A model atmosphere).
Integrating this H$_2$S number density from the pressure of the absorbing layer to the top of the atmosphere gives us the column density of H$_2$S, $N_{\rm H_2S} \sim 7.7 \times 10^{20}$ cm$^{-2}$. 
This value is 3 to 130 times higher than column amounts determined from retrievals for solar system planets: $N_{\rm H_2S}$ varies between $6 \times 10^{18}$ to $4.9 \times 10^{19}$ molecules per cm$^{2}$ across the disc of Uranus \citep{Irwin2018}, and $9 \times 10^{18}$ to $2.8 \times 10^{20}$ molecules per cm$^{2}$ across the disc of Neptune \citep{Irwin2019}.

The detection of H$_2$S also offers tentative evidence of iron rain-out in the atmosphere of 2M0817. Below temperatures of 2300~K, iron is predominantly in the form of metallic droplets that settle to deeper atmospheric layers \citep{Fegley1994,Burrows1999}. At temperatures $\lesssim$ 750~K, any iron in the atmosphere would take the form of FeS, leaving no sulphur to form H$_2$S, meaning that H$_2$S would be absent from spectra \citep{Fegley1994,Burrows2001,Lodders2006}. The surface chemical reactions for the conversion of FeS to solid iron and H$_2$S are described in \citet{Helling2019}. We detect H$_2$S in the atmosphere of 2M0817, so iron must be in the process of raining out. Rain-out chemistry is indeed assumed in the Sonora Bobcat and Bobcat Alternative A models.

\subsection{Molecular Hydrogen}
\label{sec:h2}

\begin{figure*}
	\includegraphics[trim={0cm 4.4cm 0cm 0cm}, clip,width=\textwidth]{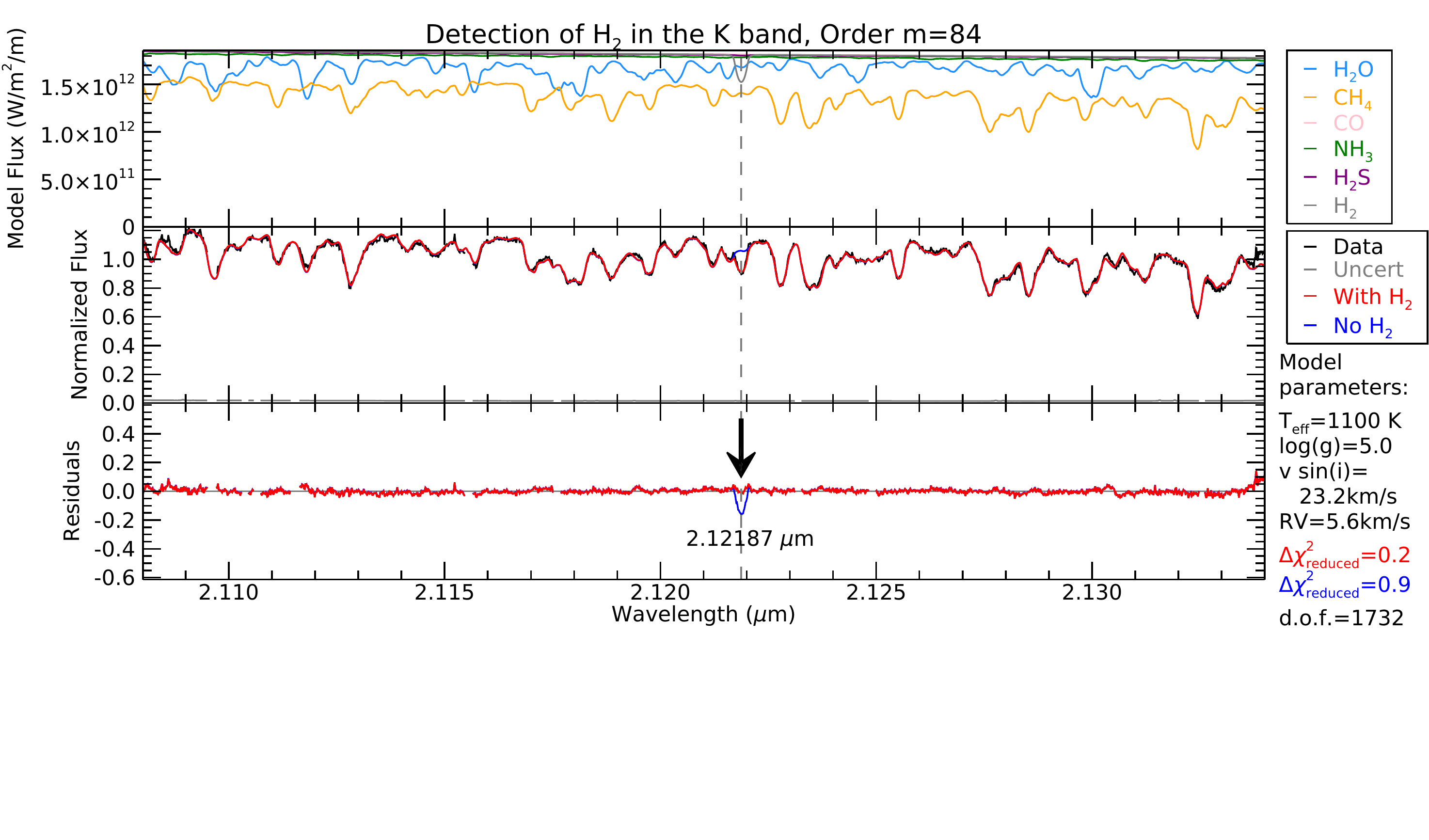}
    \caption{The same layout as Fig.~\ref{fig:NH3detection}, but showing Bobcat Alternative A models with and without H$_2$ absorption (collision-induced absorption from molecular hydrogen and helium is included in both models). This order shows a clear H$_2$ absorption feature at 2.12187 $\mu$m, indicated with a black arrow. 
    Order $m=84$ of the $K$ band 
    is very well fit by models, and we see significant improvement in the $\chi^2_{\rm reduced}$ value for the model including H$_2$ absorption.
    }
    \label{fig:missingline}
\end{figure*}

In order $m=84$ of the $K$ band, 
we identified an absorption feature at 2.12187~$\mu$m that does not appear in any model of any family. This line is indicated by a black arrow in Fig.~\ref{fig:missingline}. 
We first identified this line through our model deviation analysis (Sec.~\ref{sec:mysterylines}), and after correcting for 2M0817's radial velocity (6.1 $\pm$ 0.5 km/s, Table~\ref{table:parameters}), we found the position of this line to be 2.12183 $\pm$ 0.00005 $\mu$m (vacuum wavelength). This matches the 2.121834 $\mu$m wavelength of the molecular hydrogen (H$_2$) 1-0 S(1) transition to six significant figures \citep{Scoville1983, Roueff2019}. 
The H$_2$ 1-0 S(1) feature is among the strongest H$_2$ lines when present in emission in photon-dominated regions, shocks, planetary nebulae, young stellar objects, and starburst galaxies (e.g., \citealt{Habart2005}). 
It has never before been detected in absorption in an extra-solar atmosphere, although it has been seen in Jupiter, Saturn, and Neptune \citep{Kim1995,Trafton1997}.

We added the HITRAN 2016 \citep{Gordon2017} line list for H$_2$ to the  Bobcat Alternative A model and found excellent agreement between order $m=84$ of our observed spectrum and the model. 
As we did for our NH$_3$ and H$_2$S identification, we present models with and without H$_2$ in Fig.~\ref{fig:missingline}, and we see the clear signature of the H$_2$ line in the residuals. We investigated each of the other wavelength regions where the model showed strong H$_2$ absorption features, in particular  
the 1-0 S(3) transition (1.957559 $\mu$m), the 1-0 S(2) transition (2.033758 $\mu$m), 
and the strongest line of the $Q$-branch: the 1-0 Q(1) transition (2.406592 $\mu$m;  \citealt{Gautier1976, Roueff2019}). 
We were unable to detect additional H$_2$ features, due to either the much stronger absorption by other molecules (mostly CH$_4$), the lower SNR in other parts of the spectrum, or both.

The first detection of a molecular hydrogen absorption feature in a brown dwarf atmosphere gives a new semi-empirical upper limit on the atmospheric dust concentration in this T6 dwarf. We attain this limit by comparing to the concentration content of the interstellar medium (ISM), where an atomic hydrogen column density of $N_{\ion{H}{i}}=2.2\times10^{21}$~cm$^{-2}$ corresponds to a visual extinction of $A_V=1.0$~mag, given a 100:1 gas-to-dust ratio \citep{Gorenstein1975}. We expect the ISM gas-to-dust ratio to be orders of magnitude lower than in the upper atmosphere of this brown dwarf, which unlike the ISM is gravitationally differentiated. Silicates (e.g., SiO$_3$, MgSiO$_3$, Mg$_2$SiO$_4$) have $\sim$38--70 times higher molecular weights than molecular hydrogen, and should have settled mostly below the H$_2$-dominated layers of the atmosphere.

The $K$-band extinction in the ISM is such that $A_K/A_V=0.114$ \citep[Table 3 of][]{Cardelli1989}, so in the ISM $N_{\ion{H}{i}}/A_K=1.94\times10^{22}$~cm$^{-2}$~mag$^{-1}$. Virtually all of the hydrogen in the brown dwarf atmosphere is expected to be bound in H$_2$ (e.g, \citealt{Burrows2001}), so the projected H$_2$ column density per magnitude of ISM-like $K$-band extinction is $N_{\rm H_2}/A_K=0.97\times10^{22}$~cm$^{-2}$~mag$^{-1}$.

Following the same prescription as for our calculation of the H$_2$S column density (Section~\ref{sec:hydrogensulfide}), we obtain that the H$_2$ column density in the visible atmosphere of 2M0817 is $N_{\rm H_2} \sim 5.1 \times 10^{24}$~cm$^{-2}$ (the centre of the strongest line corresponds to $P=4$~bars and $T_b=1070$~K, and H$_2$ has an approximately constant equilibrium VMR of 0.836 throughout the Bobcat Alternative A model atmosphere). If the gas-to-dust ratio in the atmosphere of 2M0817 were ISM-like, this H$_2$ column density would correspond to $A_K\sim500$ mag of extinction. Instead, the H$_2$ line is readily detectable, so $A_K$ must be less than 1 mag. This implies that the amount of dust in the atmosphere is $>$500 times less than the interstellar value, and that the atmosphere of 2M018 is almost completely dust-free, as expected for a late-T dwarf. 

Finally, the presence of the H$_2$ line in absorption instead of in emission indicates that the H$_2$ layer is cooler than the layers underneath it. Hence, the upper atmosphere of 2M0817 does not have a strong thermal inversion, as might otherwise be expected in the presence of hot eddies \citep{Showman2020}.

\subsection{Shortcomings of the Models and Unidentified Lines}
\label{sec:mysterylines}

\begin{figure*}
	\includegraphics[trim={0cm 4.4cm 0cm 0cm}, clip,width=\textwidth]{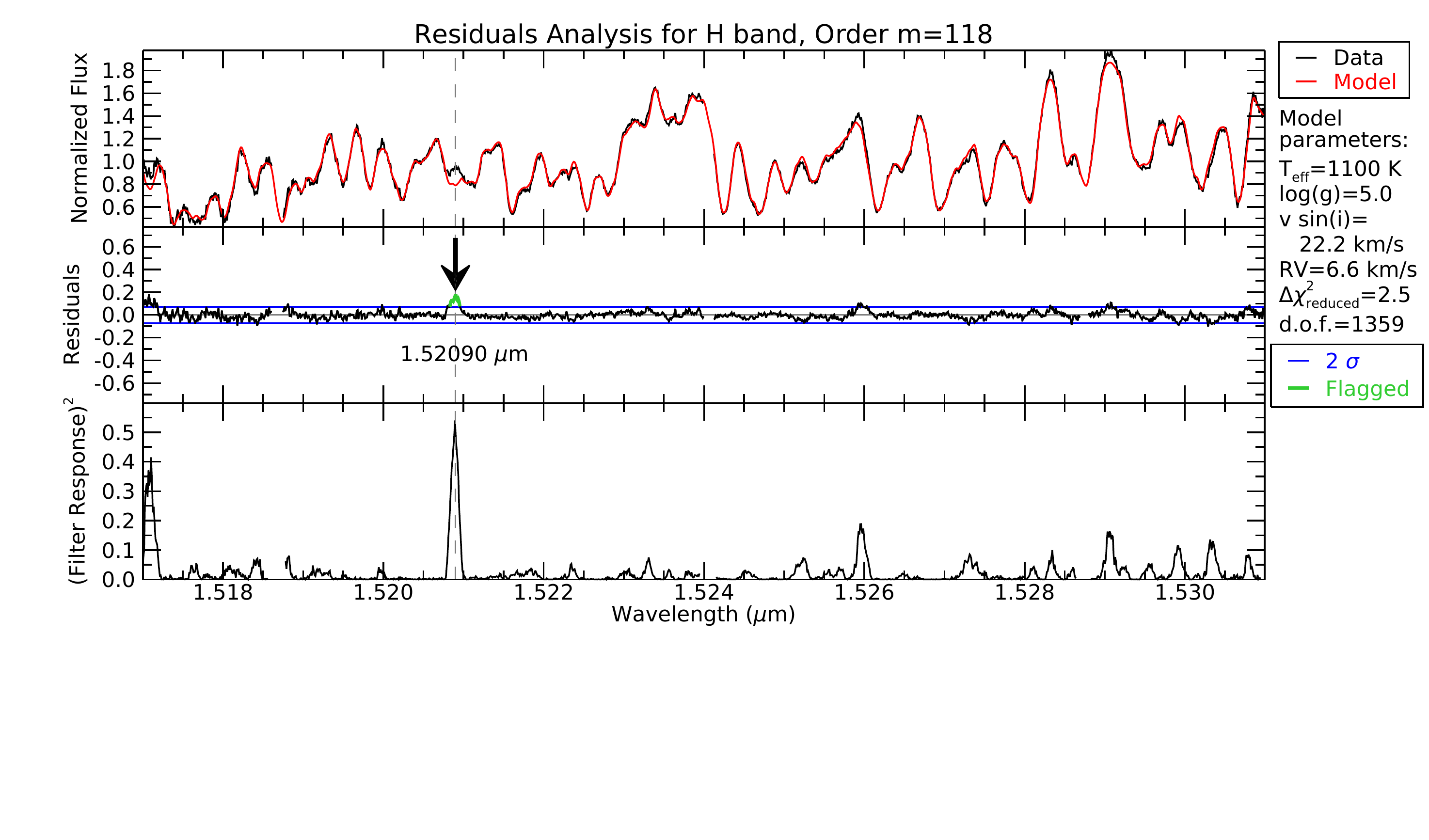}
    \caption{An example of the analysis done on the residuals to identify discrepancies between the models and data. The top panel shows the IGRINS spectrum with the best-fitting Bobcat Alternative A model for this order. The middle panel shows the residuals on the same y-scale as the top panel. Horizontal blue lines delineate $2 \sigma$ threshold, and regions with more than five consecutive pixels beyond $2 \sigma$ are highlighted with green. The filter response of a matched filter using a clean telluric line surrounded by a flat continuum is shown in the bottom panel. There is a clear outlier region at 1.52094 $\mu$m flagged by both the residuals analysis, and also giving a high filter response. Other regions with a high filter response (e.g., 1.51714 $\mu$m and 1.52602 $\mu$m) don't meet our residuals criteria, and are therefore more likely due to noise in the data. The dominant absorber in 
    order $m=118$ of the $H$ band 
    is H$_2$O, which also appears to be the molecule responsible for the flagged feature.
    }
    \label{fig:residualsanalysis}
\end{figure*}

A goal of this work is to identify regions where the photospheric models do not completely reproduce the features in the observed spectra. To identify regions and specific absorption lines in the data which are not well reproduced with the models, we performed two checks. First, we measured the standard deviation, $\sigma$, of the residuals in each order, and then selected regions with at least five consecutive pixels more than $2 \sigma$ away from zero. Second, we applied a matched filter to the residuals by convolving the residuals with a template of an inverted absorption feature. For each order, we selected a telluric absorption feature in the reduced A0 V standard’s spectrum to use as our filter template. We selected features surrounded by a flat continuum and avoided lines with greater than 65 per cent absorption (the threshold for our continuum mask, Section~\ref{sec:reductions}). In orders which had very few, or very weak telluric features, we selected a line from a neighbouring order. We then identified regions in the spectra where both the pixel values were outside of two standard deviations, and the matched filter response was higher than the surrounding pixels. This helped to eliminate false detections due to noise. We performed these checks only for the Bobcat Alternative A models, as they are the most up-to-date and provide the best fits to the data. We show an example of this analysis in Fig.~\ref{fig:residualsanalysis}, and we summarize the regions of interest in Table~\ref{table:mysterylines}, with a brief description of the potential issue affecting the model in each case. These discrepancies can be seen in Fig.~\ref{fig:Horders} and \ref{fig:Korders}, indicated with with vertical grey dashed lines for individual features, and black brackets for wider discrepant regions. The H$_2$ feature (Section~\ref{sec:h2}) was initially identified through this type of analysis.

\begin{figure*}
	\includegraphics[trim={0cm 4.4cm 0cm 0cm}, clip,width=\textwidth]{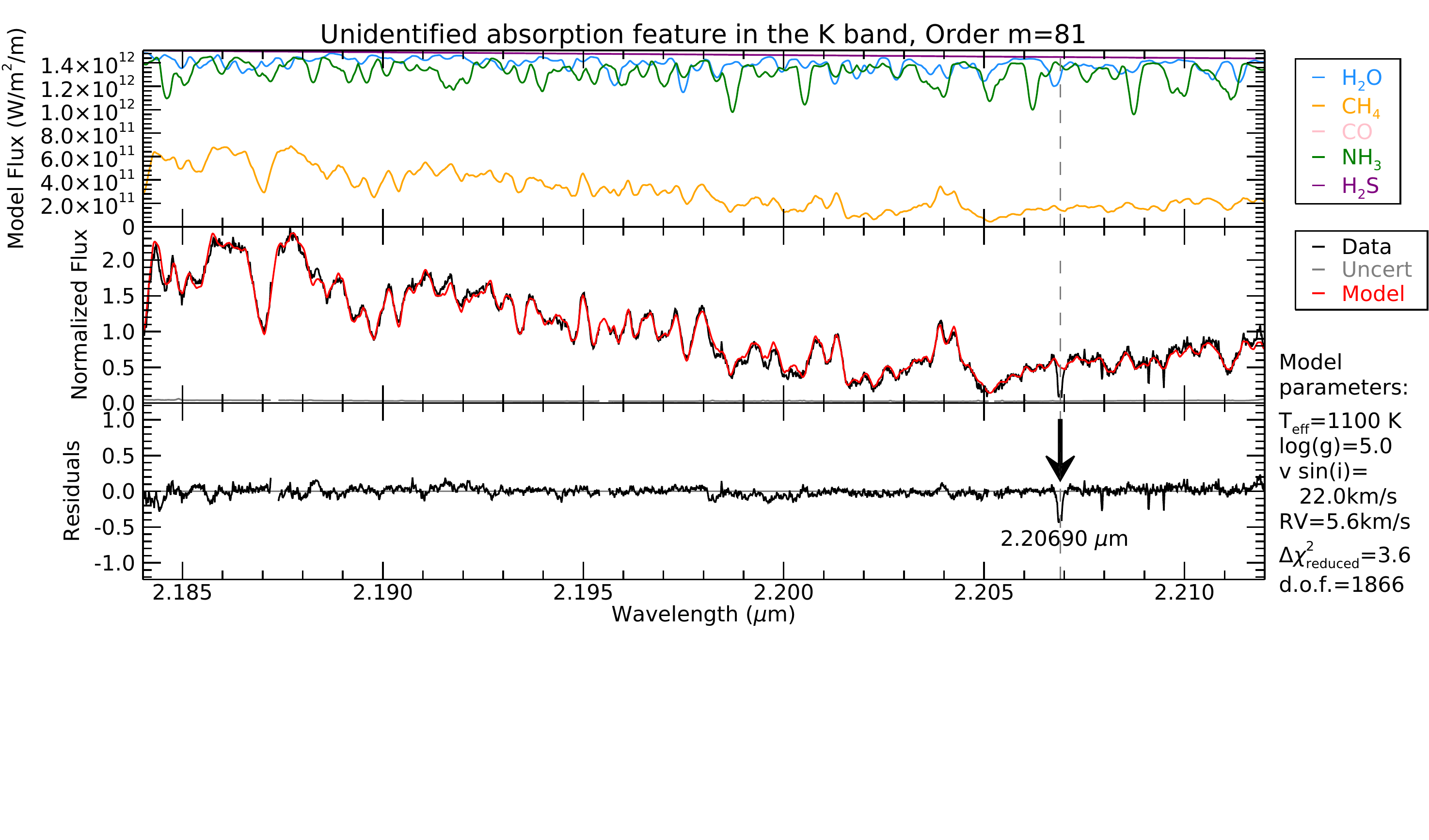}
    \caption{The same layout as Fig.~\ref{fig:missingline}, but showing the best-fitting Bobcat Alternative A model for this order. 
    Order $m=81$ of the $K$ band 
    is fit well by models, and shows an unknown absorption feature that doesn't appear in any model of any family at 2.20690 $\mu$m. This line is indicated with a black arrow.
    }
    \label{fig:missingline2}
\end{figure*}

Most notably, a line is clearly missing from the models at 2.20695~$\mu$m in order $m=81$ of the $K$ band, 
shown in Fig.~\ref{fig:missingline2}. 
None of the models includes a line at this wavelength, and we have not identified the element or molecule responsible for this feature. Additionally, we find no absorption or emission in the A0 V stars at the wavelengths given in Table~\ref{table:mysterylines} which could introduce these unidentified features to our T6 spectrum. The feature in $H$ band order $m$=121 does line up with a weak telluric H$_2$O feature, but given the difference in the line widths, we believe this discrepancy between the data and model is not caused by the telluric line.

The Bobcat Alternative A models we use to anlayze our data are comprised of the five most abundant molecules (H$_2$O, CH$_4$, CO, NH$_3$, and H$_2$S), plus collision-induced absorption from molecular hydrogen and helium. The Sonora Bobcat, Morley, and BT-Settl models consist of more complete sets of molecules. We have confirmed that the lines listed in Table~\ref{table:mysterylines} are indeed missing in all families of models. We cannot eliminate all molecules that are included in the more complete Sonora Bobcat, Morley, and BT-Settl model families as being responsible for these missing lines, as the line lists could be incomplete or inaccurate, or there could be disequilibrium chemistry taking place, as we observed with CO (Section~\ref{sec:carbonmonoxide}). We also confirmed that H$_2$ is not responsible for any of the unidentified lines.

Disequilibrium chemistry could imply that other mixing-sensitive gases such as phosphine (PH$_3$; the next most abundant molecule in these cold atmospheres) could also be present at higher abundances than expected for chemical equilibrium \citep{Fegley1996}. We generated a Bobcat Alternative A model with a greatly over-estimated abundance of PH$_3$ (VMR of $1 \times 10^{-4}$, which is more than 300 times the amount expected for equilibrium chemistry, and would require far more phosphorus than would be available in a solar-composition atmosphere) to compare to our spectra, intending to match the locations of the PH$_3$ features to the unidentified lines. We use the SAlTY line list from Exomol \citep{Sousa-Silva2015} for PH$_3$. We found that the PH$_3$ features did not match with any of the unidentified lines, and PH$_3$ is likely not responsible for these features. A study by \citet{Miles2020} searched for PH$_3$ in atmospheres of cold brown dwarfs displaying disequilibrium CO absorption. This study was performed in the $L$ and $M$ bands (centred at 3.45~$\mu$m and 4.75~$\mu$m, respectively), where H$_2$O, CH$_4$, and NH$_3$ absorb less strongly, but PH$_3$ absorbs much more strongly, and so should give the best chance at detecting PH$_3$. Unfortunately, they were also unable to detect PH$_3$.

Among the list of unidentified lines in Table~\ref{table:mysterylines}, we list nearly the full wavelength coverage of 
orders $m=113$ through $m=107$ of the $H$ band. 
These orders cover 1.596--1.681~$\mu$m and the dominant absorber in these orders is CH$_4$. As discussed in Section~\ref{sec:methane}, while the strongest absorption features are very well modelled in the Bobcat Alternative A models, the weaker features and continua in the models deviate significantly from the observations. We observe some bumpiness in the residuals throughout the entire IGRINS wavelength coverage, especially in these CH$_4$-dominant regions of the $H$-band. There are a host of weaker features that are not being taken into account in the models, and these features contribute a non-trivial amount to the atmospheric opacity.

\begin{table}
	\centering
	\caption{The wavelengths of discrepancies in the models and unidentified absorption features. These regions are identified in Fig.~\ref{fig:Horders} and \ref{fig:Korders}, with vertical grey dashed lines for individual features, and black brackets for regions. }
	\label{table:mysterylines}
	\begin{tabular}{cclp{0.45\linewidth}} 
		\hline
		Band & Order & Wavelength & Notes \\
		& ($m$) & ($\mu$m) & \\
		\hline
        $H$ & 121 &	1.48463	&	Potential issue with blended line or something missing in the model	\\ 
        $H$ & 118 &	1.52090	&	A line appears in the model which is missing in the data (see Fig. \ref{fig:residualsanalysis}). The model line appears to be a water/ammonia blend. \\ 
        $H$ & 116 &	1.55120	&	Potential issue with blended line (H$_2$O and NH$_3$?)	\\ 
        $H$ & 115 &	1.56396	&	Potential issue with blended line (H$_2$O and NH$_3$ or H$_2$O and H$_2$S?)	\\ 
        $H$ & 113 &	1.5875--1.5960	&	CH$_4$ region \tablenotemark{a}	\\ 
        $H$ & 112 &	1.5980--1.6090	&	CH$_4$ region \tablenotemark{a}	\\ 
        $H$ & 111 &	1.6080--1.6240	&	CH$_4$ region \tablenotemark{a}	\\ 
        $H$ & 110 &	1.6244--1.6390	&	CH$_4$ region \tablenotemark{a}	\\ 
        $H$ & 109 &	1.6375--1.6510	&	CH$_4$ region \tablenotemark{a}	\\ 
        $H$ & 108 &	1.6515--1.6650	&	CH$_4$ region \tablenotemark{a}	\\  
        $H$ & 108 &	1.65355	&	Potential issue with blended line (CH$_4$ and H$_2$O?)	\\  
        $H$ & 108 &	1.65446	&	Model over-estimates flux	\\  
        $H$ & 108 &	1.66319	&	Model over-estimates flux	\\  
        $H$ & 107 &	1.6675--1.6810	&	CH$_4$ region \tablenotemark{a}	\\  
        $H$ & 107 &	1.66960	&	Line too weak in model	\\ 
        $H$ & 107 &	1.67380	&	Model under-estimates flux \\ 
        $H$ & 106 &	1.68443	&	Potential issue with blended line (CH$_4$ and H$_2$O?) \\ 
        $H$ & 106 &	1.68672	&	Potential issue with blended line (CH$_4$ and H$_2$O?) \\ 
        $H$ & 106 &	1.69600	&	Potential issue with blended line (CH$_4$ and H$_2$O?)	\\ 
        $K$ & 87 &	2.04020	&	Model over-estimates flux	\\  
        $K$ & 87 &	2.05478	&	Model under-estimates flux	\\  
        $K$ & 81 &	2.20690	&	Line missing from model (see Fig.~\ref{fig:missingline2})	\\   
        \hline
	\end{tabular}
    \tablenotetext{a}{In these CH$_4$ regions the model accurately represents the deepest features, but appears to be incorrect or incomplete in the weaker features and continuum. Given the accuracy of the H$_2$O lines elsewhere in the spectrum, we suspect these discrepancies are due to weak CH$_4$ lines, and not H$_2$O.
    }
\end{table}


\section{Lessons Learned}
\label{sec:lessons}

We find that atmospheric models that use state-of-the-art line lists represent observations well. We are now able to extract more precise information from our data than merely detect the most abundant molecules: we can detect absorption by trace species that have never been seen before (like H$_2$S and H$_2$), see low abundance species, and more readily detect abundances of species (as we have done for CO here). 

In all cases we recommend using the most-up-to-date models available with the most recent molecular line lists. 
We have found that the line lists used in the Bobcat Alternative A models (Table~\ref{table:linelists}) give the most reliable and consistent estimates of all physical parameters across all wavelength regions of this study. More generally, we have found that all models do an adequate job fitting the data in regions where H$_2$O is the dominant absorber. 

We summarize our main recommendations and warnings, organized by the information of interest in the following two subsections.

\subsection{Fitted Spectroscopic Parameters}

\textbf{Effective Temperature ($T_{\rm eff}$)} measurements are most accurate and consistent in the $K$ band, in regions where H$_2$O is the dominant opacity source. We recommend using the Bobcat Alternative A models for measuring $T_{\rm eff}$ anywhere in the $H$ and $K$ bands. The Morley models over-estimate $T_{\rm eff}$ in the $H$ band and the Sonora Bobcat models over-estimate $T_{\rm eff}$ in the $K$ band, while The BT-Settl models under-estimate $T_{\rm eff}$ in the $K$ band. If disequilibrium chemistry effects are not taken into consideration, $T_{\rm eff}$ may also be over-estimated.

\textbf{Surface Gravity ($\log{g}$)} measurements are the most accurate and consistent in regions where H$_2$O is the dominant opacity source in the $H$ band. We recommend using the Bobcat Alternative A models for measuring $\log{g}$ anywhere in the $H$ and $K$ bands. The Sonora Bobcat models over-estimate $\log{g}$ in the $K$ band. $\log{g}$ may also be over estimated in regions where the dominant molecule switches from H$_2$O to CH$_4$, near the peaks of $H$ and $K$ bands (1.6 $\mu$m and 2.1 $\mu$m, respectively).

\textbf{Projected Rotation Velocity ($v \sin{i}$)} measurements are the most accurate and consistent in regions where H$_2$O is the dominant opacity source in both the $H$ and $K$ bands. 
We recommend using the Bobcat Alternative A models for measuring $v \sin{i}$ anywhere in the $H$ and $K$ bands. We recommend using the region from 1.45 to 1.57 $\mu$m in the $H$ band, or 1.89 to 2.10 $\mu$m in the $K$ band if measuring $v \sin{i}$ with any other model. $v \sin{i}$ may be over estimated in regions where the dominant molecule switches from H$_2$O to CH$_4$, near the peaks of $H$ and $K$ bands (1.6 $\mu$m and 2.1 $\mu$m, respectively).

\textbf{Radial Velocity (RV)} measurements are the most accurate and consistent in regions where H$_2$O is the dominant opacity source in both the $H$ and $K$ bands. We recommend using the Bobcat Alternative A models for measuring RV anywhere in the $H$ and $K$ bands. 
We recommend using the region from 1.45 to 1.58 $\mu$m in the $H$ band if measuring RV with any other model. RV measurements demonstrate a blueshift with wavelength when measured from the Sonora Bobcat and BT-Settl models between 1.894 and 2.060 $\mu$m.

\subsection{Specific Molecules}

\textbf{Water (H$_2$O)} is the dominant opacity source between 1.45 and 1.58 $\mu$m in the $H$ band, and between 1.89 and 2.10 $\mu$m in the $K$ band. The H$_2$O-dominant region of the $H$ band (1.45--1.58 $\mu$m) gives consistent results for all parameters across all model families. We recommend using the ExoMol/POKAZATEL \citep{Polyansky2018} line list when studying water.

\textbf{Methane (CH$_4$)} is the dominant opacity source between 1.60 and 1.73~$\mu$m in the $H$ band, and between 2.10 and 2.48~$\mu$m in the $K$ band. The CH$_4$-dominant region of the $K$ band (2.10--2.48 $\mu$m) gives consistent results for all parameters for the Bobcat Alternative A models. Weak CH$_4$ lines between 1.59 and 1.67~$\mu$m are poorly matched to data in all model families and in all line lists. We recommend using the HITEMP \citep{Hargreaves2020} line list when studying CH$_4$.

\textbf{Carbon monoxide (CO)} bands occur between 1.55 and 1.57~$\mu$m in the $H$ band, and 2.29 to 2.45~$\mu$m in the $K$ band. To measure accurate and consistent parameters, especially $T_{\rm eff}$, disequilibrium chemistry may need to be considered for CO. We recommend using the HITEMP 2010 \citep{Rothman2010} line list when studying CO.

The strongest \textbf{ammonia (NH$_3$)} features occur between 1.50 and 1.52~$\mu$m in the $H$ band, and 1.95 to 2.09~$\mu$m and 2.18 to 2.21~$\mu$m in the $K$ band. The choice of NH$_3$ line list does not appear to significantly impact the measured parameters, and we recommend using the ExoMol/CoYuTe \citep{Coles2019} line list when studying NH$_3$.

The strongest \textbf{hydrogen sulfide (H$_2$S)} features occur in the $H$ band between 1.58 and 1.60~$\mu$m. The choice of H$_2$S line list does not appear to impact the measured parameters, and we recommend using the combinations of line lists from ExoMol \citep{Tennyson2012}, \citet{Azzam2015}, and HITRAN 2012 \citep{Rothman2013}, or the updated versions of HITRAN from 2016 and 2020 \citep{Gordon2017,Gordon2022}, which we have not tested here.

\section{Applications to Exoplanets}
\label{sec:exoplanets}

In high-dispersion spectroscopic observations of exoplanets, where the planet itself cannot be spatially resolved, cross-correlation is a powerful technique for detecting and characterizing the planet. In addition to the identification of specific molecules, the velocity relative to the host star, information about planetary spin ($v \sin{i}$) and atmospheric wind speeds may be determined (e.g., \citealt{Snellen2010,Snellen2014}). However, when an observed spectrum combines the star and planet, individual lines from the planet can have SNR $\ll$ 1, and the ability to recover a planet is only as good as the model. If fitting an incorrect model to a low SNR spectrum, the planet may not be recovered, or even discovered. 

\begin{figure*}
	\includegraphics[width=0.9\textwidth]{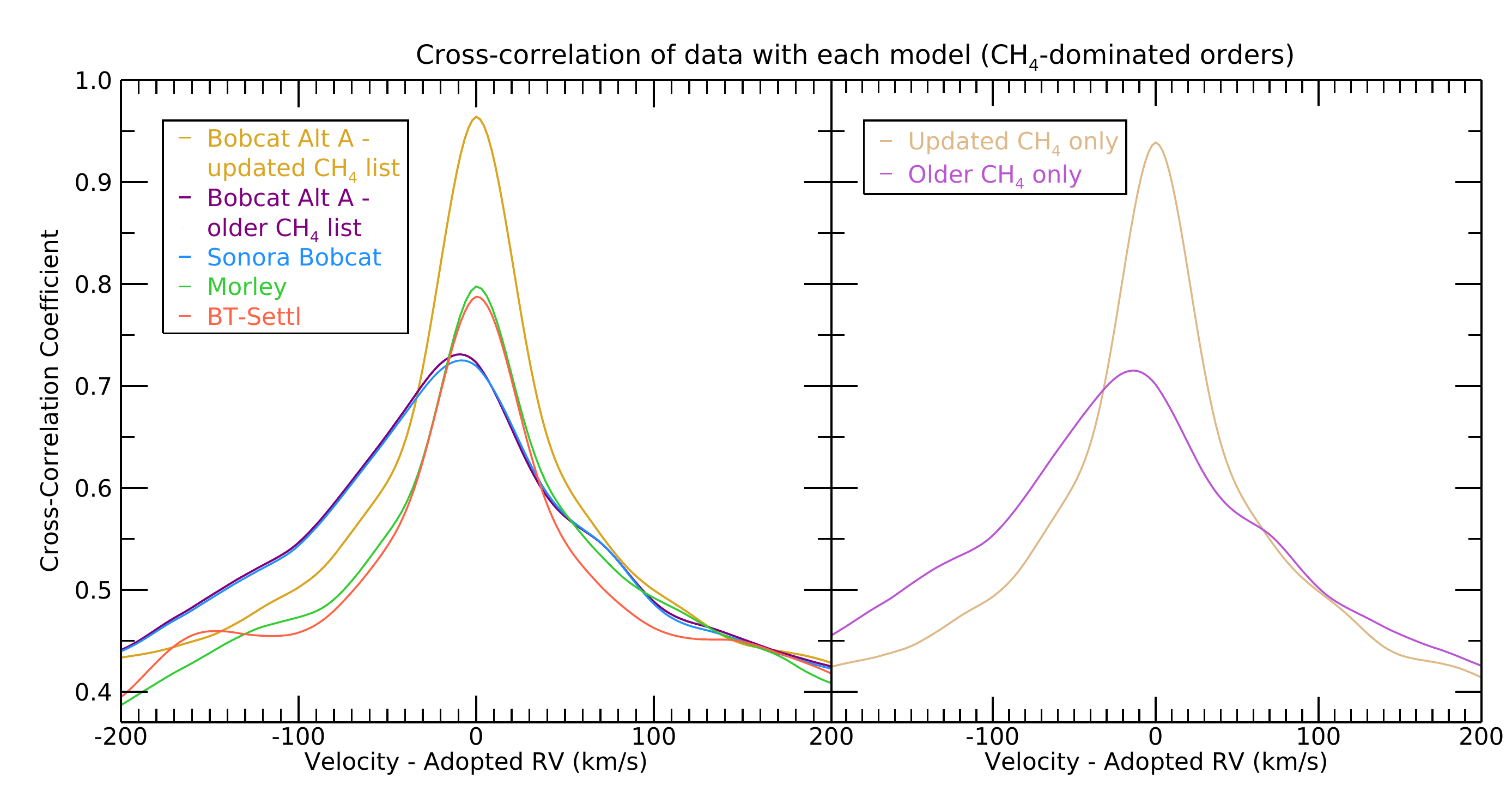}
    \caption{\textit{Left panel:} Cross-correlations of the data with each of the model families, including a version of the Bobcat Alternative A model modified to use the older CH$_4$ line list (the other molecules use the newer line lists given in Table~\ref{table:linelists}). We performed this cross correlation across the CH$_4$-dominated orders of both the $H$ and $K$ bands for models with parameters given in Table~\ref{table:parameters}. As expected, the peak of the Bobcat Alternative A model that uses the newest CH$_4$ line lists is higher than for the other models. \textit{Right panel:} The same as the left panel, but for Bobcat Alternative A models that incorporate only CH$_4$. The `Updated CH$_4$-only' model uses the line list used in the new Bobcat Alternative A model, while the `Older CH$_4$-only' model uses the line list used in the Sonora Bobcat model.}
    \label{fig:crosscorr}
\end{figure*}

We have confirmed that the older CH$_4$ line lists are inaccurate in the 1.60--1.73 $\mu$m and 2.10--2.40 $\mu$m regions (see Section~\ref{sec:methane}), and the inaccurate line positions could result in a non-detection of an exoplanet. We cross-correlated our data against the various models to showcase the improvement that the newer CH$_4$ lines offer in a cross-correlation analysis.
We used the IDL function \texttt{c\_correlate} and included only the CH$_4$-dominated orders of the $H$ and $K$ bands ($m=$ 112--104, 1.594--1.730 $\mu$m; $m=$ 85--77, 2.084--2.294 $\mu$m). The model parameters were fixed to the values given in Table~\ref{table:parameters}. The results of the cross-correlations are shown in Fig.~\ref{fig:crosscorr}. 

To isolate the effects of the choice of CH$_4$ line list 
our cross-correlation analysis includes two different Bobcat Alternative A models: one with the updated CH$_4$ line lists and one with the older CH$_4$ line lists used in the Sonora Bobcat models, while keeping the H$_2$O, CO, NH$_3$, and H$_2$S line lists the same as in the newest models (see Table~\ref{table:linelists}). As seen in the left panel of Fig.~\ref{fig:crosscorr}, the peak of the Bobcat Alternative A model with the newest CH$_4$ line lists is the highest. The peak for the Bobcat Alternative A model with the older CH$_4$ line lists is significantly lower. In fact, the latter is nearly identical to the cross-correlation peak for the Sonora Bobcat model. This is as expected, as these two models use the same underlying atmospheric model, and we showed previously that the choice of line lists for the other molecules has less impact on the quality of the model fits to the data. We see that the peak of the Bobcat Alternative A model with the older CH$_4$ line lists 
is slightly higher than the peak of the Sonora Bobcat model, likely due to the newer line lists used for the other molecules, particularly H$_2$O. We also see that the CH$_4$ line lists used in the Sonora Bobcat models result in an offset in the peak towards negative velocities. This is consistent with the lower radial velocities we found for the Sonora Bobcat model in orders where CH$_4$ was the main absorber (Fig.~\ref{fig:Hcomparemodels} and \ref{fig:Kcomparemodels}).

The importance of the choice of CH$_4$ line lists in the methane-dominated IGRINS orders is demonstrated in the right panel of Fig.~\ref{fig:crosscorr}. That figure compares the cross-correlation functions of our data with versions of the models that include only methane absorption and ignore the contributions from H$_2$O, CO, NH$_3$, and H$_2$S. The `Updated CH$_4$ only' models uses the line list incorporated in the Bobcat Alternative A models, while the `Older CH$_4$ only' uses the line lists incorporated in the Sonora Bobcat models. The `updated' and `older' cross-correlations are very similar in shape and peak strengths to the Bobcat Alternative A model and Sonora Bobcat cross-correlations, respectively. This is not surprising, as we have performed the cross-correlation specifically for the CH$_4$-dominated orders, and so models which include additional molecules provide only a small improvement. 

In their analysis of the $K$-band spectra of HD~209458~b and $\beta$~Pictoris~b, \citet{Snellen2010,Snellen2014} were successful in detecting CO through cross-correlation with atmospheric models, but failed to recover CH$_4$. Inaccurate line lists could be responsible for these non-detections, as these studies used the older HITRAN'08 \citep{Rothman2009} for their CH$_4$ line lists. The atmosphere of $\beta$~Pictoris~b may also be too hot for the detection of CH$_4$ ($T_{\rm eff} = 1724$~K; \citealt{Chilcote2017}), but CH$_4$ may be detectable in HD~209458~b ($T_{\rm eq} = 1449$~K; \citealt{Torres2008}) with an improved line list. Indeed, more recently, \citet{Guilluy2019} and \citet{Giacobbe2021} had success detecting CH$_4$ from HD~102195~b and HD~209458~b, respectively, with more up-to-date line lists. \citet{Guilluy2019} used HITRAN2012 \citep{Rothman2013}, and \citet{Giacobbe2021} used \citet{Hargreaves2020}, the same CH$_4$ line list used in the Bobcat Alternative A models. A separate demonstration of the enhanced utility of the newer \citet{Hargreaves2020} CH$_4$ line lists is evident in \citet{Line2021}, who determined the C/H, O/H, and C/O ratios of the hot Jupiter WASP-77Ab ($T_{\rm eff}=1740$~K; \citealt{Maxted2013}) using cross-correlation methods with IGRINS data.

\section{Conclusions}
\label{sec:Conclusions}

The data presented here are among the highest resolution spectra ever published for a cold brown dwarf. We found that model spectra with the most recent line lists showed significant improvement in fitting the observed spectrum of the T6 dwarf 2MASS~J08173001$-$6155158. The updated line lists for water, methane, and ammonia allow for precise empirical determinations of physical parameters. We identified the most reliable regions for measuring physical parameters of cold brown dwarfs, and we summarized our findings in Section~\ref{sec:lessons}. In particular, we highlighted the excellent fits of the Sonora Bobcat Alternative A models (Hood et al. in preparation), and the accuracy of the ExoMol/POKAZATEL line list for H$_2$O \citep{Polyansky2018} and the HITEMP line list for CH$_4$ \citep{Hargreaves2020} in matching the observed spectrum. We confirmed that like other late-type T and Y dwarfs, 2M0817 demonstrates CO disequilibrium chemistry. We identified individual NH$_3$ lines in the spectrum of 2M0817, and we presented the first unambiguous detections of H$_2$S and H$_2$ absorption in an extra-solar atmosphere. Our molecular hydrogen detection allowed us to place a semi-empirical upper limit on the atmospheric dust concentration of this brown dwarf. We found that the atmosphere of 2M0817 has $>$500 times less dust than the ISM, implying that the atmosphere is almost completely dust-free. Additionally, we identified several spectroscopic features that are missing from, or are poorly fit by the models. Finally, our cross-correlation analysis showed that the most up-to-date line lists are significantly more sensitive to CH$_4$ absorption in the atmosphere of this T6 dwarf. This will improve the detectability of CH$_4$ and other atmospheric absorbers in more challenging observations, such as the high-dispersion, low-SNR, high-background spectra of exoplanets around their host stars.


\section*{Acknowledgements}

We would like to thank the anonymous referee for their careful and constructive comments that helped us improve this paper. We thank Dr. Mark Marley and Dr. Michael Cushing for useful discussions about these data, and for their insights on which molecules to investigate. We thank Dr. Michael Line for providing the updated methane and ammonia line lists used in the Bobcat Alternative A models. We thank Dr. Genaro Su{\'a}rez for performing a secondary check of our estimates of fundamental parameters with his custom fitting algorithm. We thank Chris Wyenberg for his help with the cross-correlation analysis. 

Support for this work was provided by an Ontario Graduate Scholarship, NSERC, and the Canadian Space Agency Flights and Fieldwork for the Advancement of Science and Technology (FAST) funding initiative. 
Part of the work for this paper was completed at the Other Worlds Laboratory Exoplanet Summer Program 2019. We thank the Heising-Simons Foundation and the University of California Santa Cruz for funding this program.

This paper contains data based on observations obtained at the international Gemini Observatory (Program ID GS-2018A-Q-304), a program of NSF's NOIRLab, which is managed by the Association of Universities for Research in Astronomy (AURA) under a cooperative agreement with the National Science Foundation on behalf of the Gemini Observatory partnership: the National Science Foundation (United States), National Research Council (Canada), Agencia Nacional de Investigaci\'{o}n y Desarrollo (Chile), Ministerio de Ciencia, Tecnolog\'{i}a e Innovaci\'{o}n (Argentina), Minist\'{e}rio da Ci\^{e}ncia, Tecnologia, Inova\c{c}\~{o}es e Comunica\c{c}\~{o}es (Brazil), and Korea Astronomy and Space Science Institute (Republic of Korea). We also thank Gemini Observatory for the opportunity for the first author to visit the Gemini South facility in La Serena, Chile through their `Bring One, Get One Student Observer Support Program.'

This work used the Immersion Grating Infrared Spectrometer (IGRINS) that was developed under a collaboration between the University of Texas at Austin and the Korea Astronomy and Space Science Institute (KASI) with the financial support of the US National Science Foundation under grants AST-1229522 and AST-1702267, of the University of Texas at Austin, and of the Korean GMT Project of KASI. We thank the IGRINS team for their continued support throughout this project, the opportunity to observe in person with IGRINS under their guidance, and for including 2MASS~J08173001$-$6155158 as a commissioning target, allowing us to obtain valuable additional data.


\section*{Data Availability}

The raw IGRINS data for 2MASS~J08173001$-$6155158 are available on the Gemini Archive under Program ID GS-2018A-Q-304. The reduced spectrum and best-fitting Bobcat Alternative A spectra are available through the Harvard Dataverse (\url{https://doi.org/10.7910/DVN/DV1ZLR}) or Zenodo (\url{https://doi.org/10.5281/zenodo.6082001}).




\bibliographystyle{mnras}
\bibliography{main}



 \appendix

 \section{The Full Suite of Model Fits for Every IGRINS Order}
 \label{sec:Appendix}

 We show the best fitting Bobcat Alternative A models for all orders in the IGRINS spectrum in figures ~\ref{fig:Horders} ($H$ band) and \ref{fig:Korders} ($K$ band).
 
 \vspace{15cm}
 
\begin{figure*}
    \centering
    \begin{subfigure}{0.99\textwidth}
        \includegraphics[width=\textwidth]{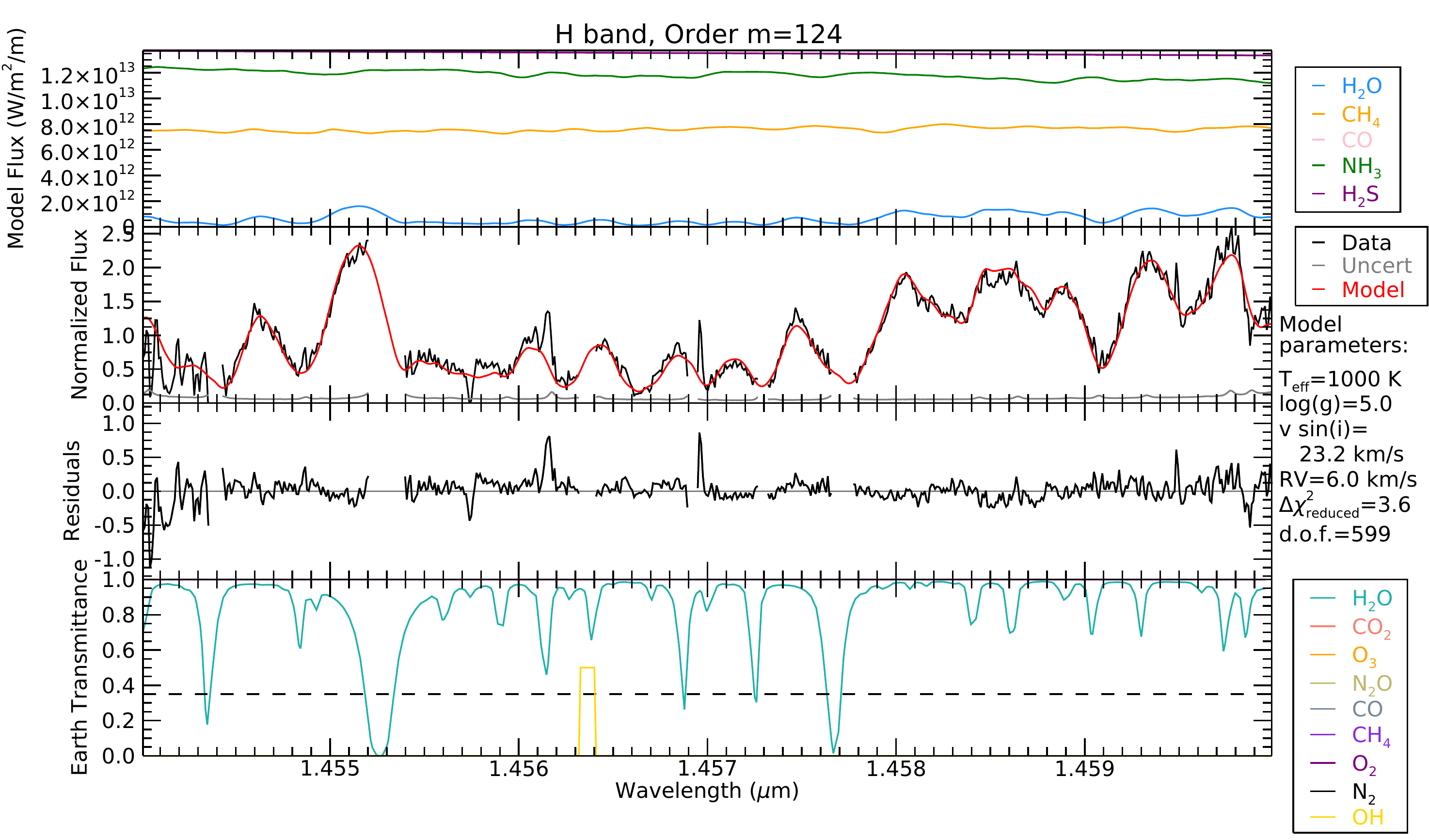} 
    \end{subfigure}
    \begin{subfigure}{0.99\textwidth}
        \includegraphics[width=\textwidth]{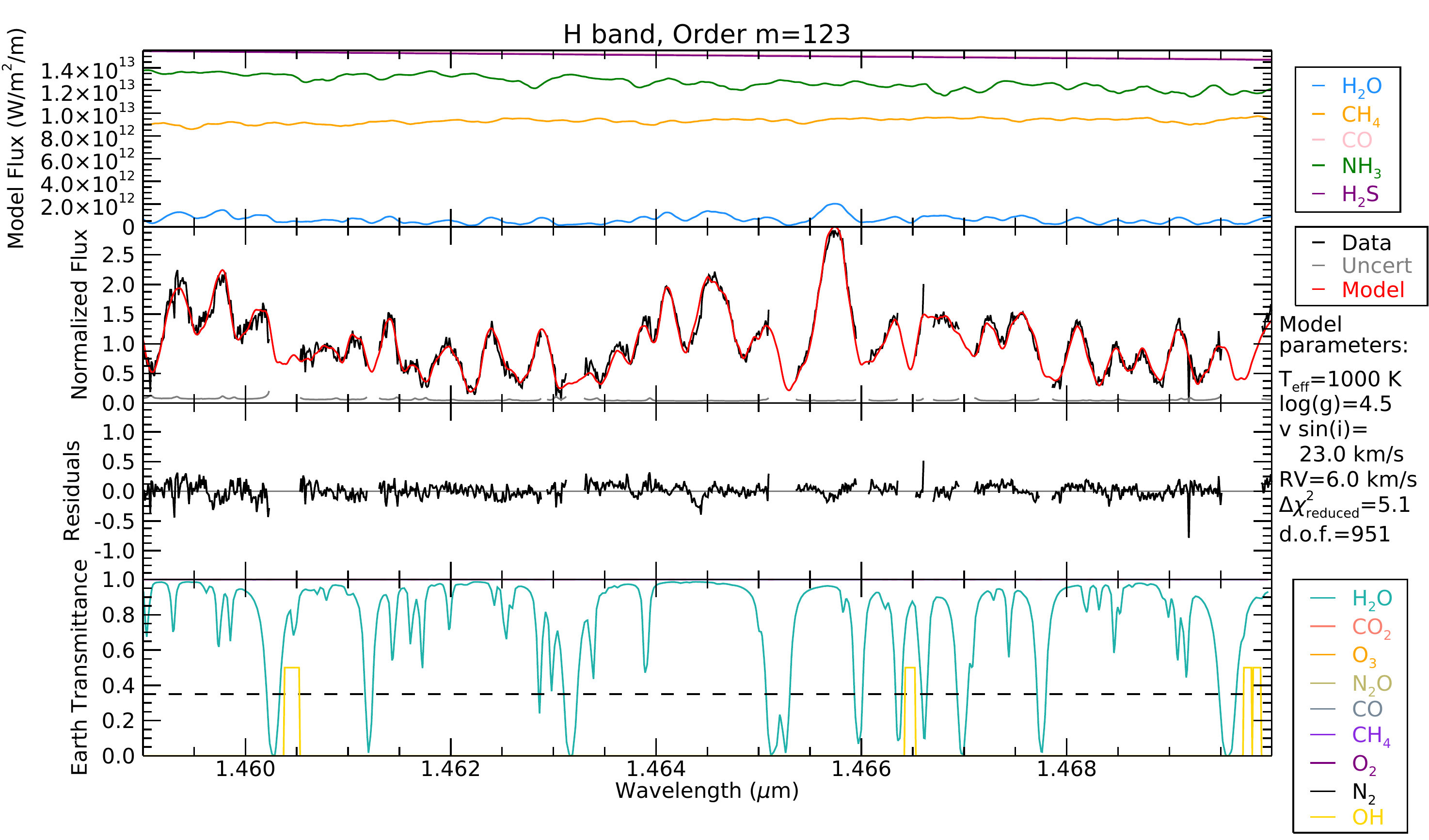} 
    \end{subfigure}
    \caption{Every order of the $H$ band. The model shown is the best fitting Bobcat Alternative A model for each order. Each order is fit independently (Section~\ref{sec:modelscomparison}), so the physical parameters may differ between orders. The top panel shows the molecule-by-molecule breakdown of the model. The second panel from the top shows the IGRINS data with the full model. The second panel from the bottom shows the residuals on the same $y$-scale as the panel above it. The model discrepancies listed in Table~\ref{table:mysterylines} are indicated in these figures with black arrows for discrepant lines, and black brackets for discrepant regions. The bottom panel shows the PSG Earth's transmittance to help assess the telluric lines in our spectra. The OH emission lines are also shown as boxes and indicate position only, not line strength. Wider boxes indicate blended OH emission lines. A dashed horizontal line indicates the 65 per cent absorption threshold used for our telluric mask. The $\Delta \chi^2_{\rm reduced}$ is the difference between the displayed order's $\chi^2_{\rm reduced}$, and the $\chi^2_{\rm reduced}$ of the best fitting order ($m=85$, Fig.~\ref{fig:Korders}). This figure continues for many pages, with two orders per page, to show all 26 orders of the $H$ band.
    }\label{fig:Horders}
\end{figure*}

\begin{figure*}
    \ContinuedFloat 
    \centering
    \begin{subfigure}{0.99\textwidth}
        \includegraphics[width=\textwidth]{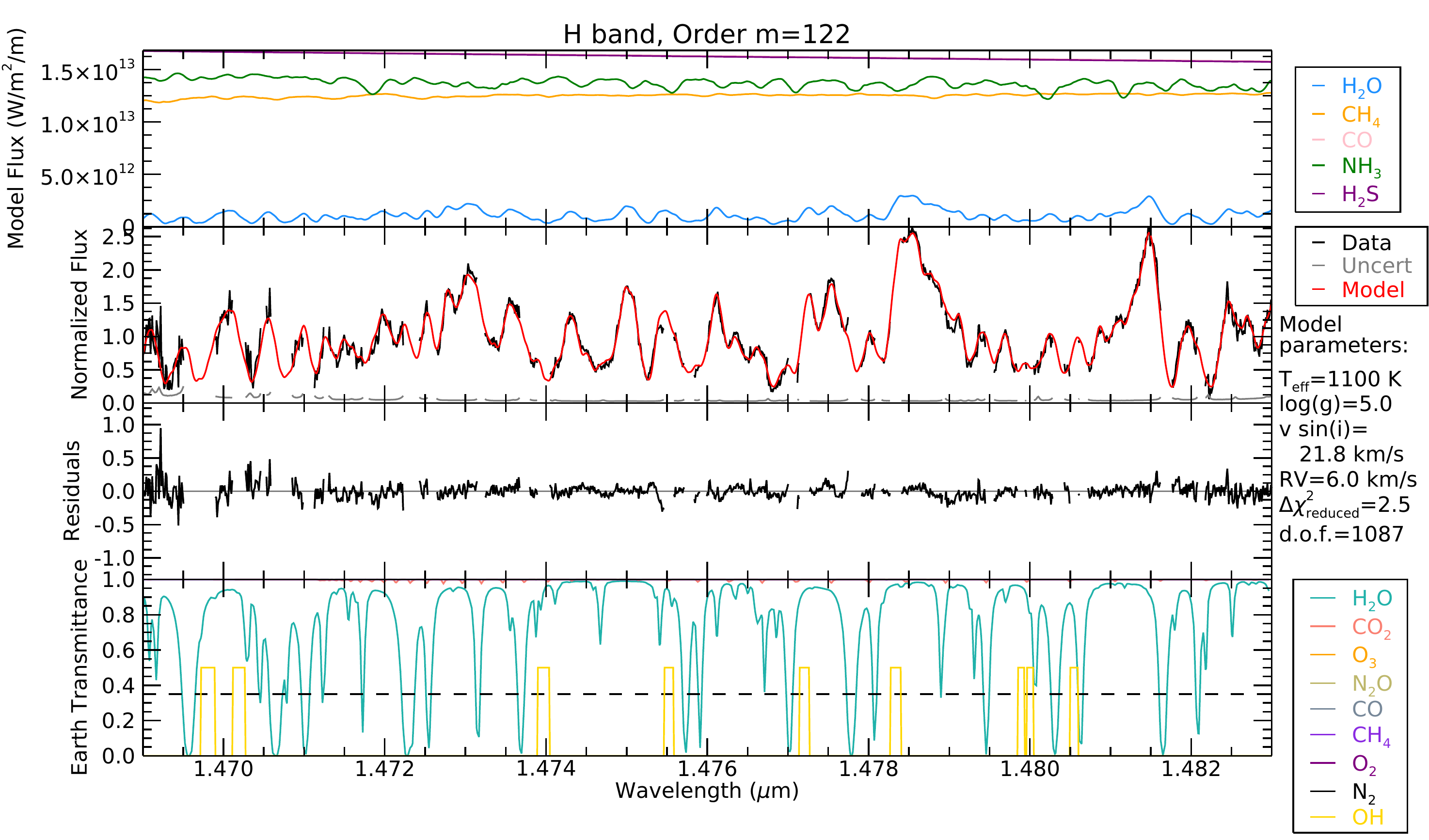} 
    \end{subfigure}
    \begin{subfigure}{0.99\textwidth}
        \includegraphics[width=\textwidth]{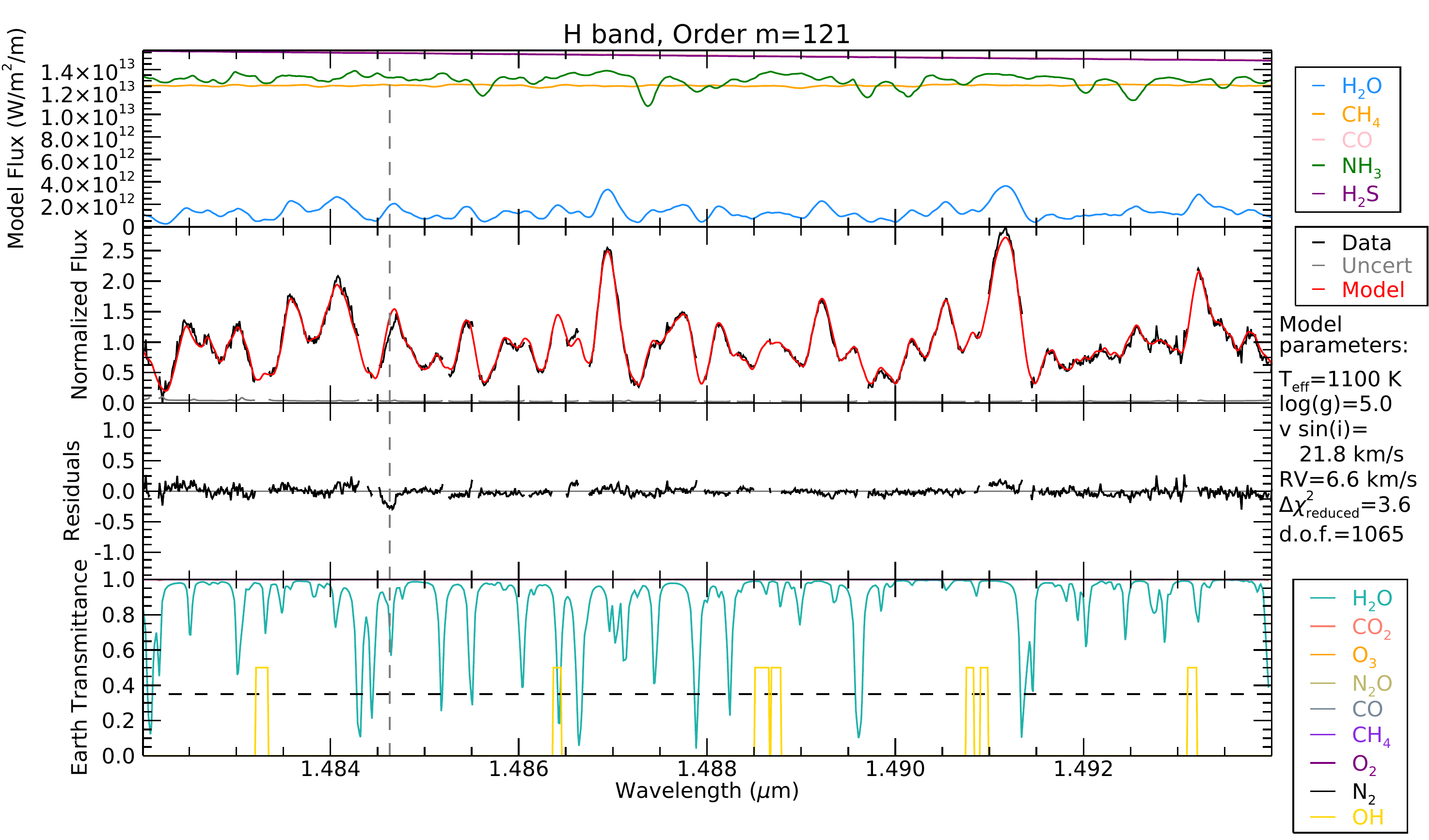} 
    \end{subfigure}
    \caption{Continued.}
\end{figure*}

\begin{figure*}
    \ContinuedFloat 
    \centering
    \begin{subfigure}{0.99\textwidth}
        \includegraphics[width=\textwidth]{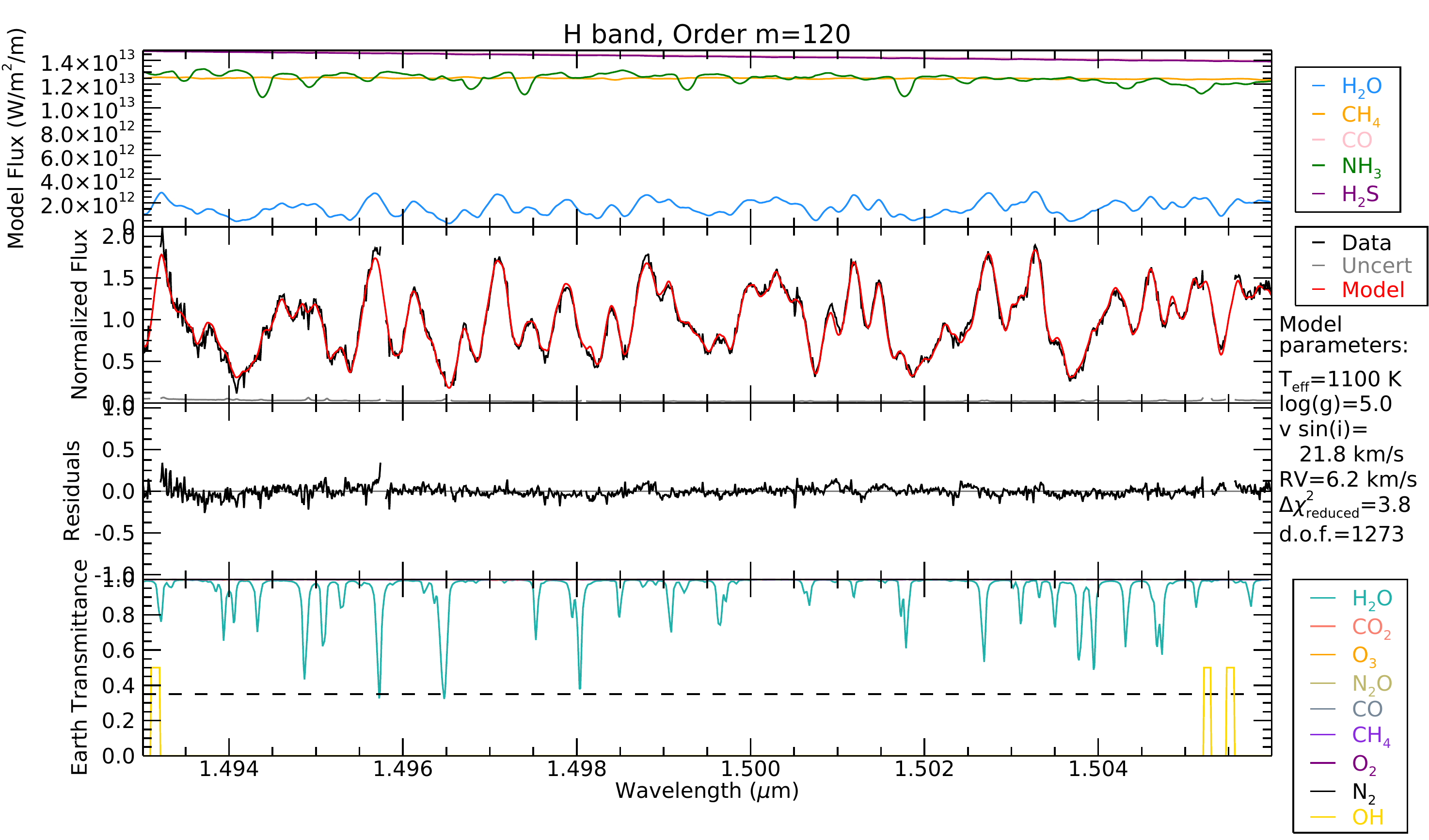} 
    \end{subfigure}
    \begin{subfigure}{0.99\textwidth}
        \includegraphics[width=\textwidth]{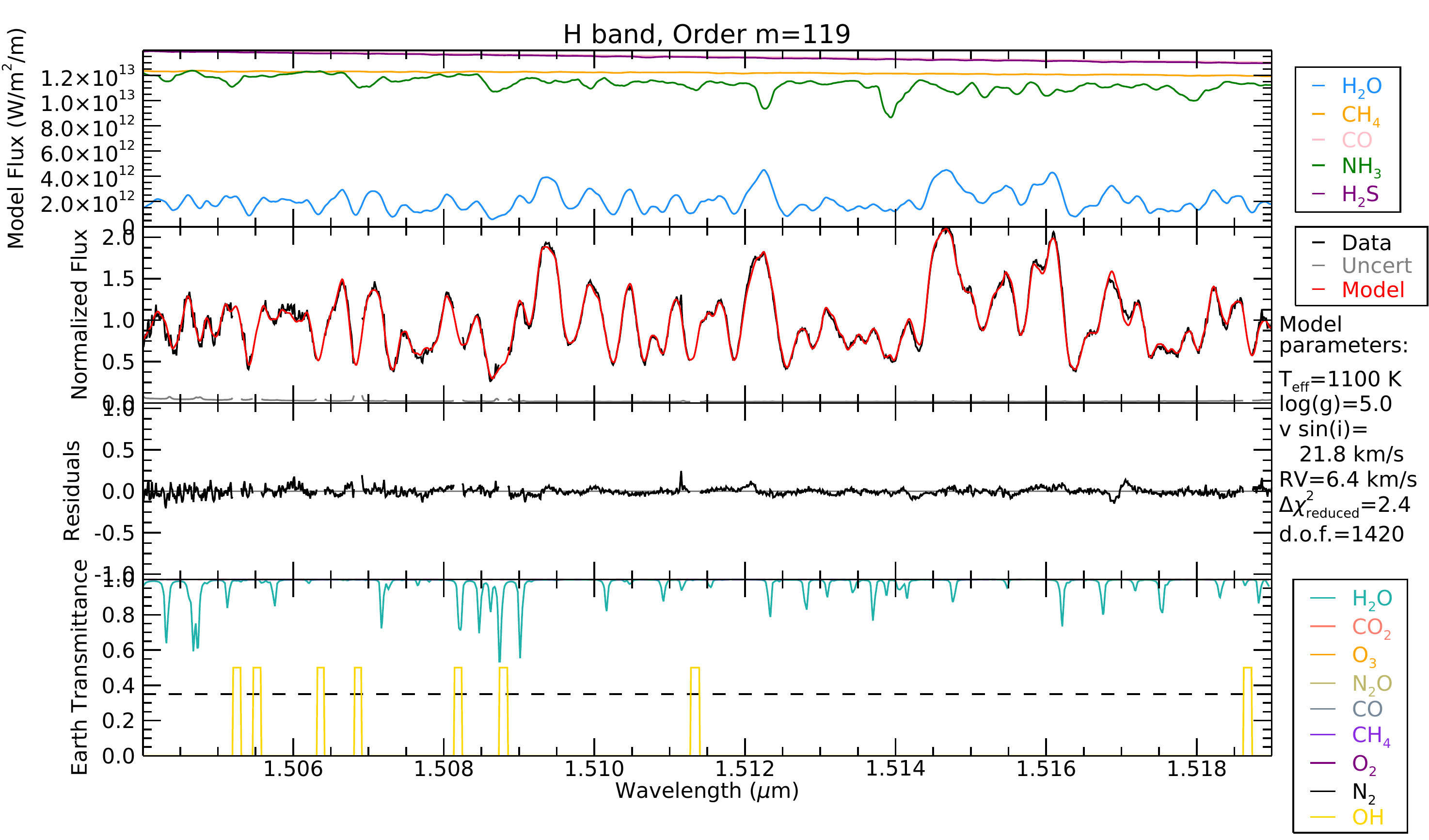} 
    \end{subfigure}
    \caption{Continued.}
\end{figure*}

\begin{figure*}
    \ContinuedFloat 
    \centering
    \begin{subfigure}{0.99\textwidth}
        \includegraphics[width=\textwidth]{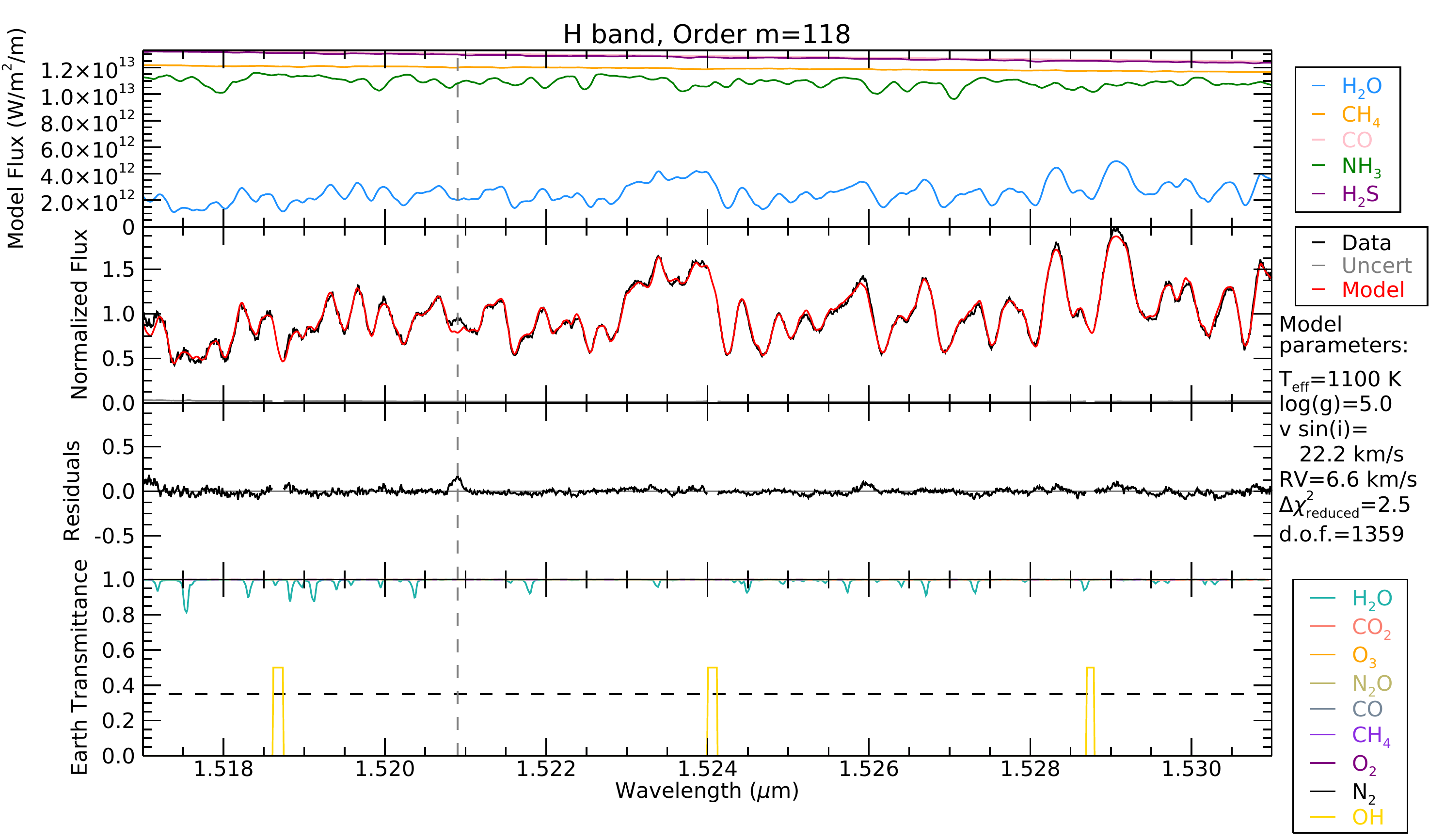} 
    \end{subfigure}
    \begin{subfigure}{0.99\textwidth}
        \includegraphics[width=\textwidth]{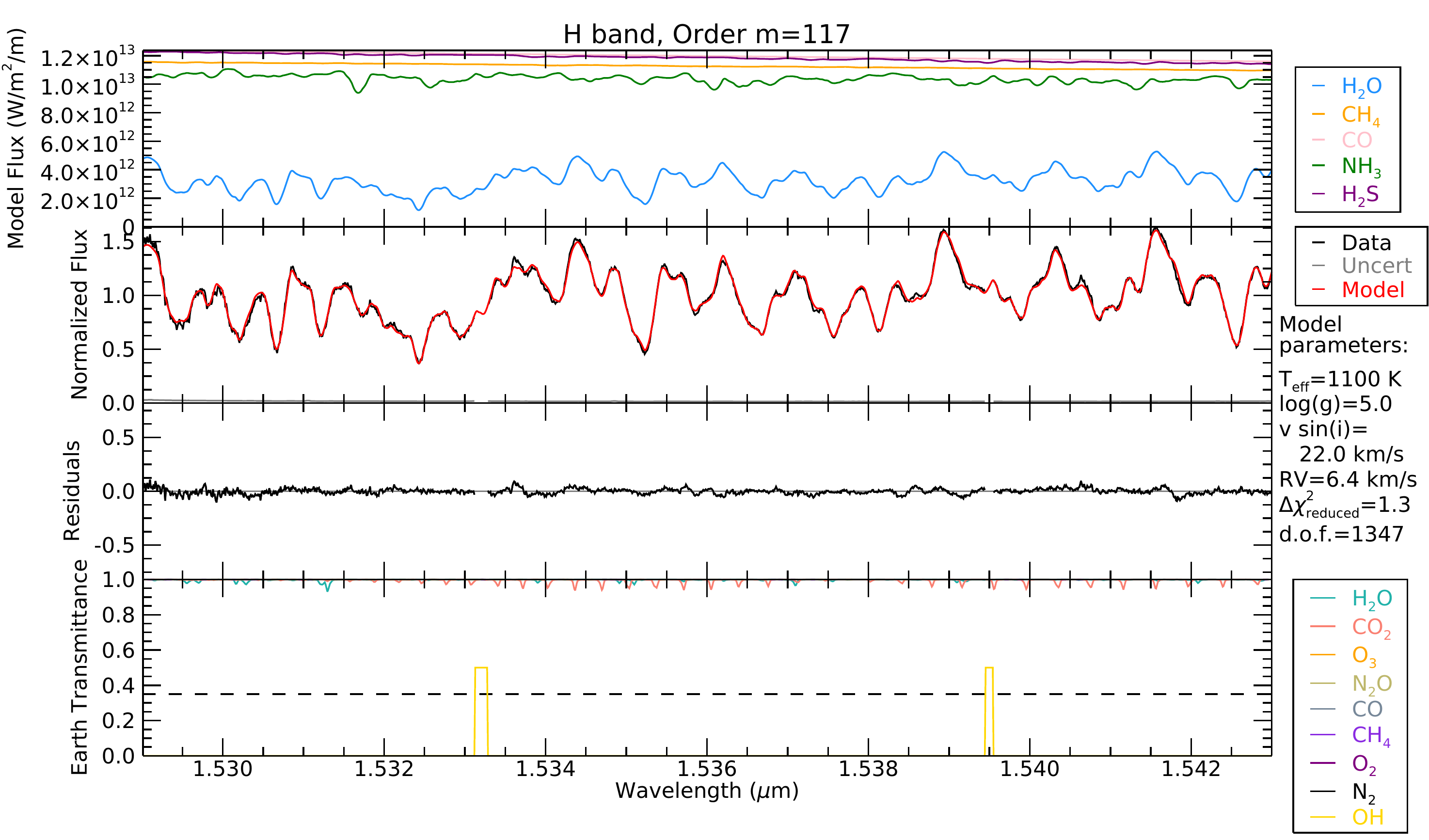} 
    \end{subfigure}
    \caption{Continued.}
\end{figure*}

\begin{figure*}
    \ContinuedFloat 
    \centering
    \begin{subfigure}{0.99\textwidth}
        \includegraphics[width=\textwidth]{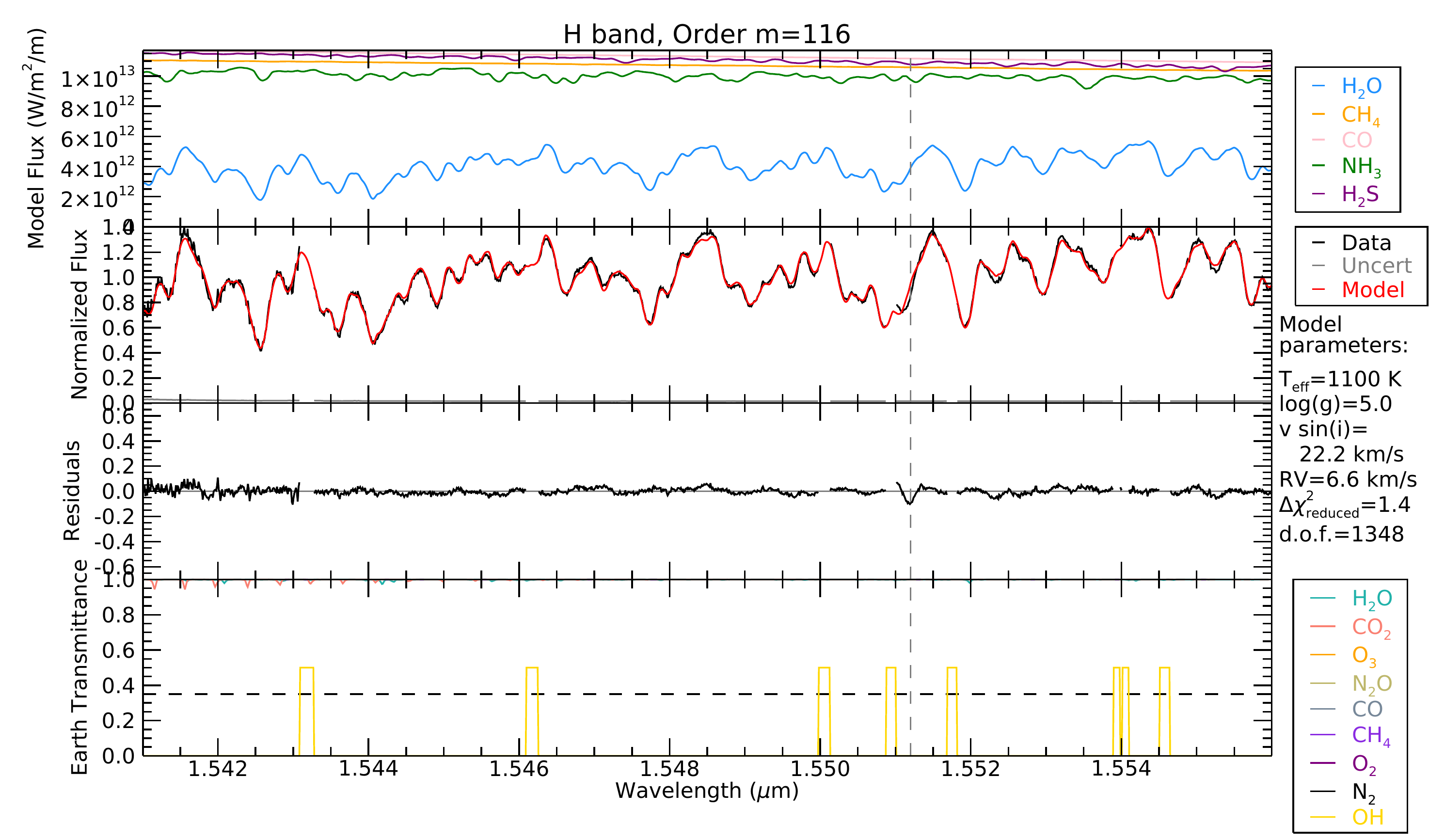} 
    \end{subfigure}
    \begin{subfigure}{0.99\textwidth}
        \includegraphics[width=\textwidth]{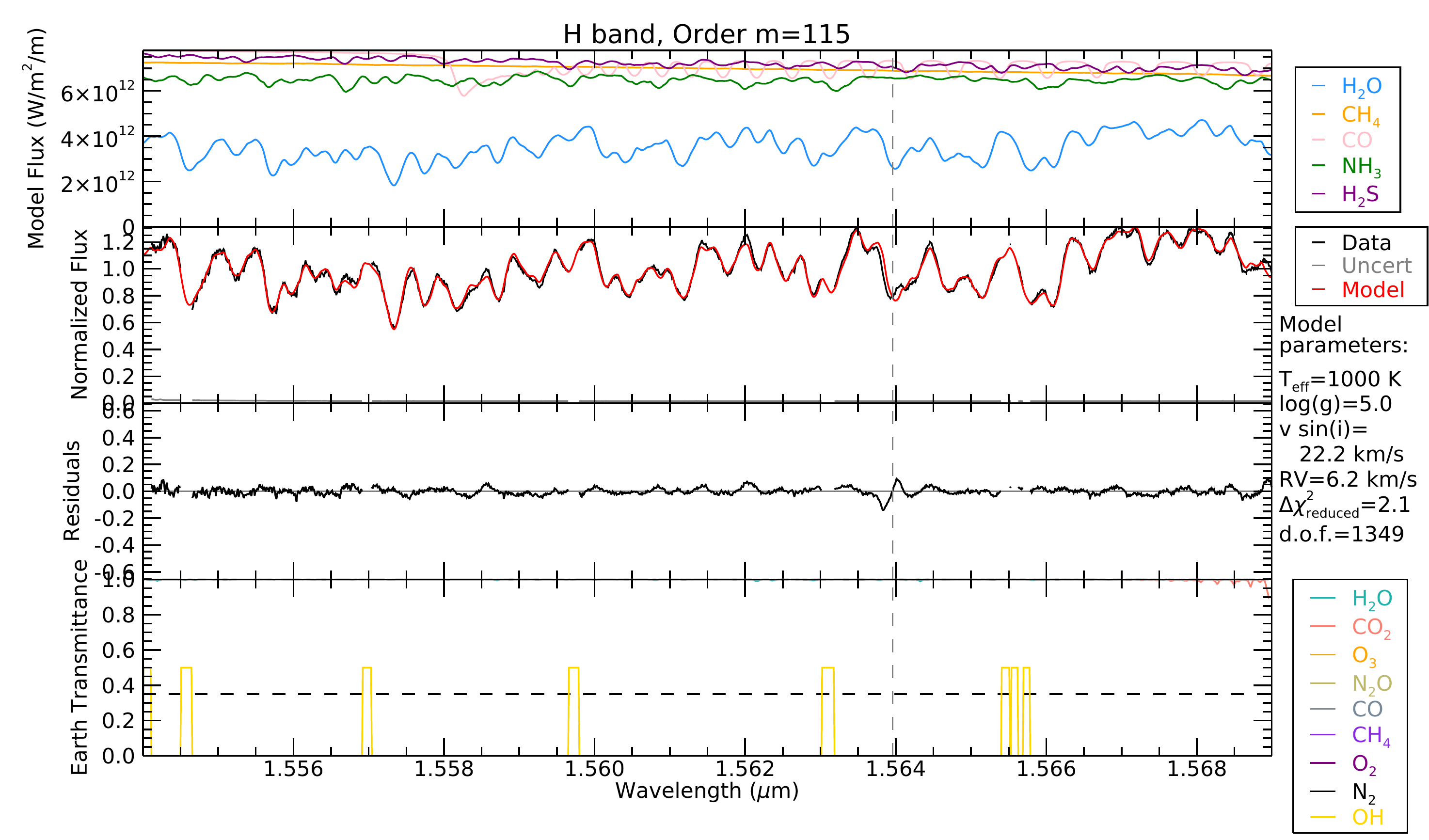} 
    \end{subfigure}
    \caption{Continued.}
\end{figure*}

\begin{figure*}
    \ContinuedFloat 
    \centering
    \begin{subfigure}{0.99\textwidth}
        \includegraphics[width=\textwidth]{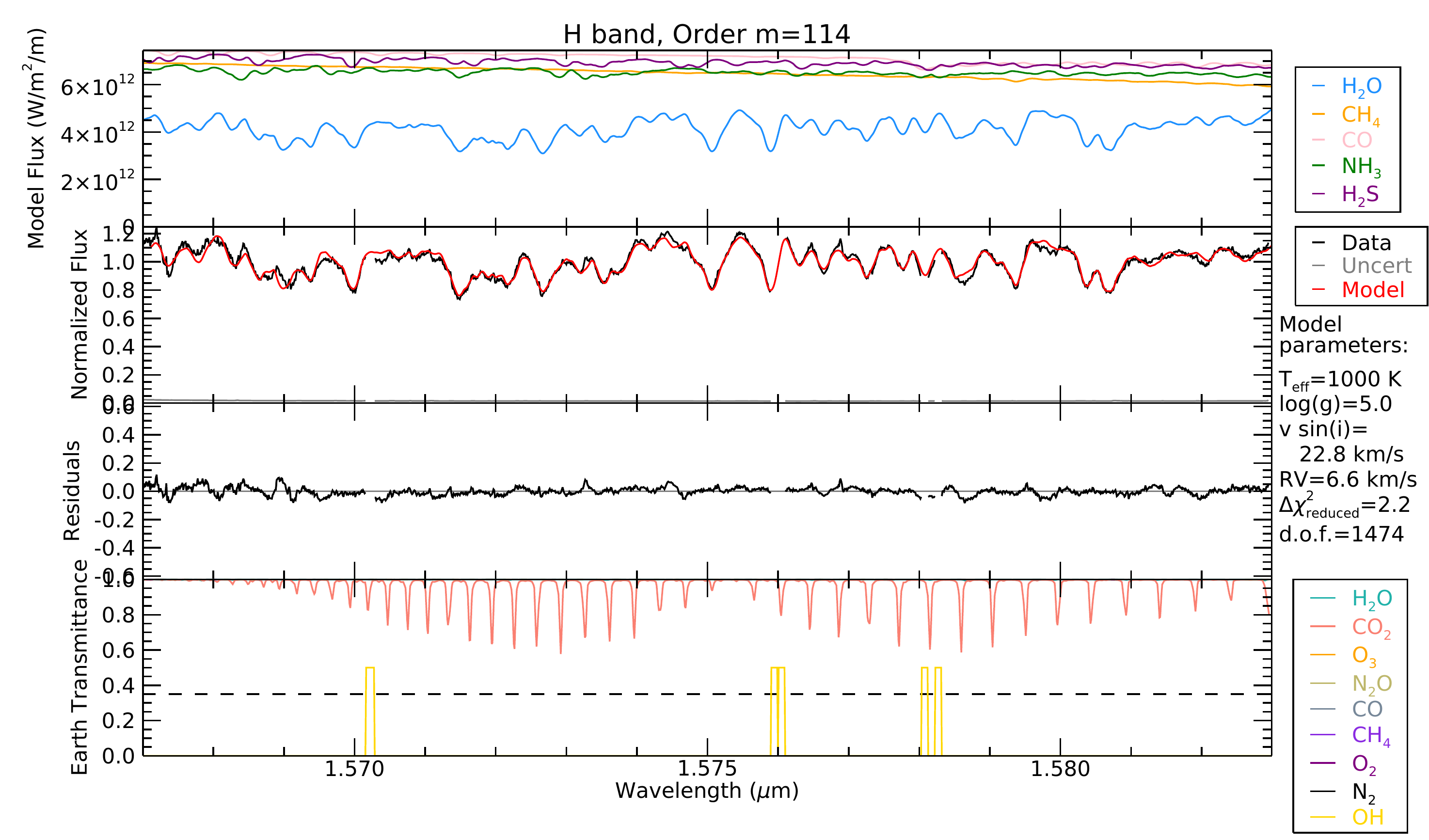} 
    \end{subfigure}
    \begin{subfigure}{0.99\textwidth}
        \includegraphics[width=\textwidth]{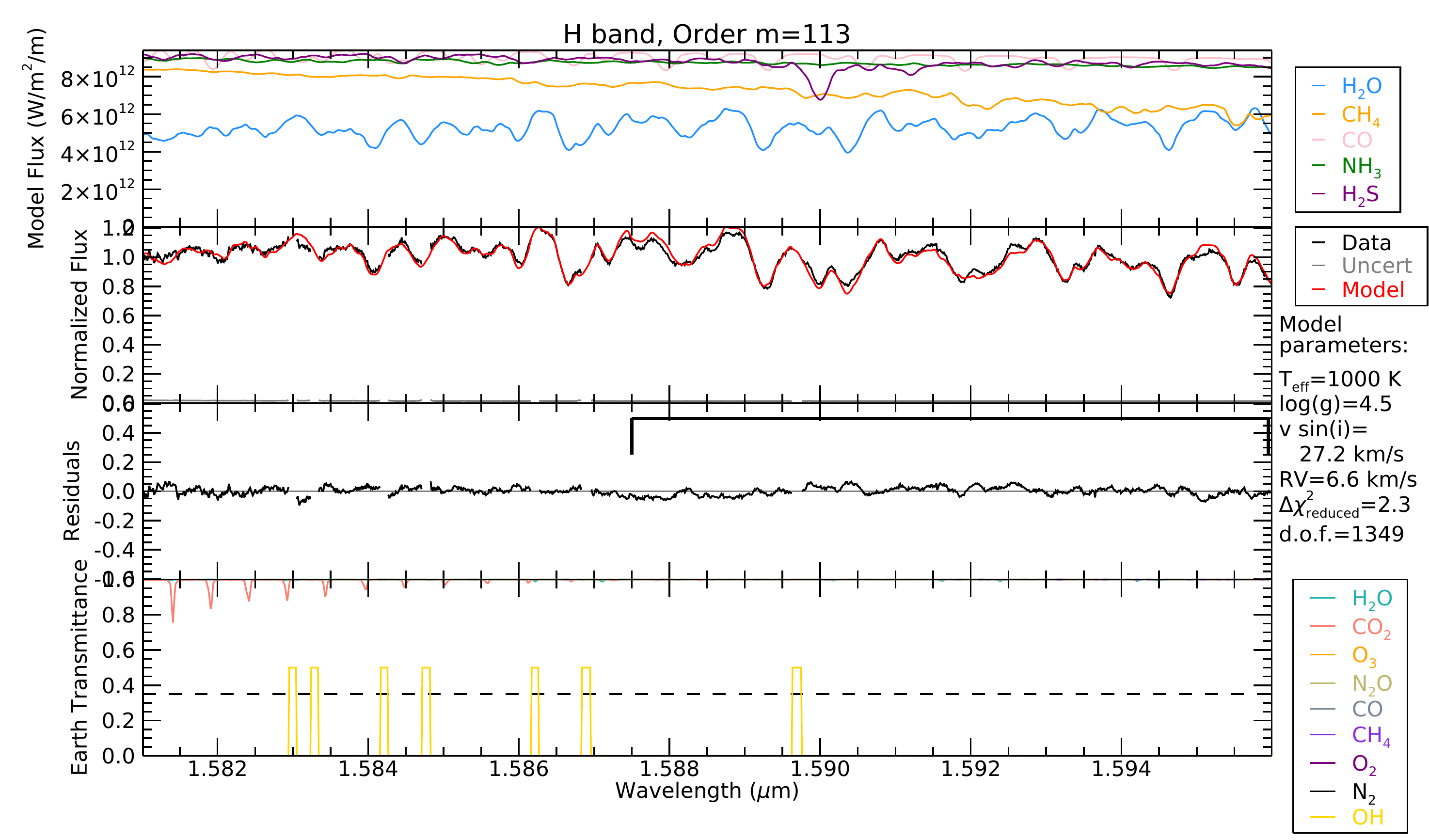} 
    \end{subfigure}
    \caption{Continued.}
\end{figure*}

\begin{figure*}
    \ContinuedFloat 
    \centering
    \begin{subfigure}{0.99\textwidth}
        \includegraphics[width=\textwidth]{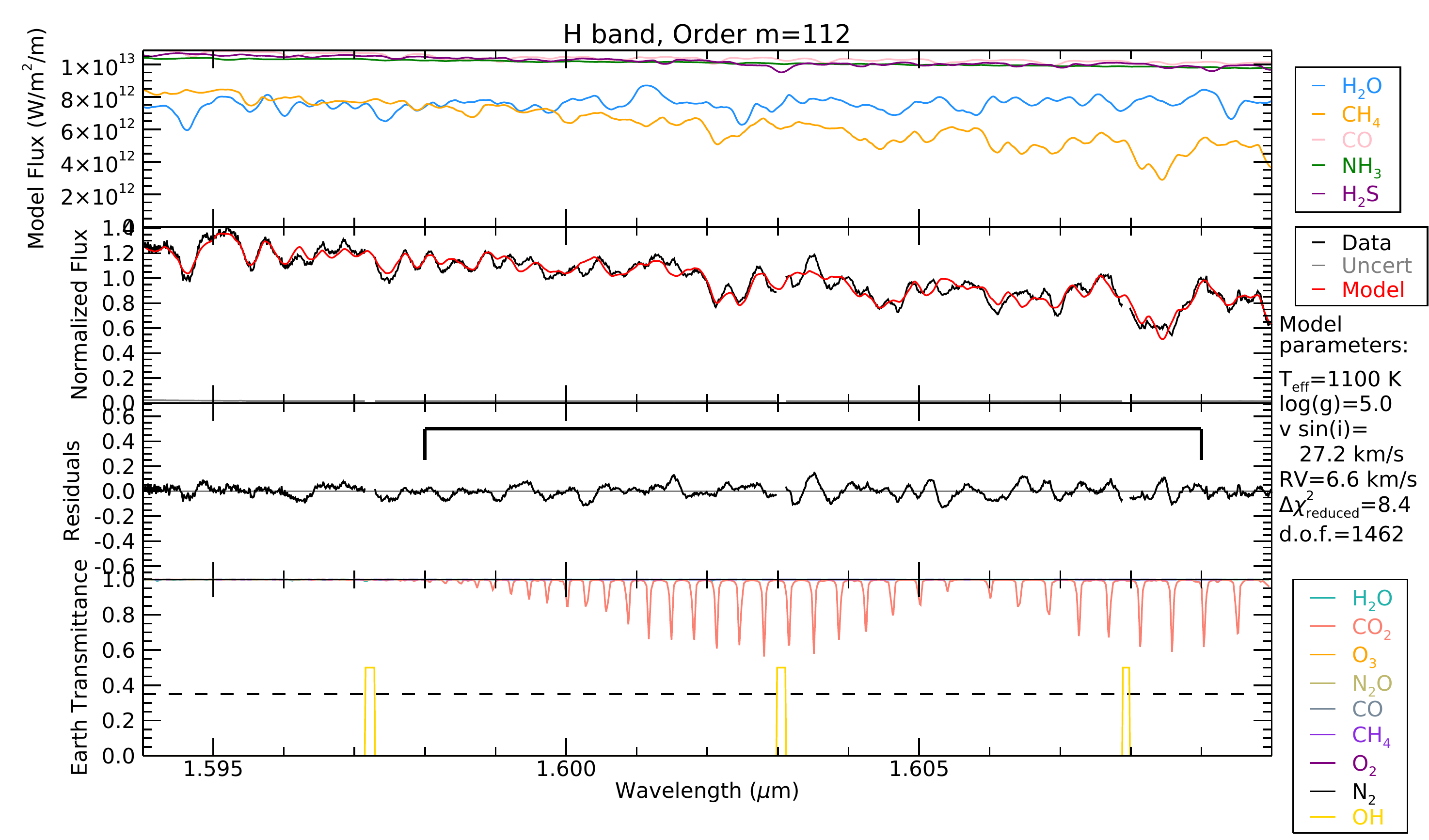} 
    \end{subfigure}
    \begin{subfigure}{0.99\textwidth}
        \includegraphics[width=\textwidth]{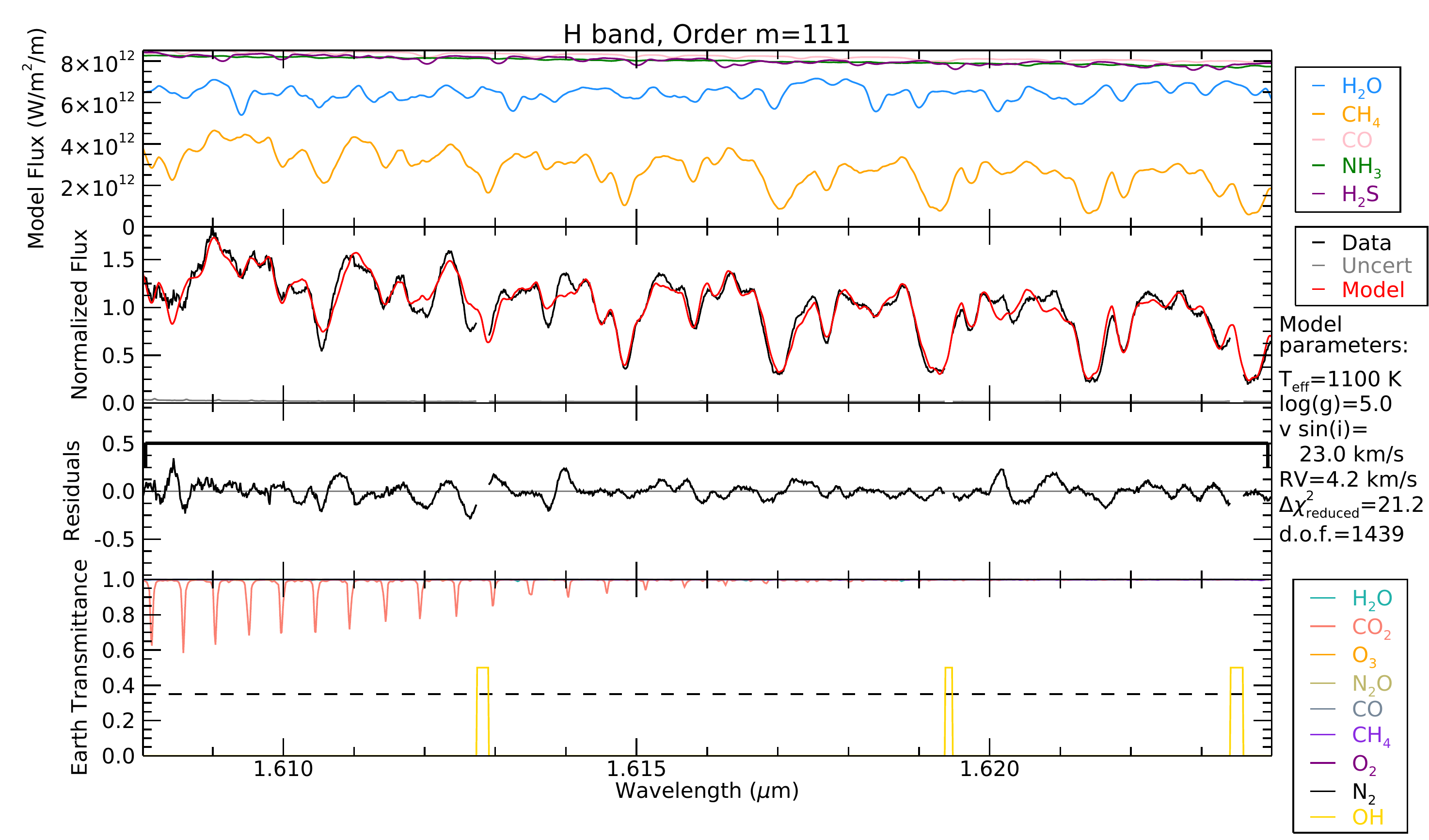} 
    \end{subfigure}
    \caption{Continued.}
\end{figure*}

\begin{figure*}
    \ContinuedFloat 
    \centering
    \begin{subfigure}{0.99\textwidth}
        \includegraphics[width=\textwidth]{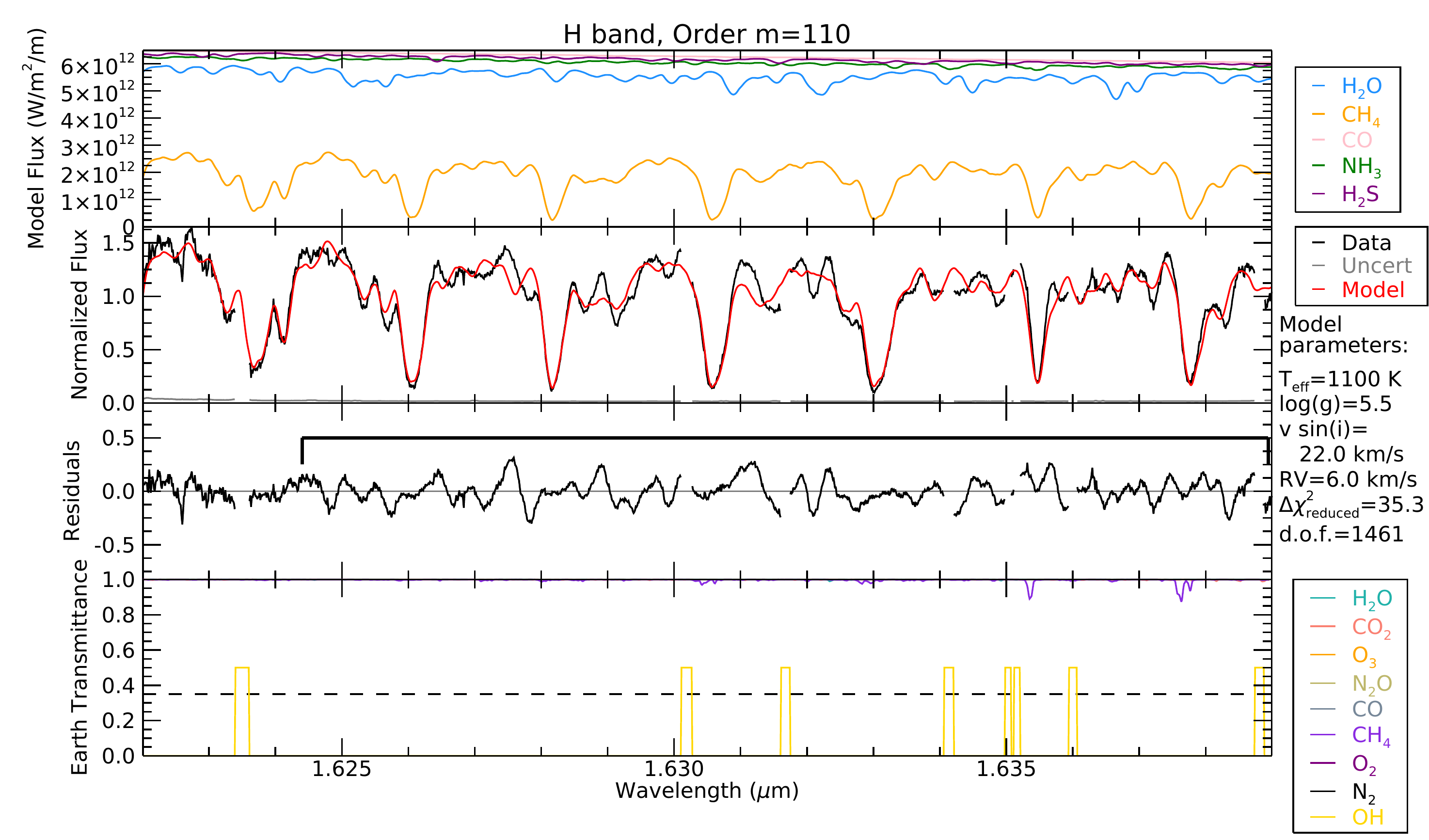} 
    \end{subfigure}
    \begin{subfigure}{0.99\textwidth}
        \includegraphics[width=\textwidth]{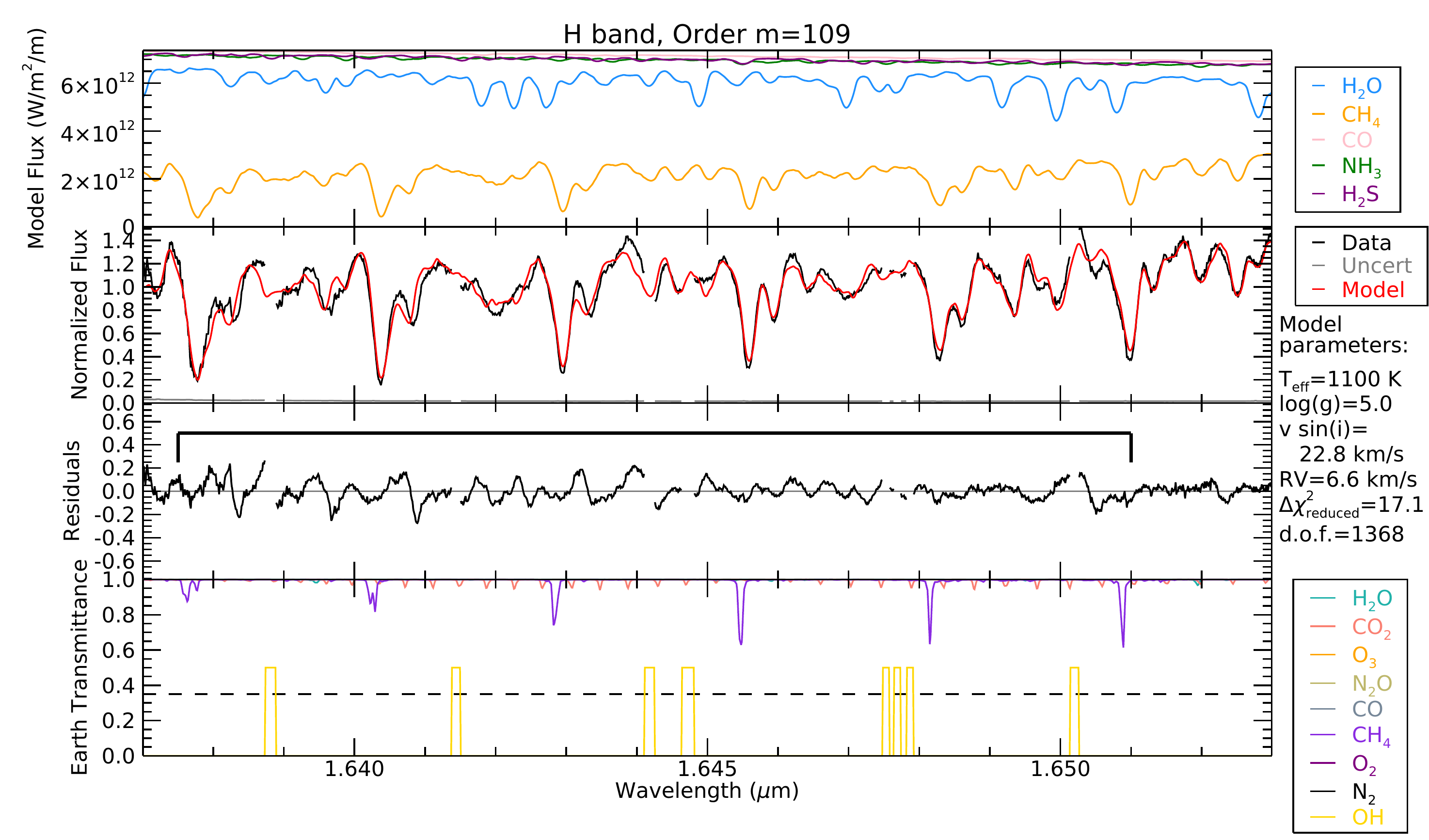} 
    \end{subfigure}
    \caption{Continued.}
\end{figure*}

\begin{figure*}
    \ContinuedFloat 
    \centering
    \begin{subfigure}{0.99\textwidth}
        \includegraphics[width=\textwidth]{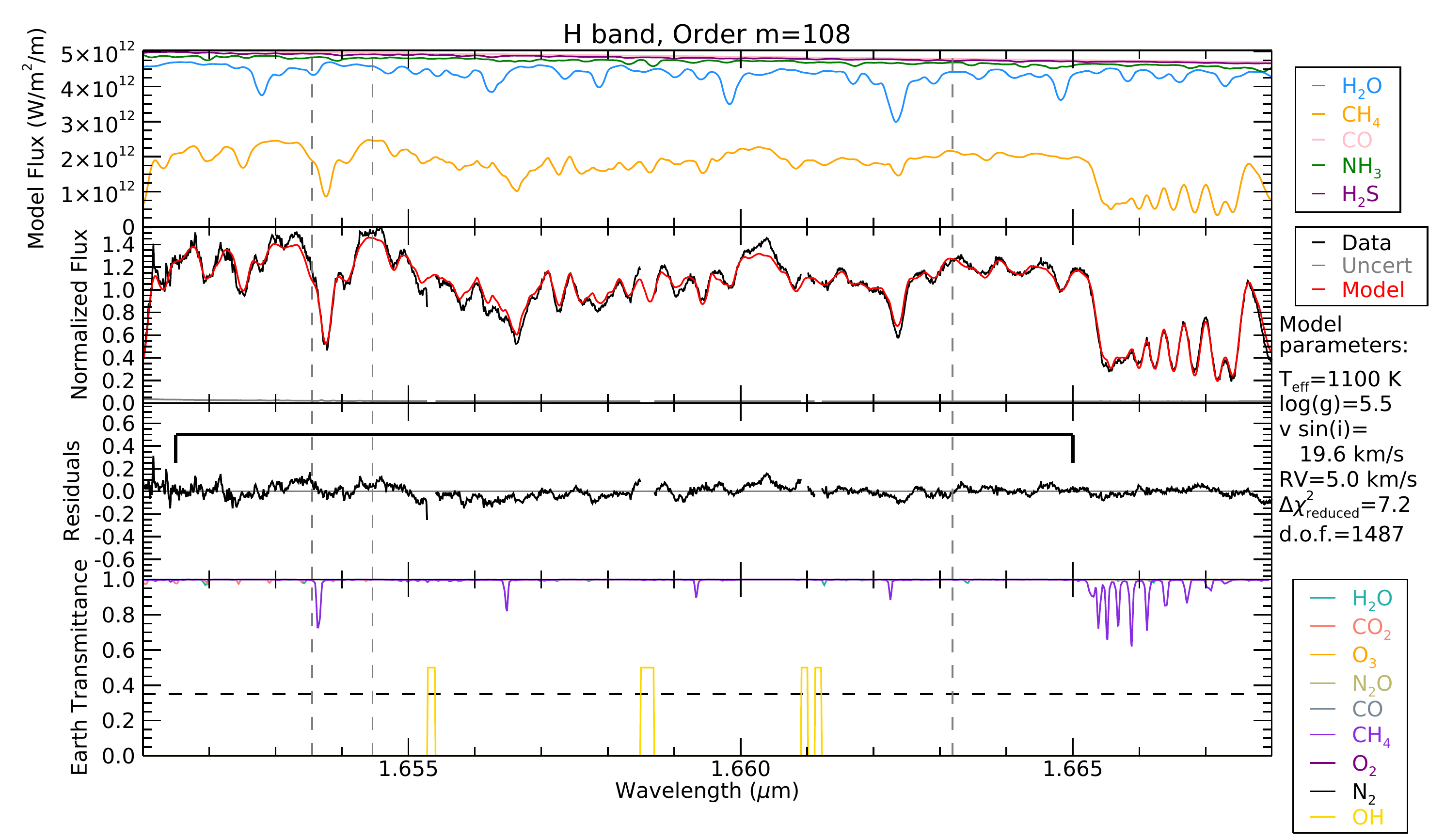} 
    \end{subfigure}
    \begin{subfigure}{0.99\textwidth}
        \includegraphics[width=\textwidth]{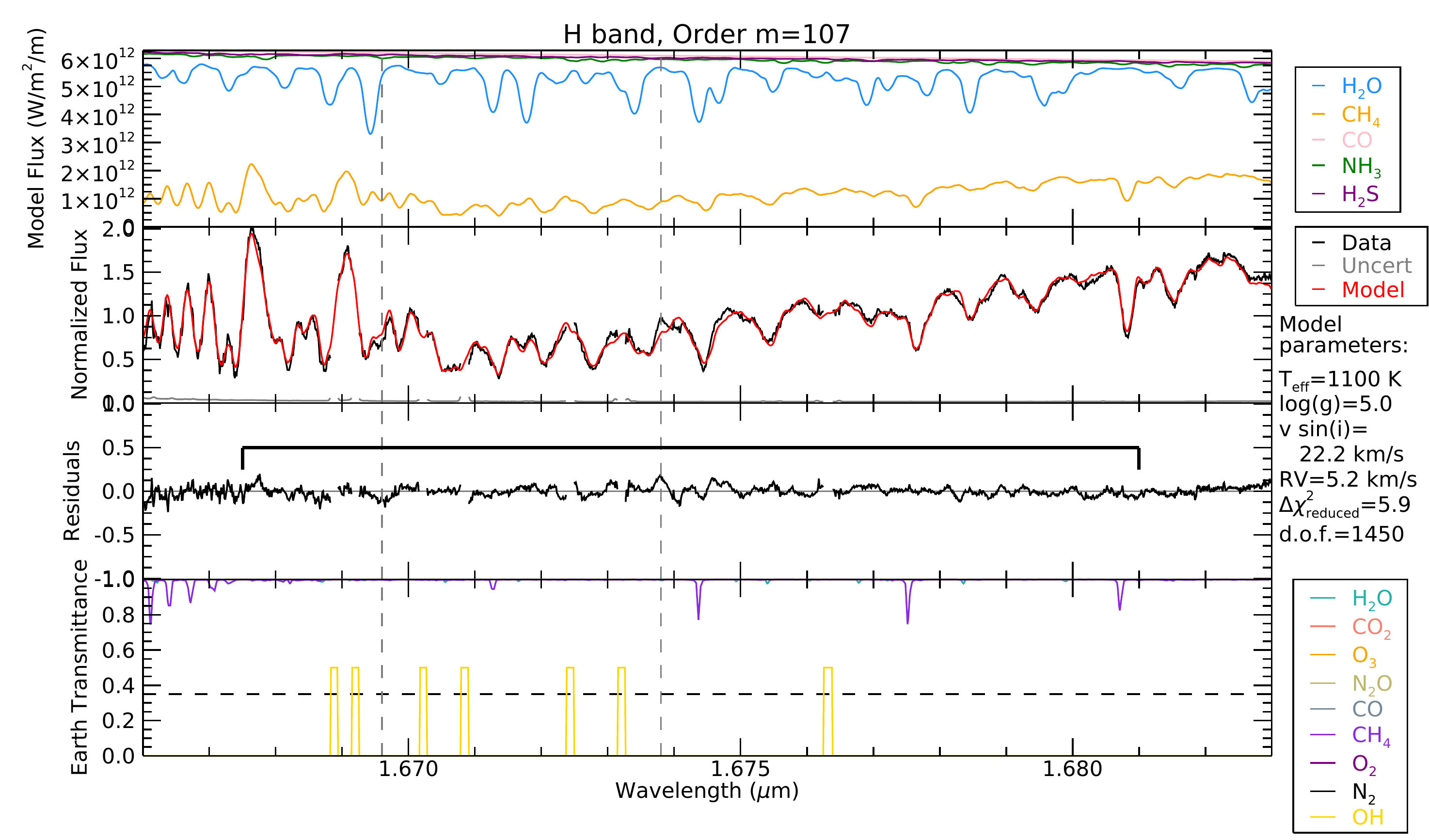} 
    \end{subfigure}
    \caption{Continued.}
\end{figure*}

\begin{figure*}
    \ContinuedFloat 
    \centering
    \begin{subfigure}{0.99\textwidth}
        \includegraphics[width=\textwidth]{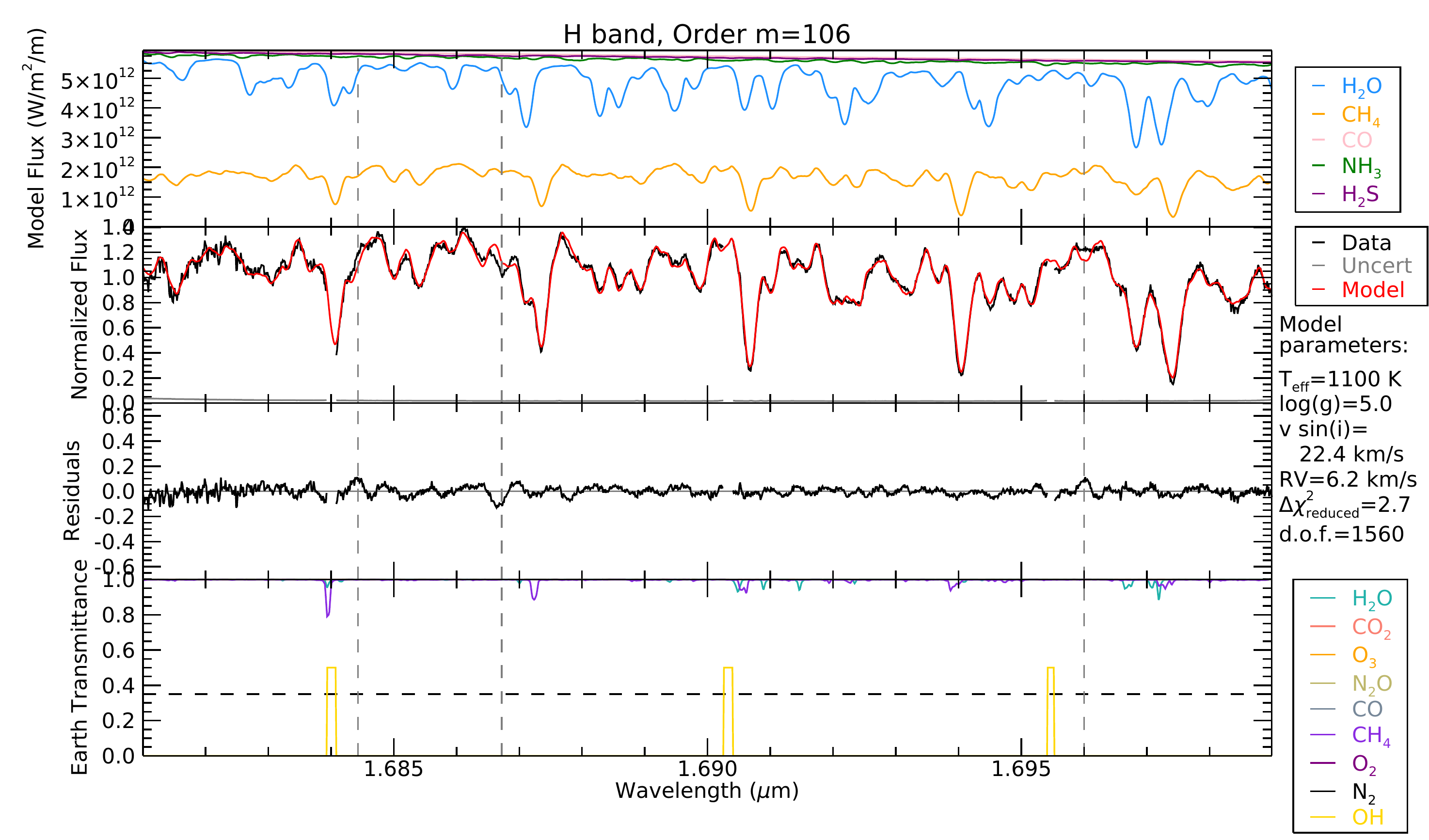} 
    \end{subfigure}
    \begin{subfigure}{0.99\textwidth}
        \includegraphics[width=\textwidth]{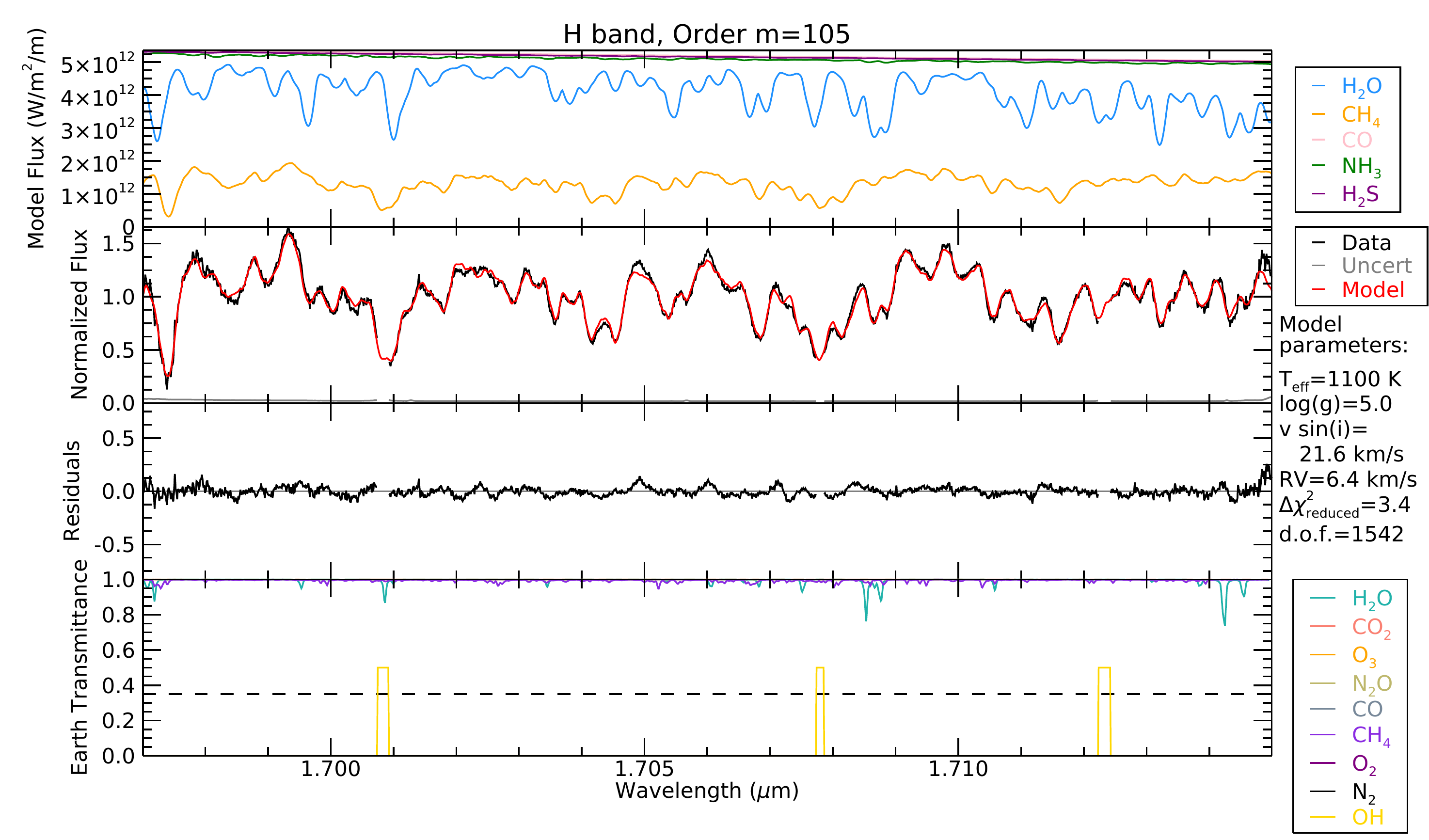} 
    \end{subfigure}
    \caption{Continued.}
\end{figure*}

\begin{figure*}
    \ContinuedFloat 
    \centering
    \begin{subfigure}{0.99\textwidth}
        \includegraphics[width=\textwidth]{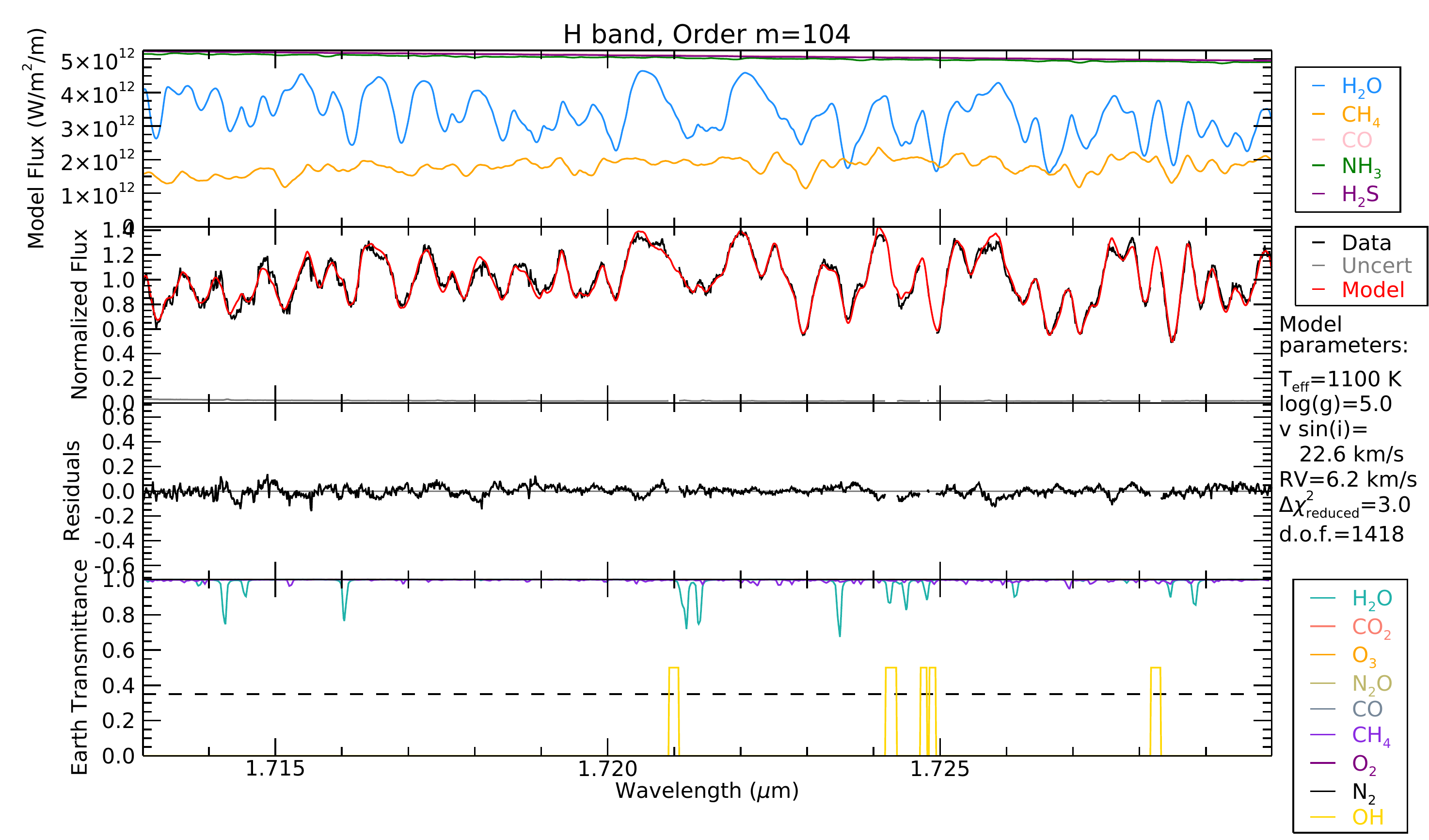} 
    \end{subfigure}
    \begin{subfigure}{0.99\textwidth}
        \includegraphics[width=\textwidth]{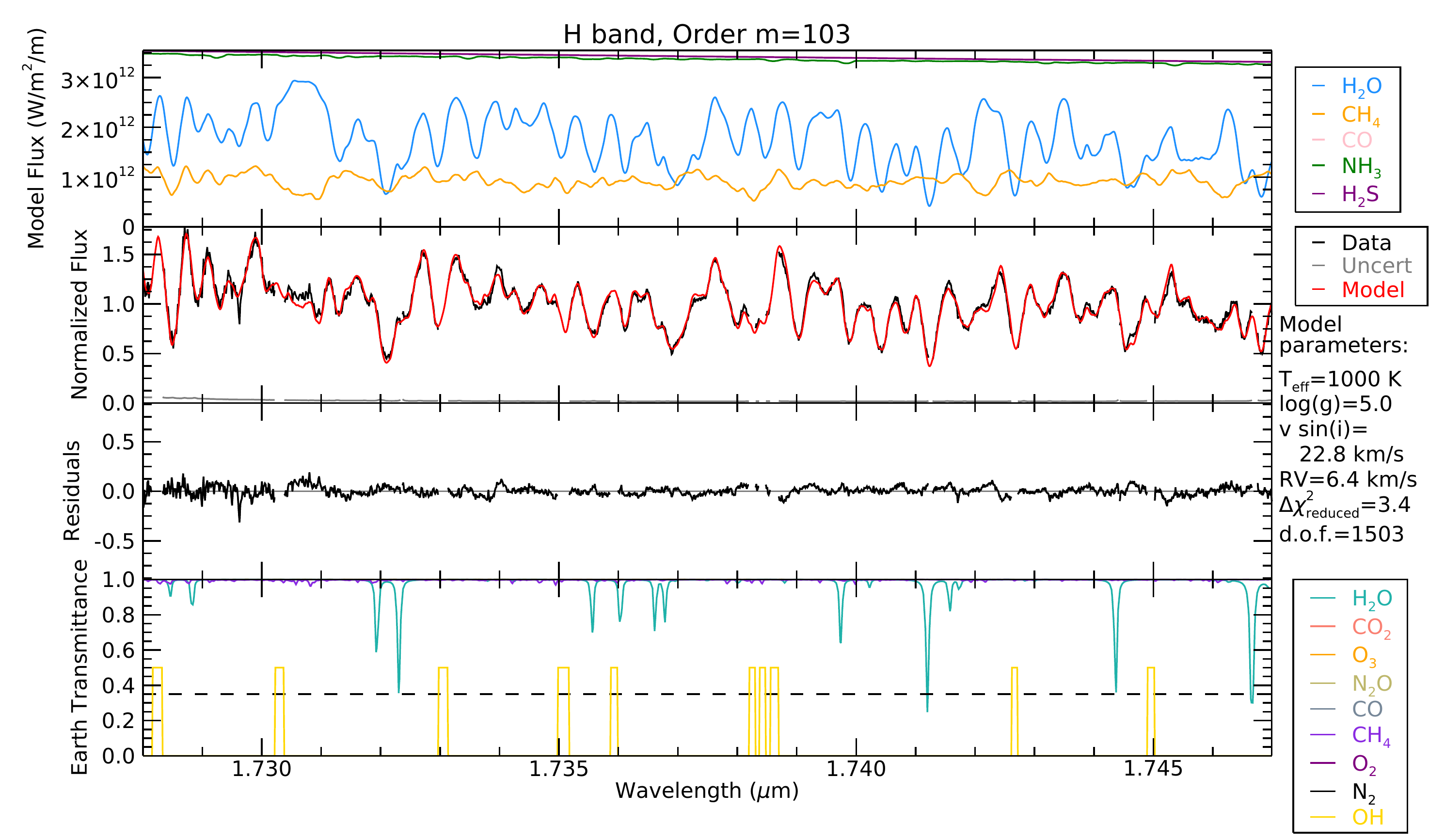} 
    \end{subfigure}
    \caption{Continued.}
\end{figure*}

\begin{figure*}
    \ContinuedFloat 
    \centering
    \begin{subfigure}{0.99\textwidth}
        \includegraphics[width=\textwidth]{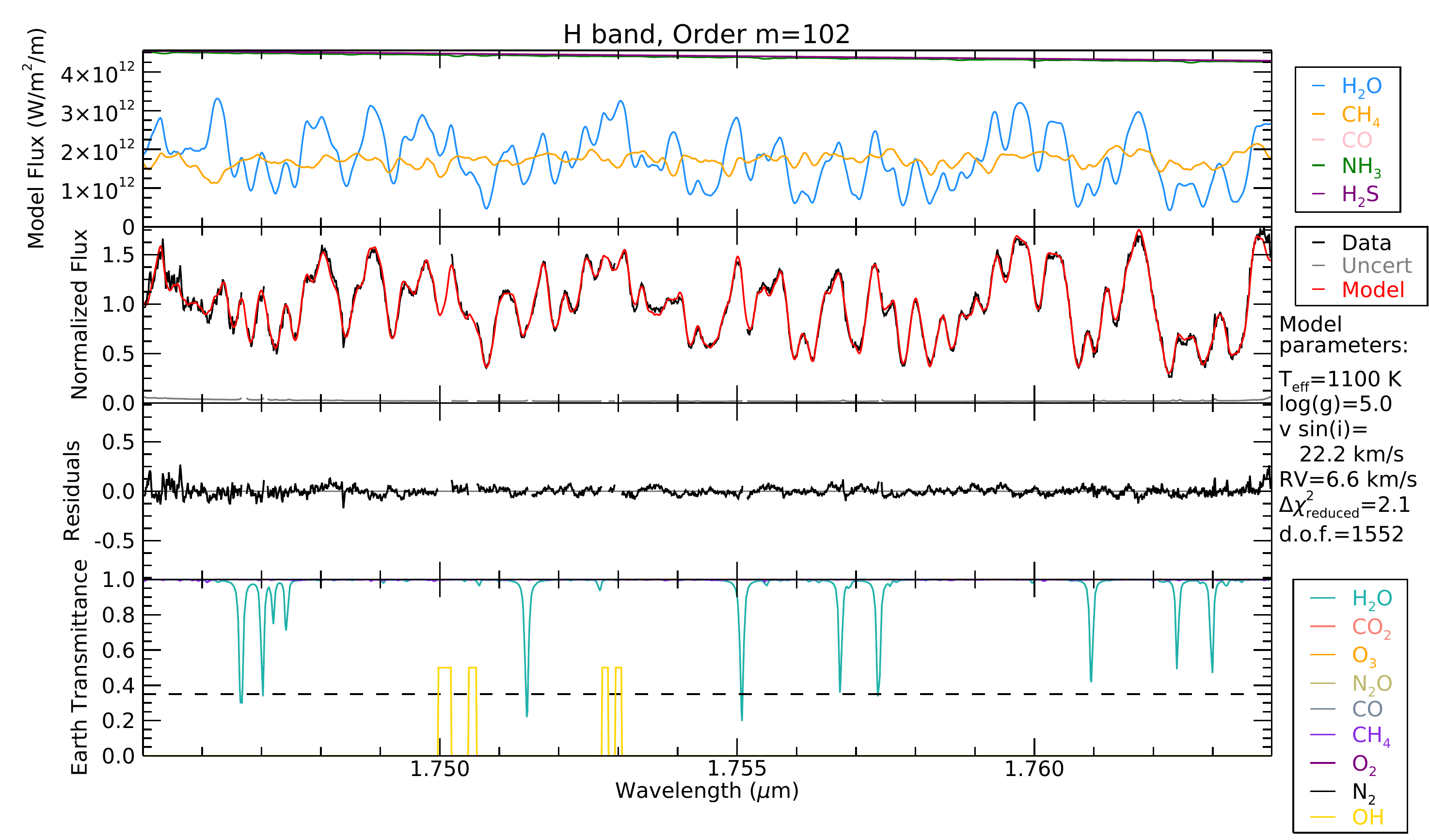} 
    \end{subfigure}
    \begin{subfigure}{0.99\textwidth}
        \includegraphics[width=\textwidth]{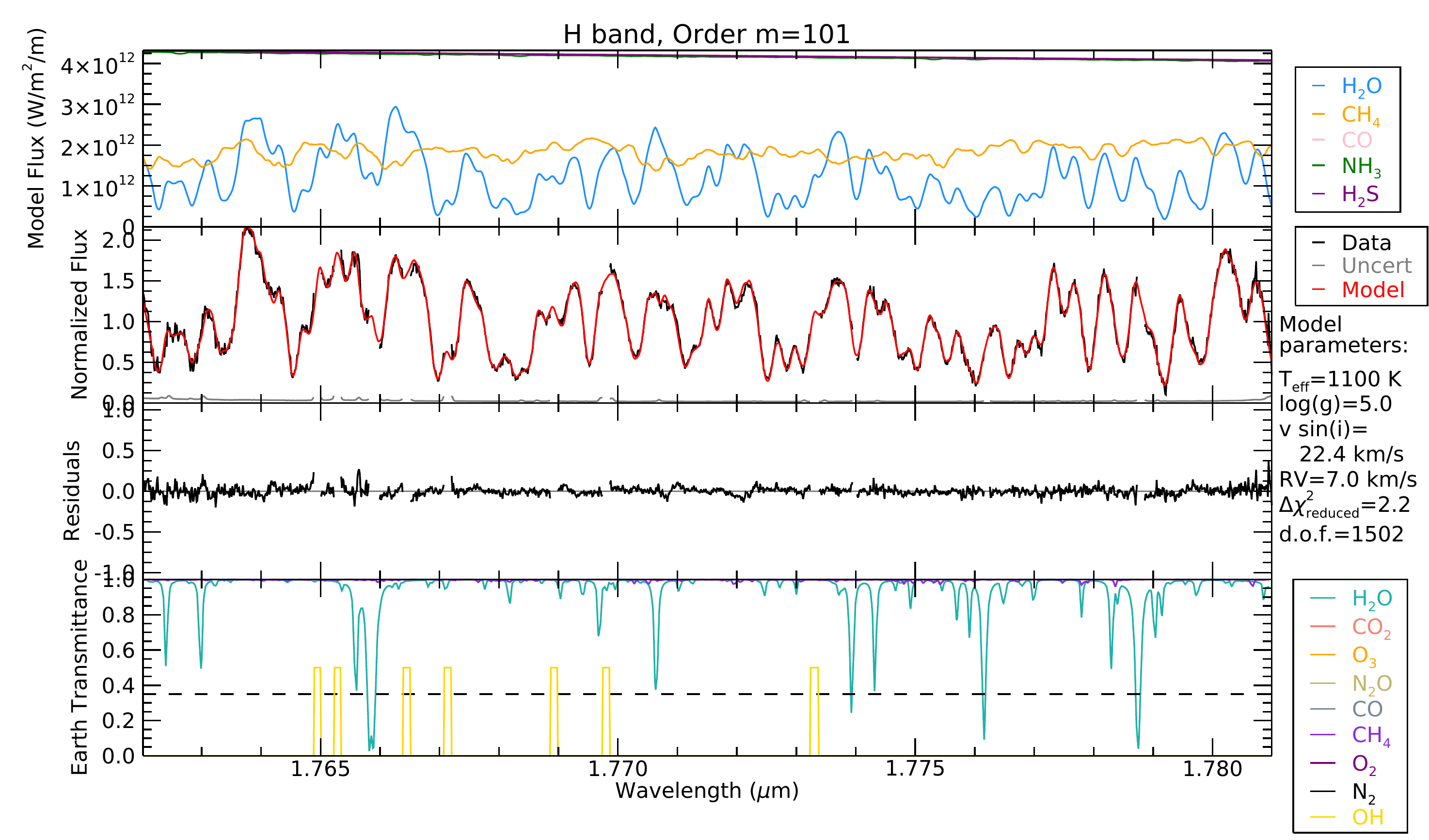} 
    \end{subfigure}
    \caption{Continued.}
\end{figure*}

\begin{figure*}
    \ContinuedFloat 
    \centering
    \begin{subfigure}{0.99\textwidth}
        \includegraphics[width=\textwidth]{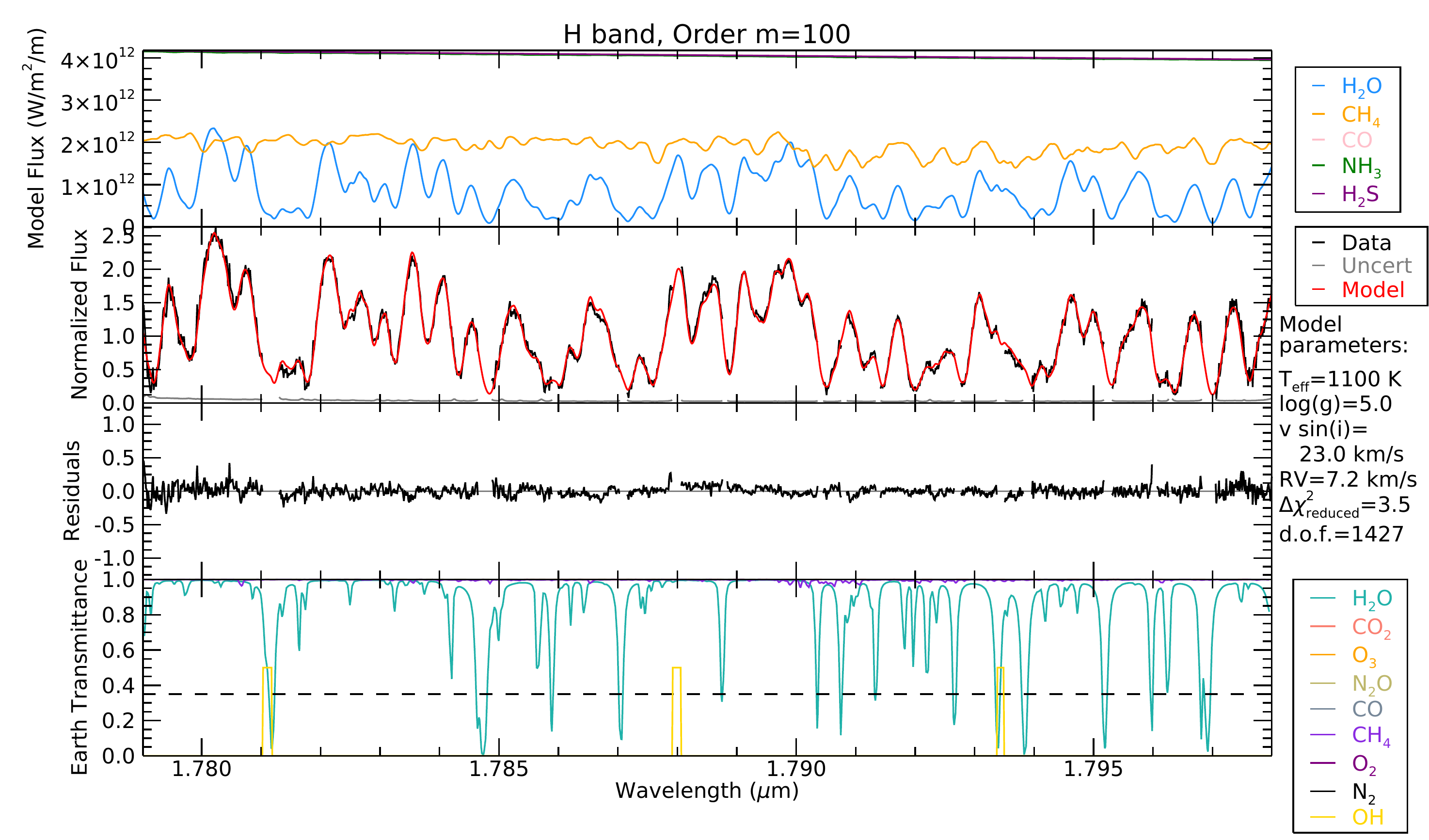} 
    \end{subfigure}
    \begin{subfigure}{0.99\textwidth}
        \includegraphics[width=\textwidth]{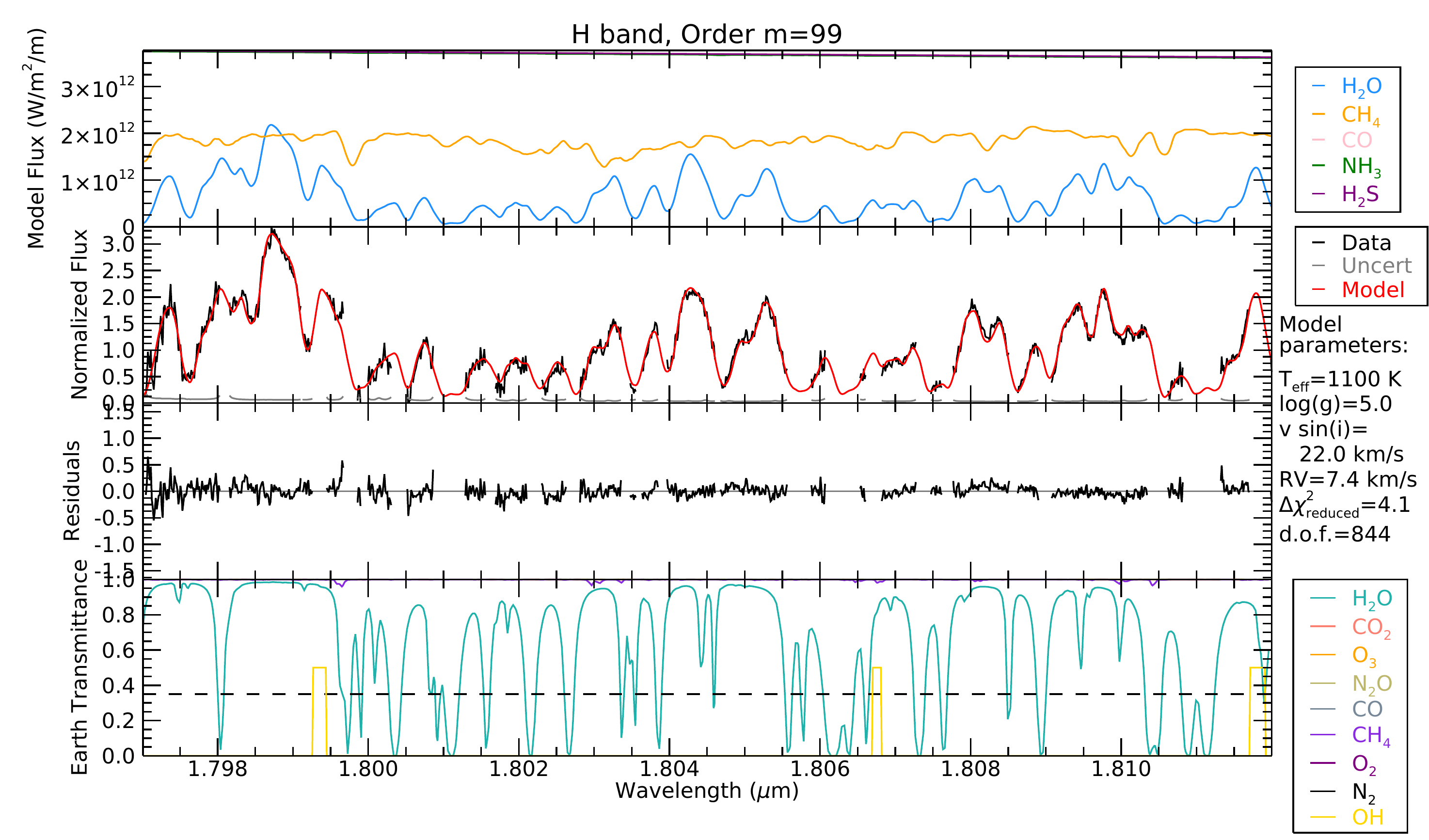} 
    \end{subfigure}
    \caption{Continued.}
\end{figure*}


\begin{figure*}
    \centering
    \begin{subfigure}{0.99\textwidth}
        \includegraphics[width=\textwidth]{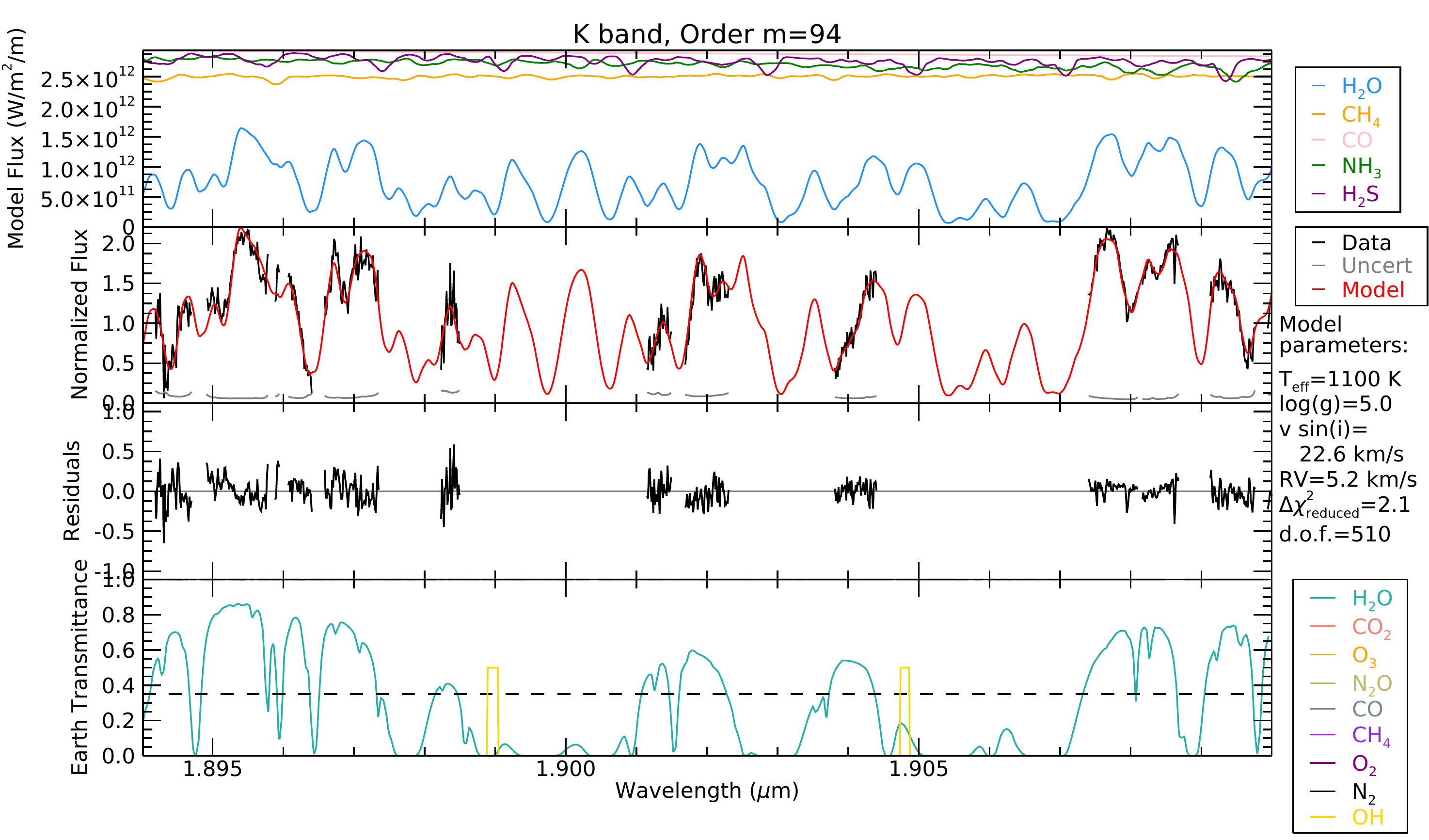} 
    \end{subfigure}
    \begin{subfigure}{0.99\textwidth}
        \includegraphics[width=\textwidth]{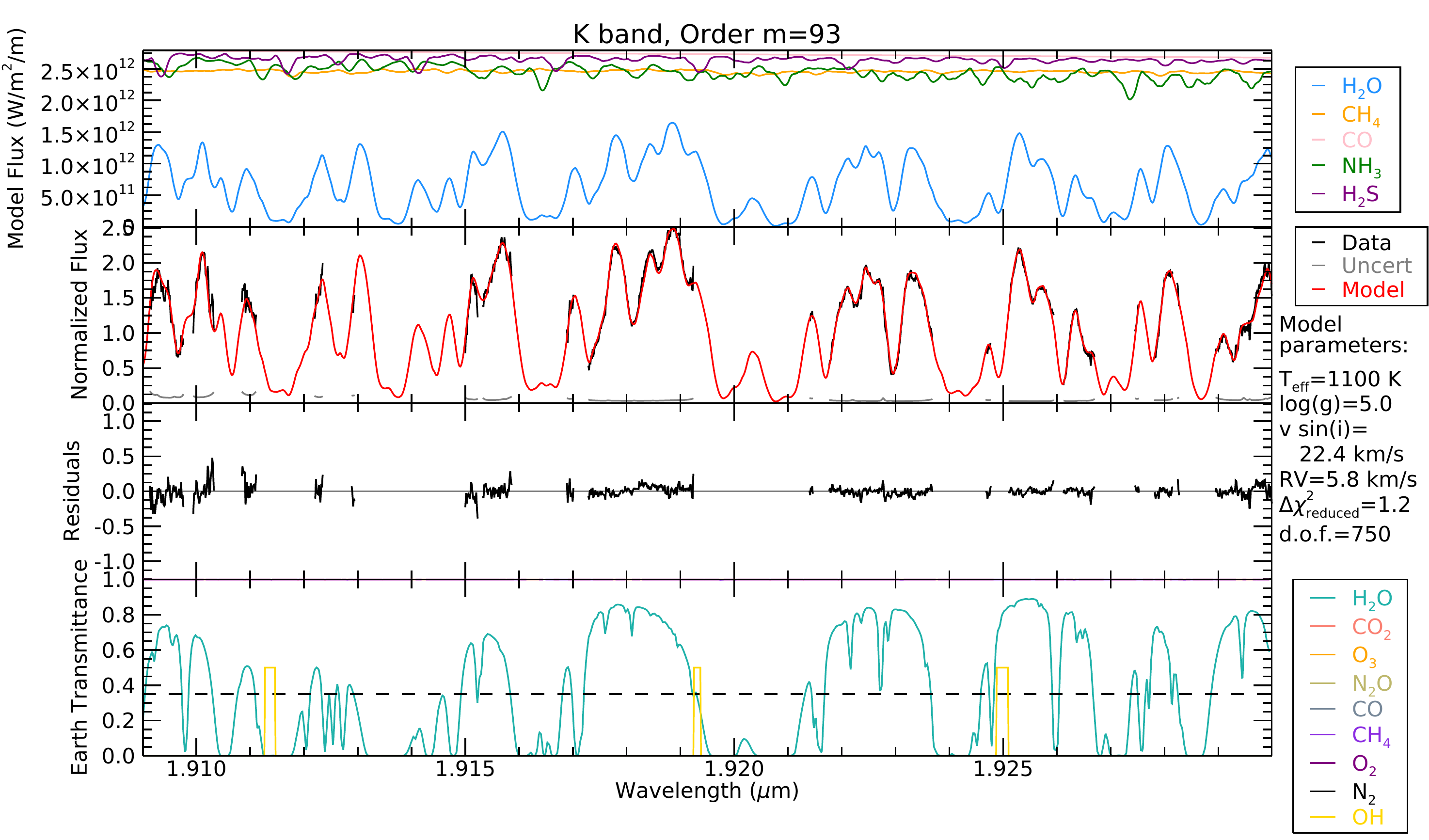} 
    \end{subfigure}
\caption{The same as Fig.~\ref{fig:Horders}, but for the $K$ band. This figure continues for many pages, with two orders per page, to show all 23 orders of the $K$ band. }
\label{fig:Korders}
\end{figure*}

\begin{figure*}
    \ContinuedFloat 
    \centering
    \begin{subfigure}{0.99\textwidth}
        \includegraphics[width=\textwidth]{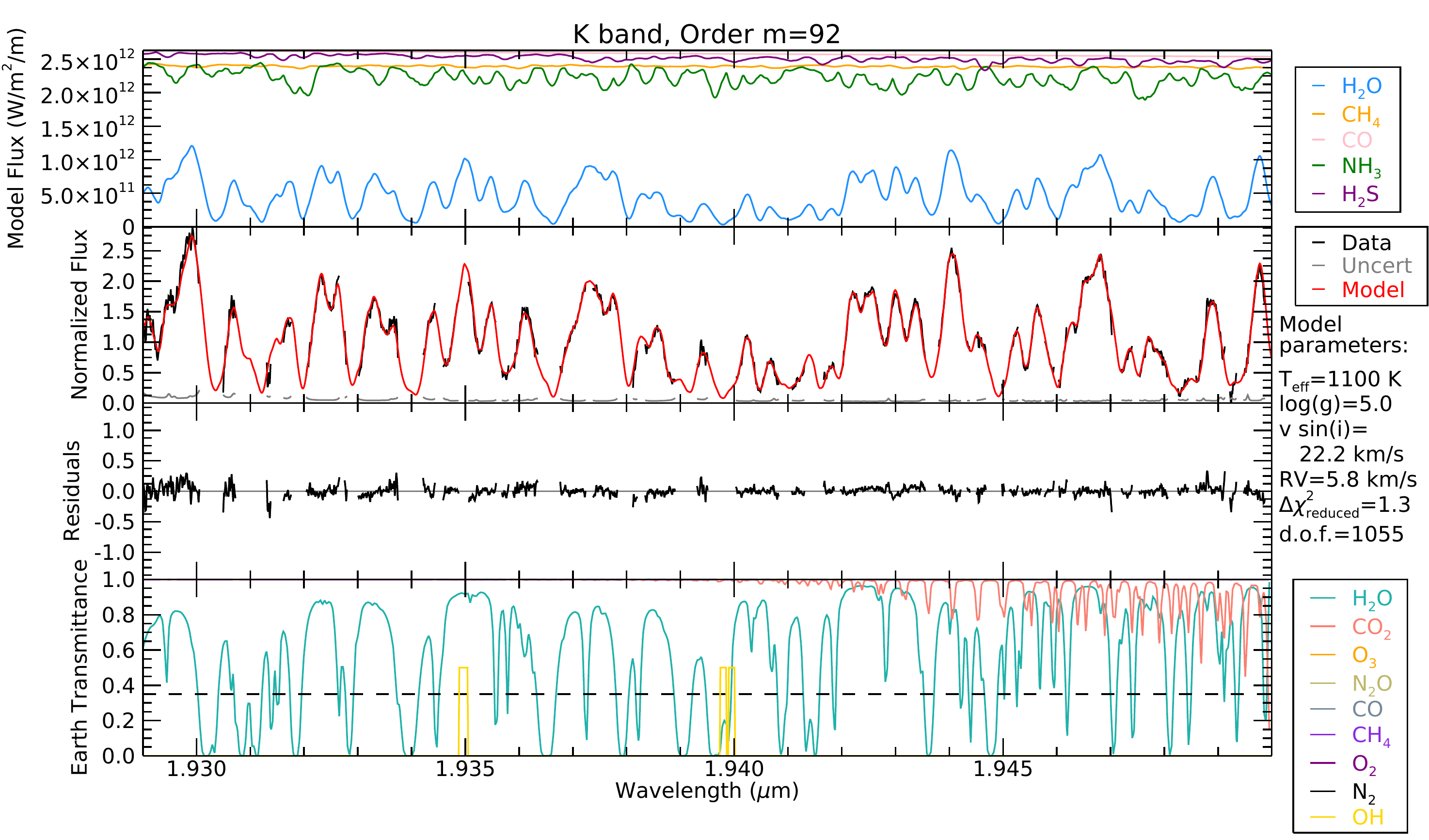} 
    \end{subfigure}
    \begin{subfigure}{0.99\textwidth}
        \includegraphics[width=\textwidth]{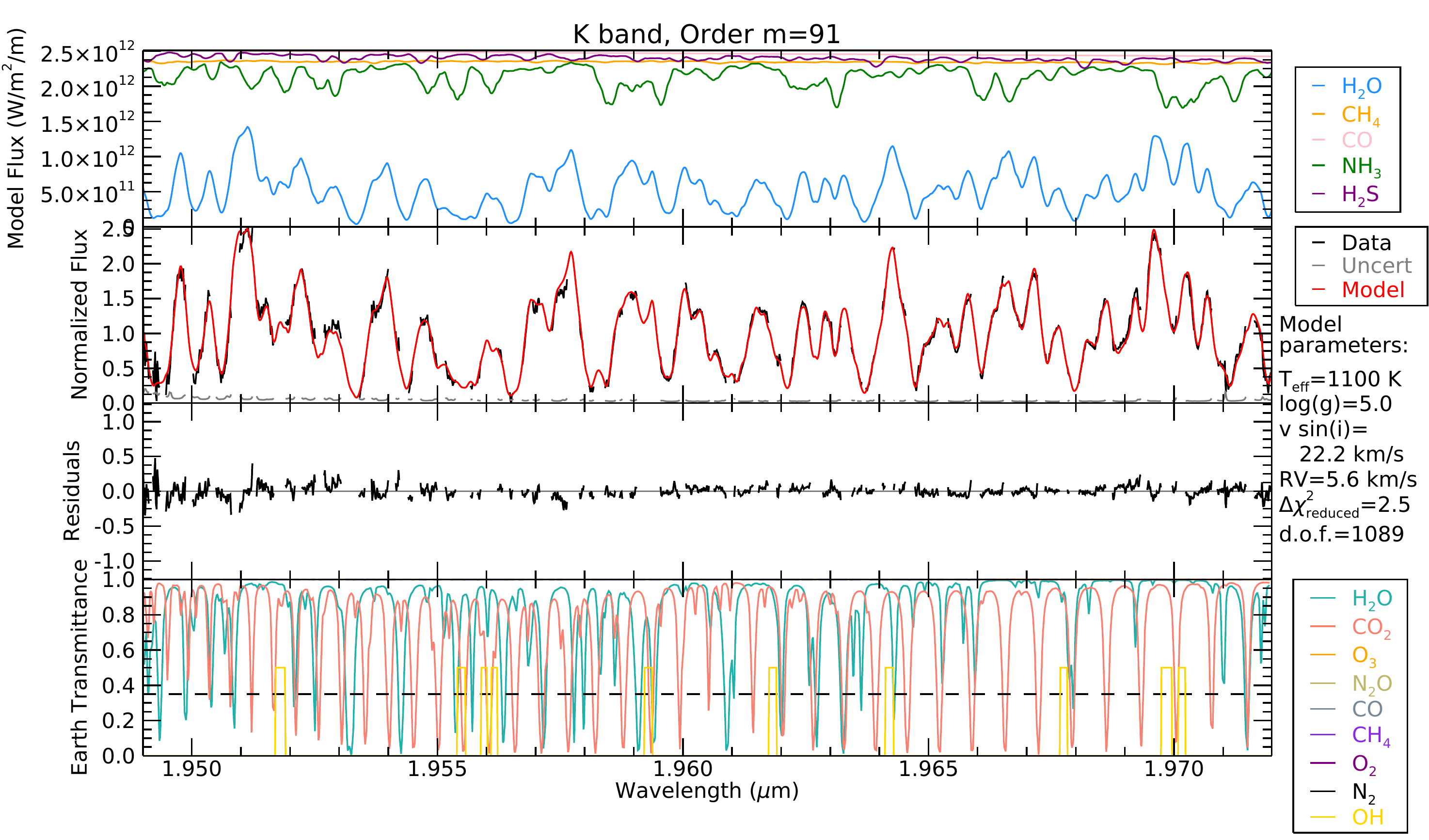} 
    \end{subfigure}
    \caption{Continued.}
\end{figure*}

\begin{figure*}
    \ContinuedFloat 
    \centering
    \begin{subfigure}{0.99\textwidth}
        \includegraphics[width=\textwidth]{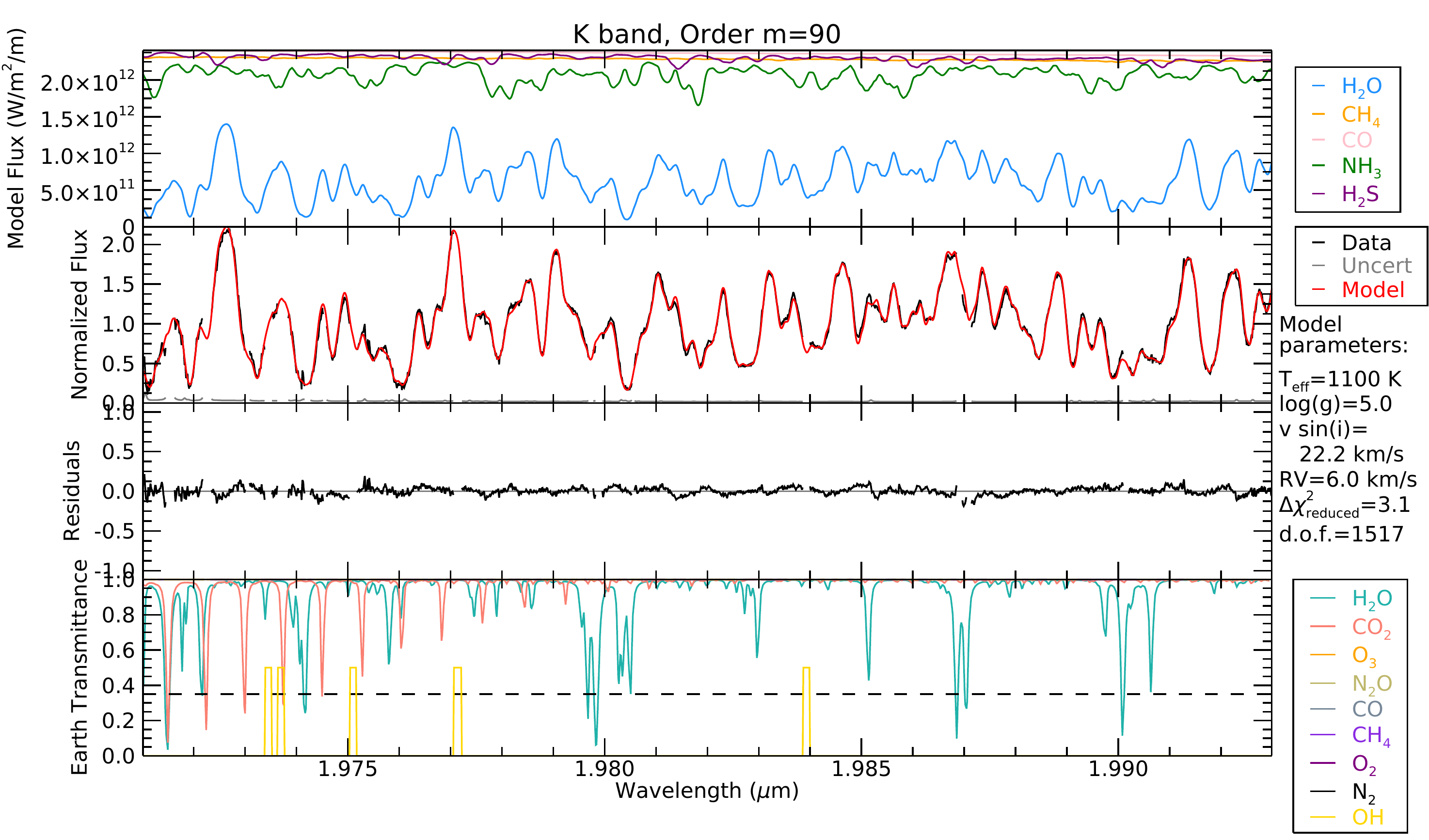} 
    \end{subfigure}
    \begin{subfigure}{0.99\textwidth}
        \includegraphics[width=\textwidth]{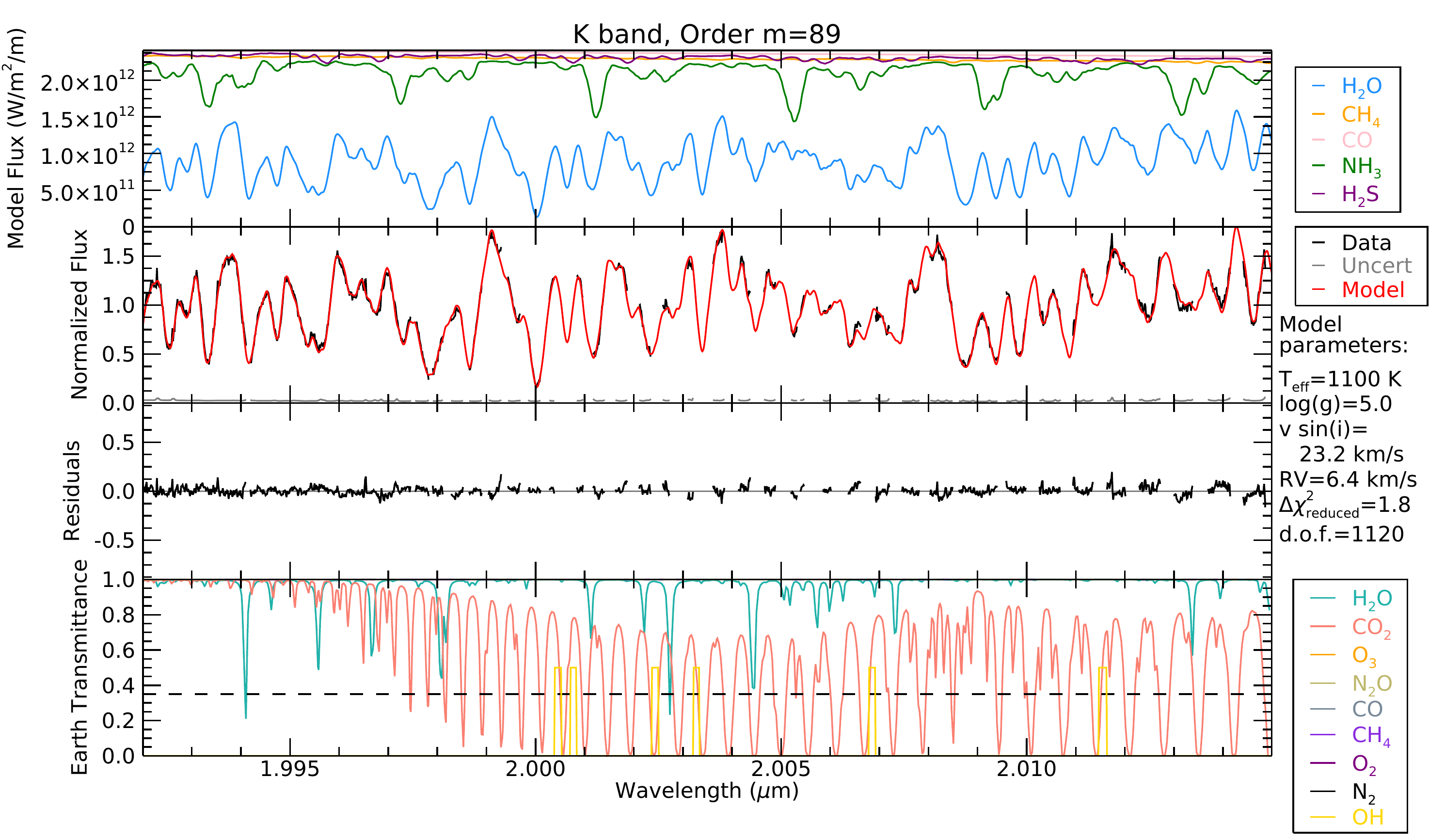} 
    \end{subfigure}
    \caption{Continued.}
\end{figure*}

\begin{figure*}
    \ContinuedFloat 
    \centering
    \begin{subfigure}{0.99\textwidth}
        \includegraphics[width=\textwidth]{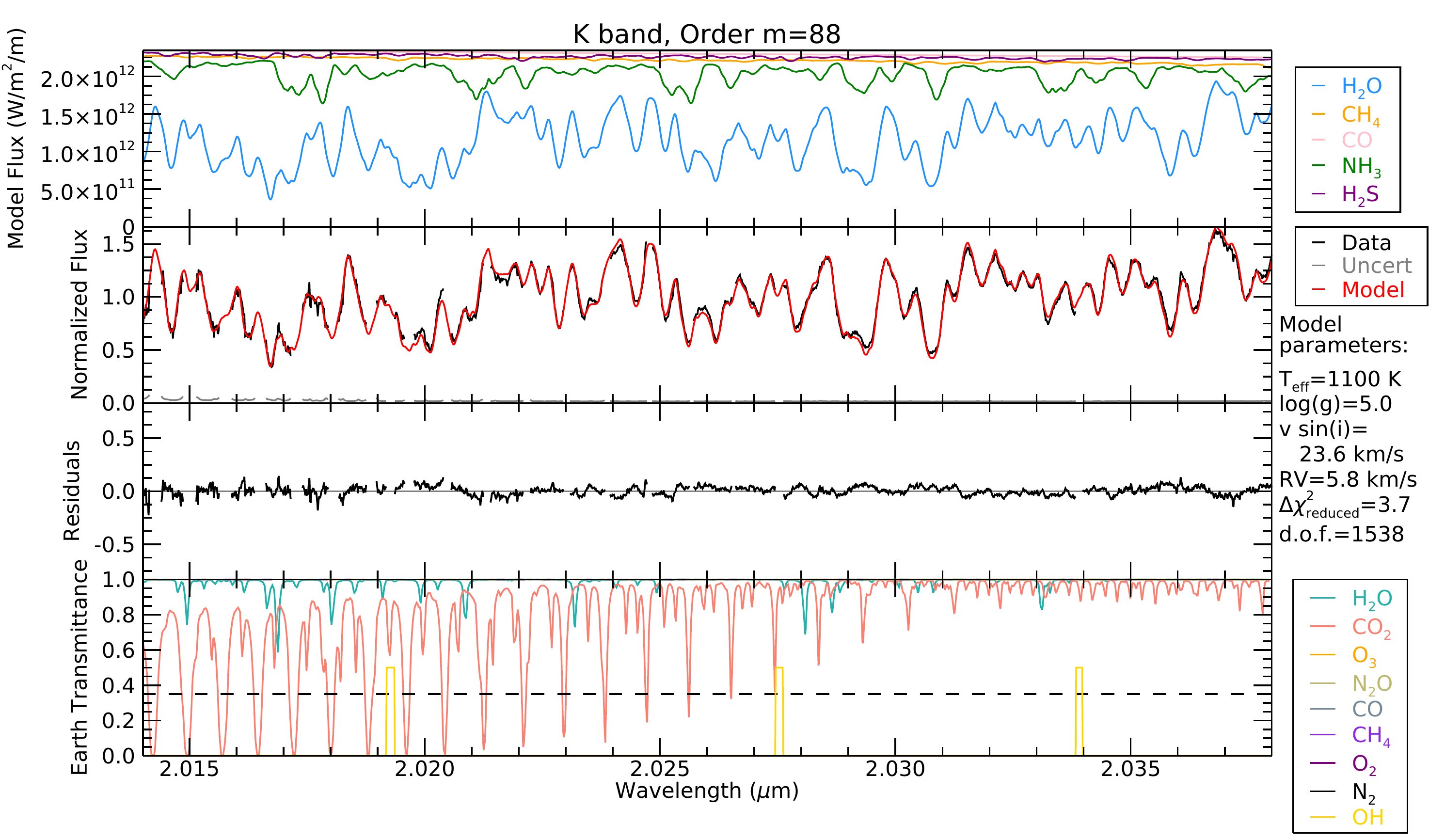} 
    \end{subfigure}
    \begin{subfigure}{0.99\textwidth}
        \includegraphics[width=\textwidth]{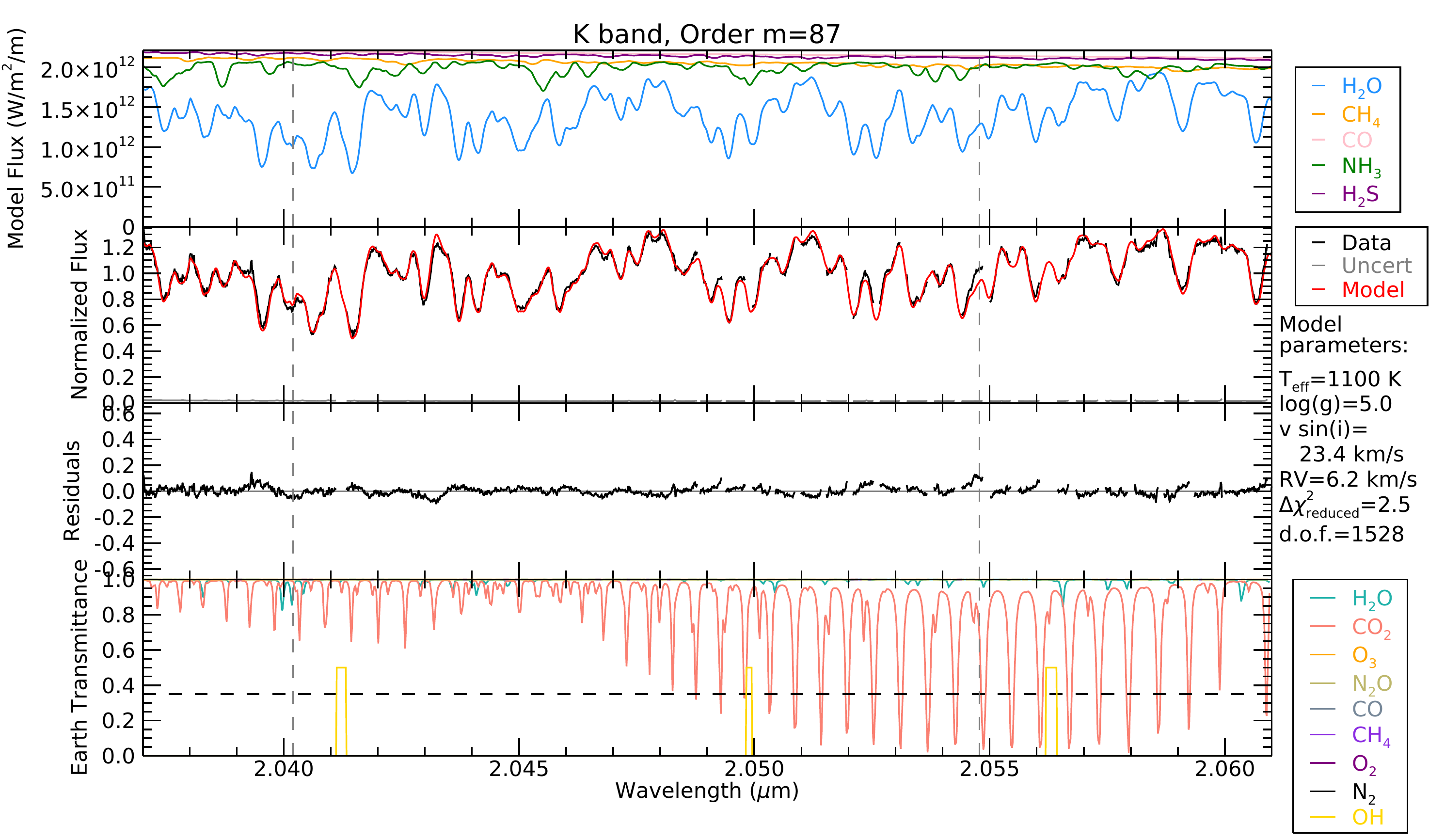} 
    \end{subfigure}
    \caption{Continued.}
\end{figure*}

\begin{figure*}
    \ContinuedFloat 
    \centering
    \begin{subfigure}{0.99\textwidth}
        \includegraphics[width=\textwidth]{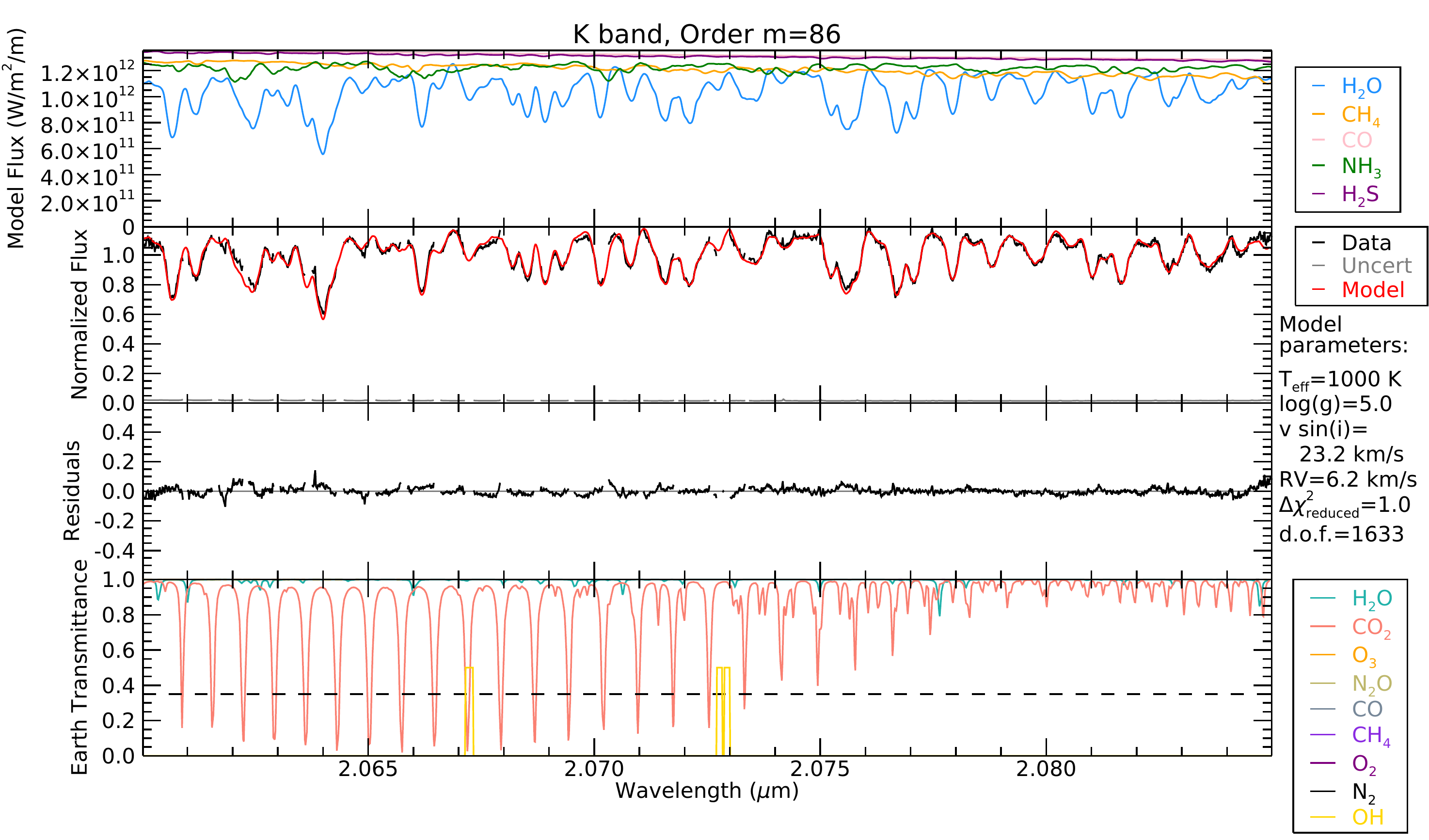} 
    \end{subfigure}
    \begin{subfigure}{0.99\textwidth}
        \includegraphics[width=\textwidth]{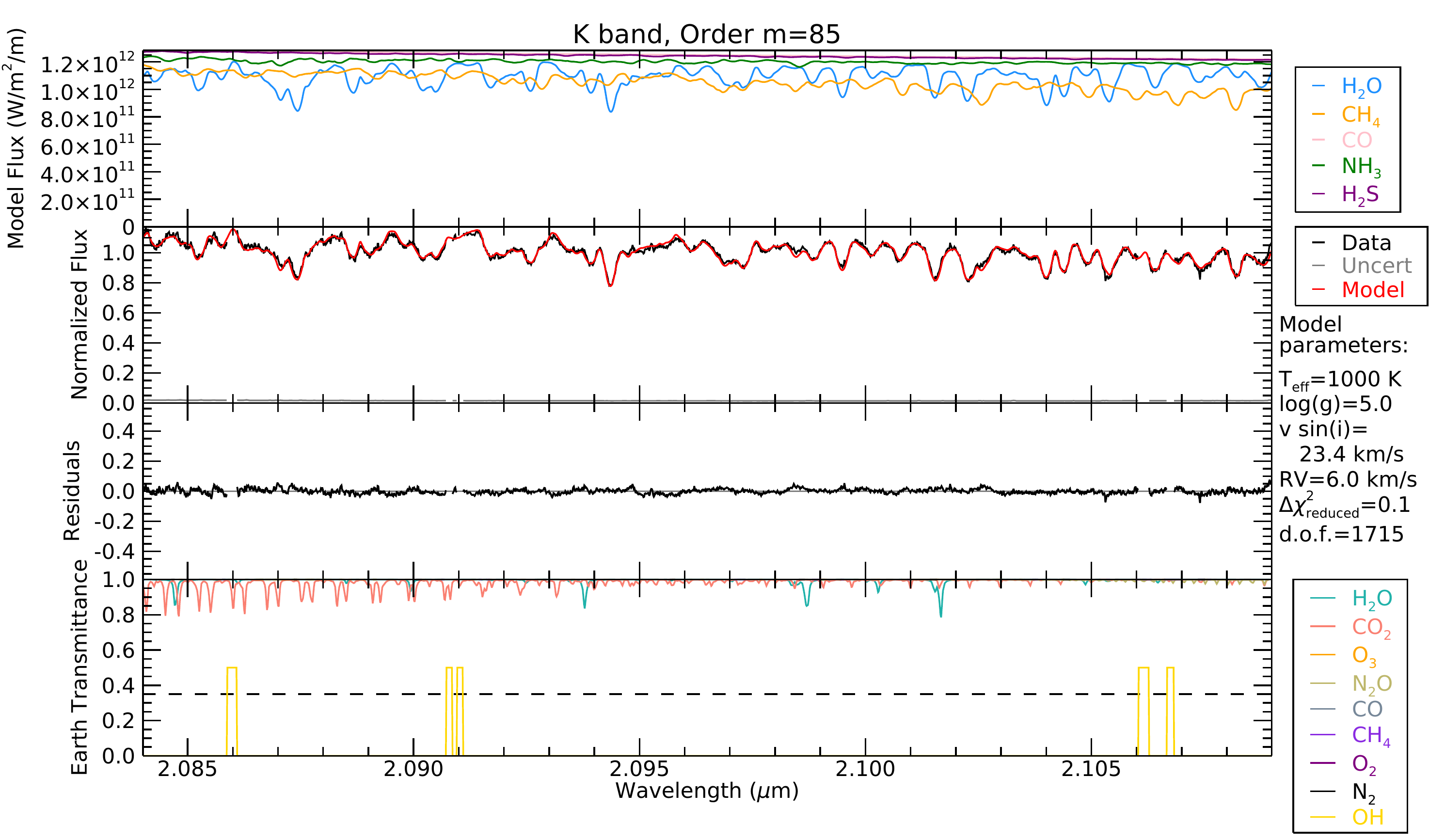} 
    \end{subfigure}
    \caption{Continued.}
\end{figure*}

\begin{figure*}
    \ContinuedFloat 
    \centering
    \begin{subfigure}{0.99\textwidth}
        \includegraphics[width=\textwidth]{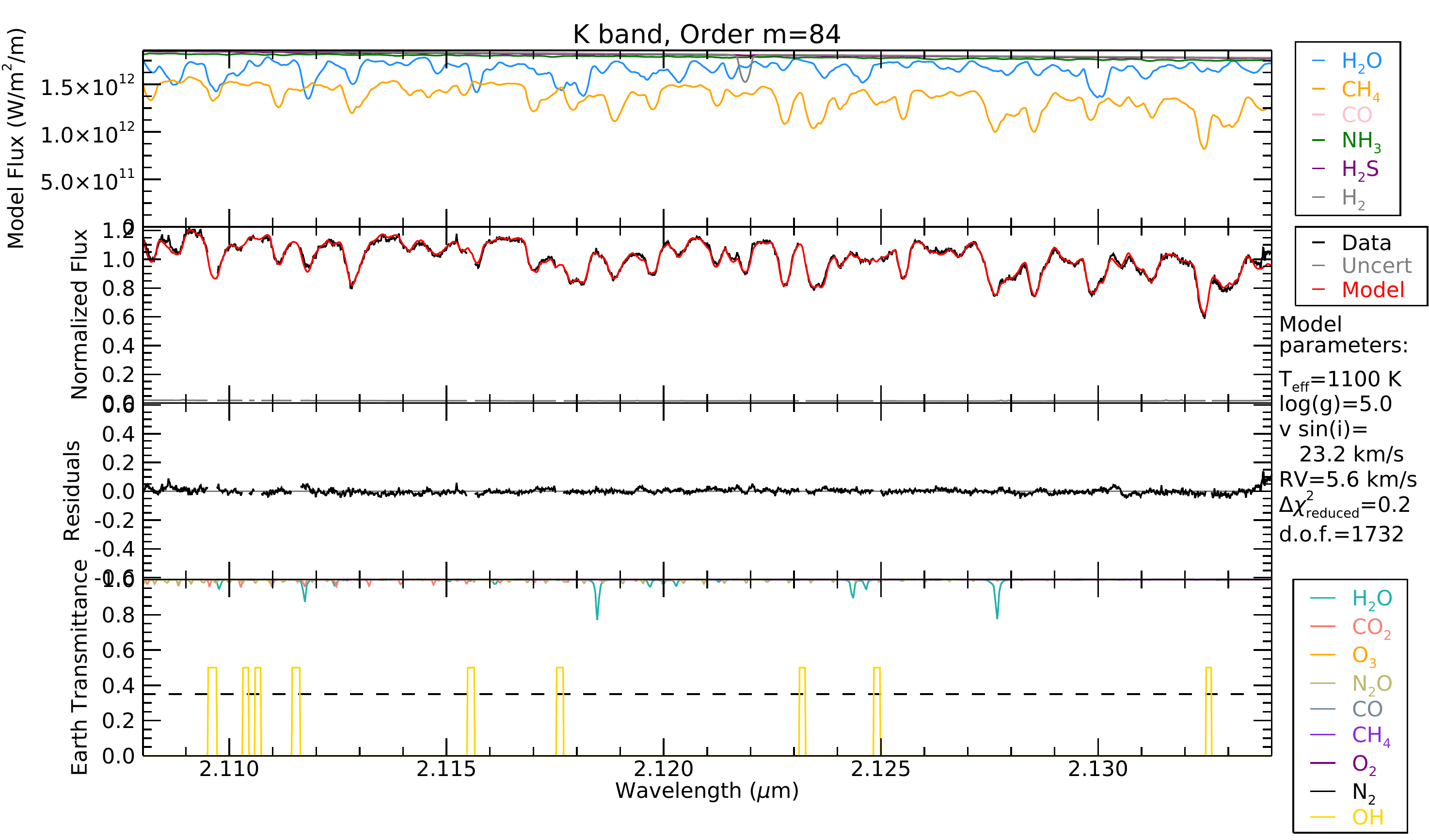} 
    \end{subfigure}
    \begin{subfigure}{0.99\textwidth}
        \includegraphics[width=\textwidth]{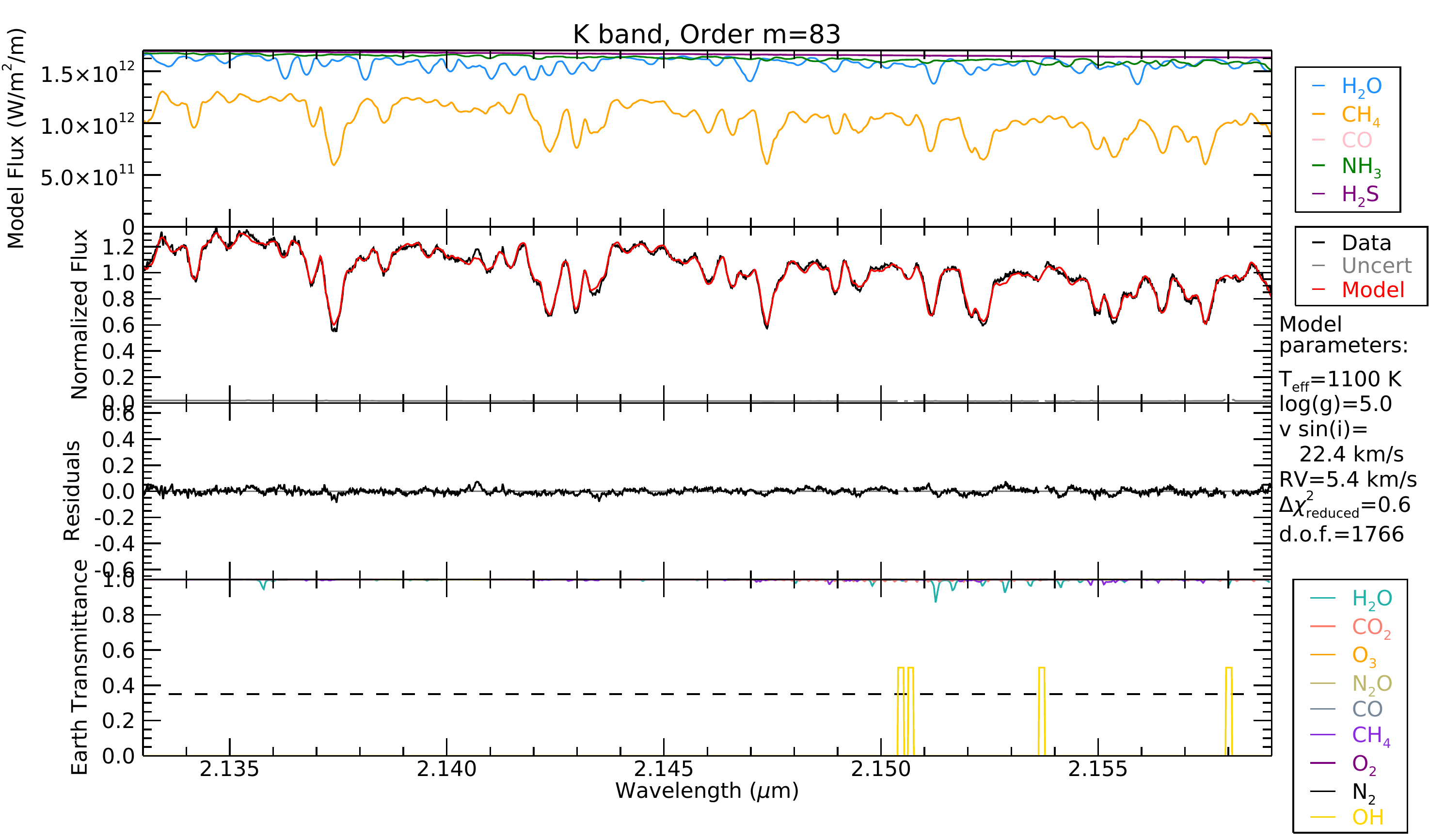} 
    \end{subfigure}
    \caption{Continued.}
\end{figure*}

\begin{figure*}
    \ContinuedFloat 
    \centering
    \begin{subfigure}{0.99\textwidth}
        \includegraphics[width=\textwidth]{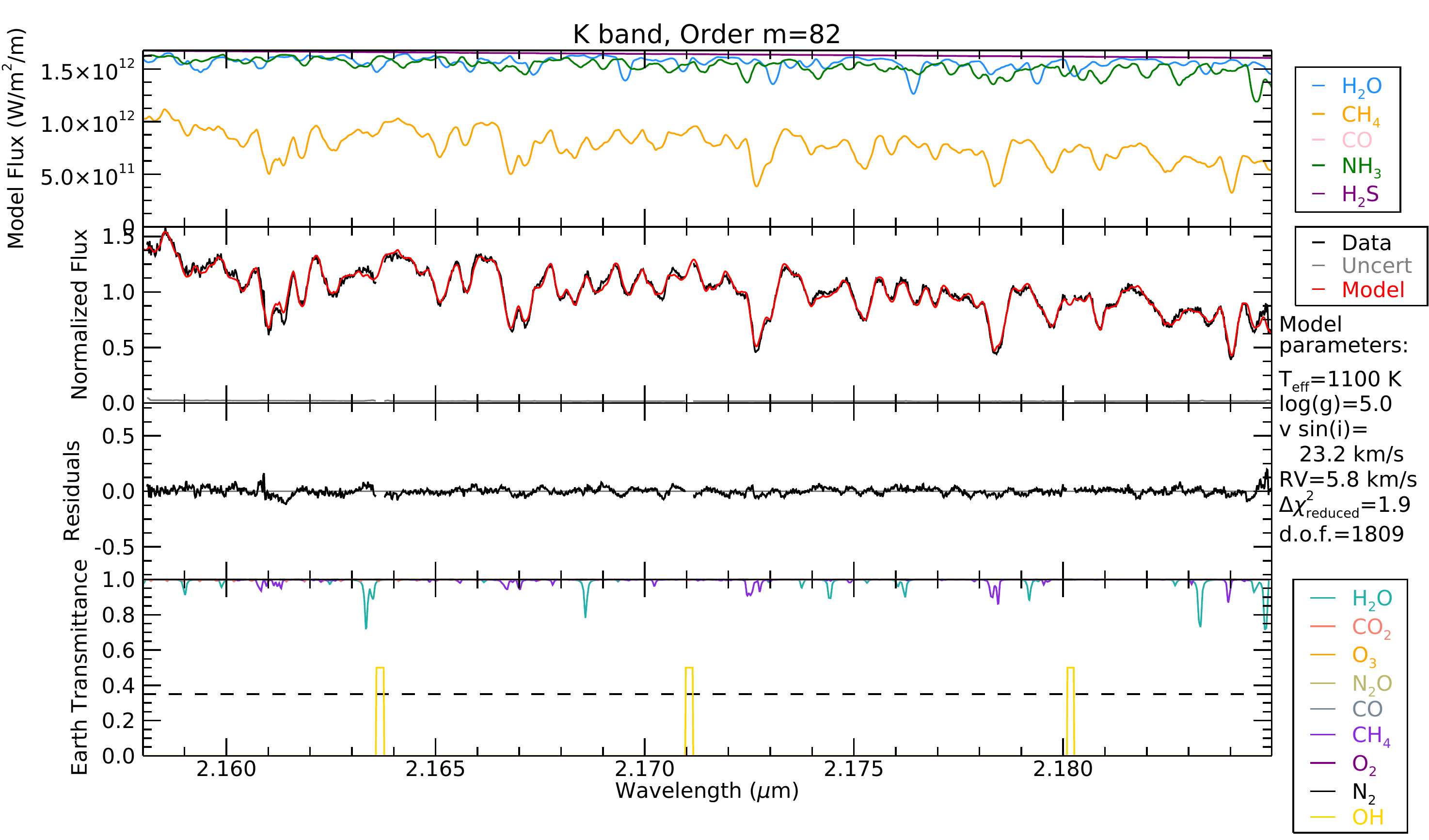} 
    \end{subfigure}
    \begin{subfigure}{0.99\textwidth}
        \includegraphics[width=\textwidth]{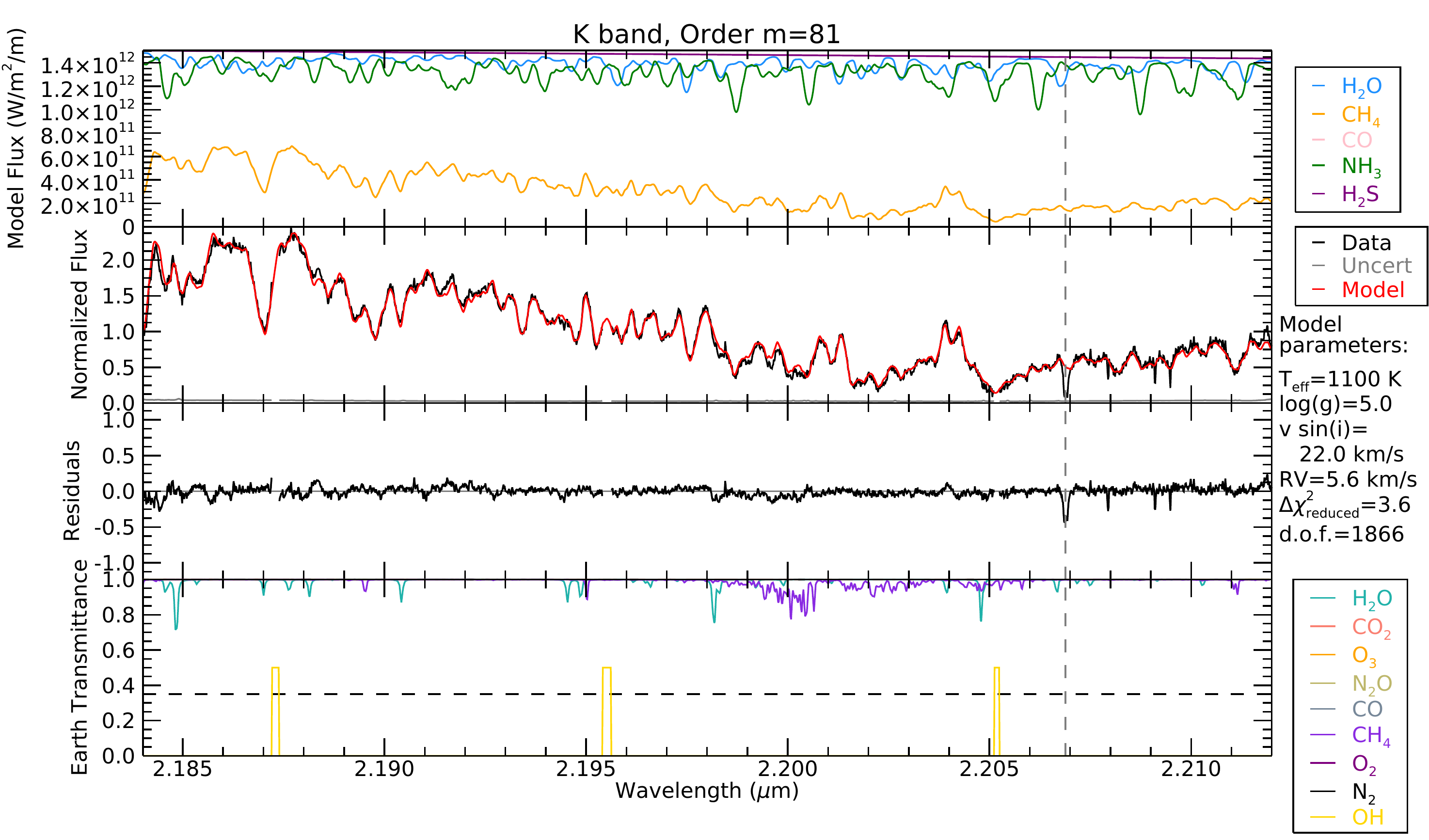} 
    \end{subfigure}
    \caption{Continued.}
\end{figure*}

\begin{figure*}
    \ContinuedFloat 
    \centering
    \begin{subfigure}{0.99\textwidth}
        \includegraphics[width=\textwidth]{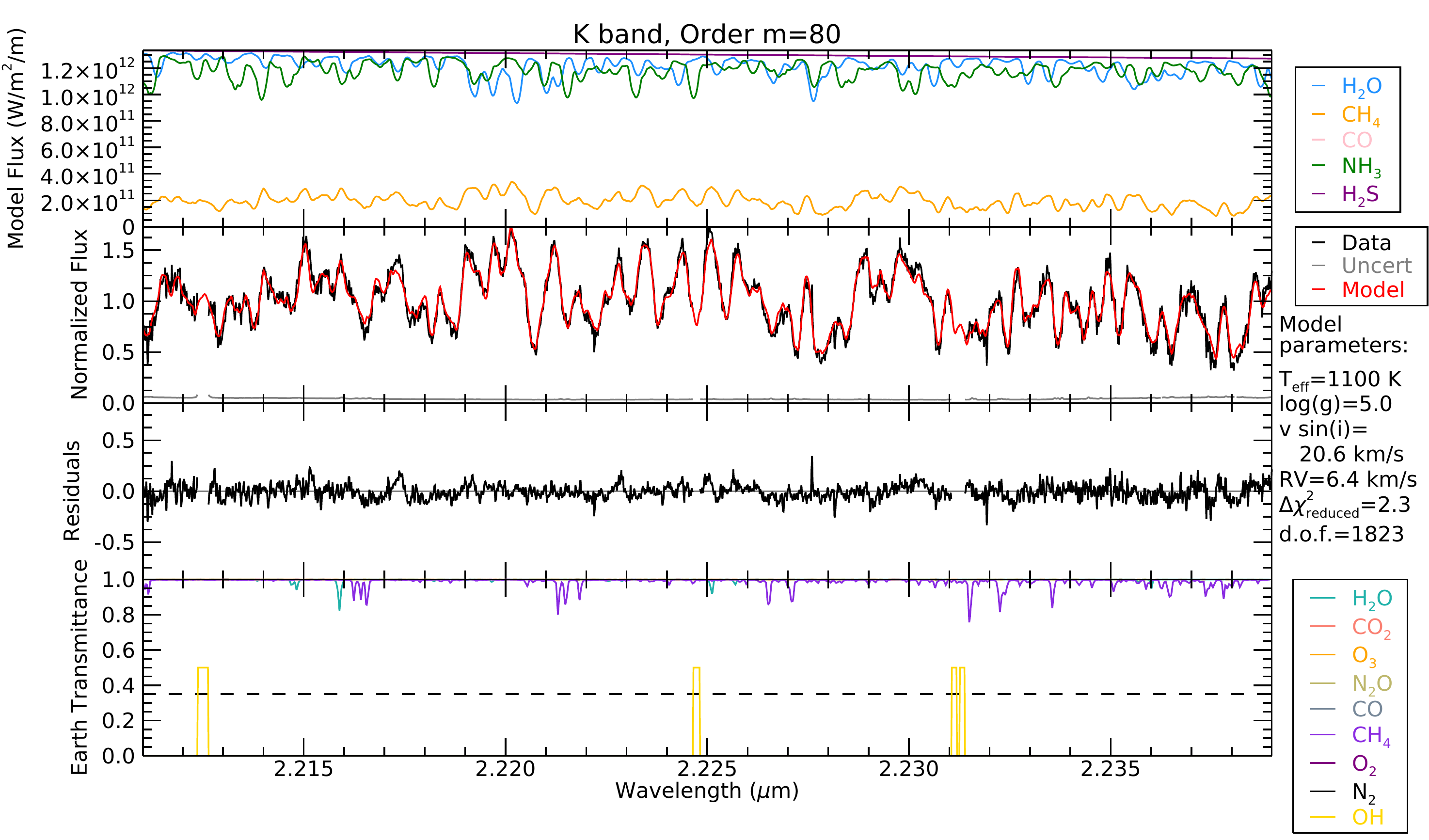} 
    \end{subfigure}
    \begin{subfigure}{0.99\textwidth}
        \includegraphics[width=\textwidth]{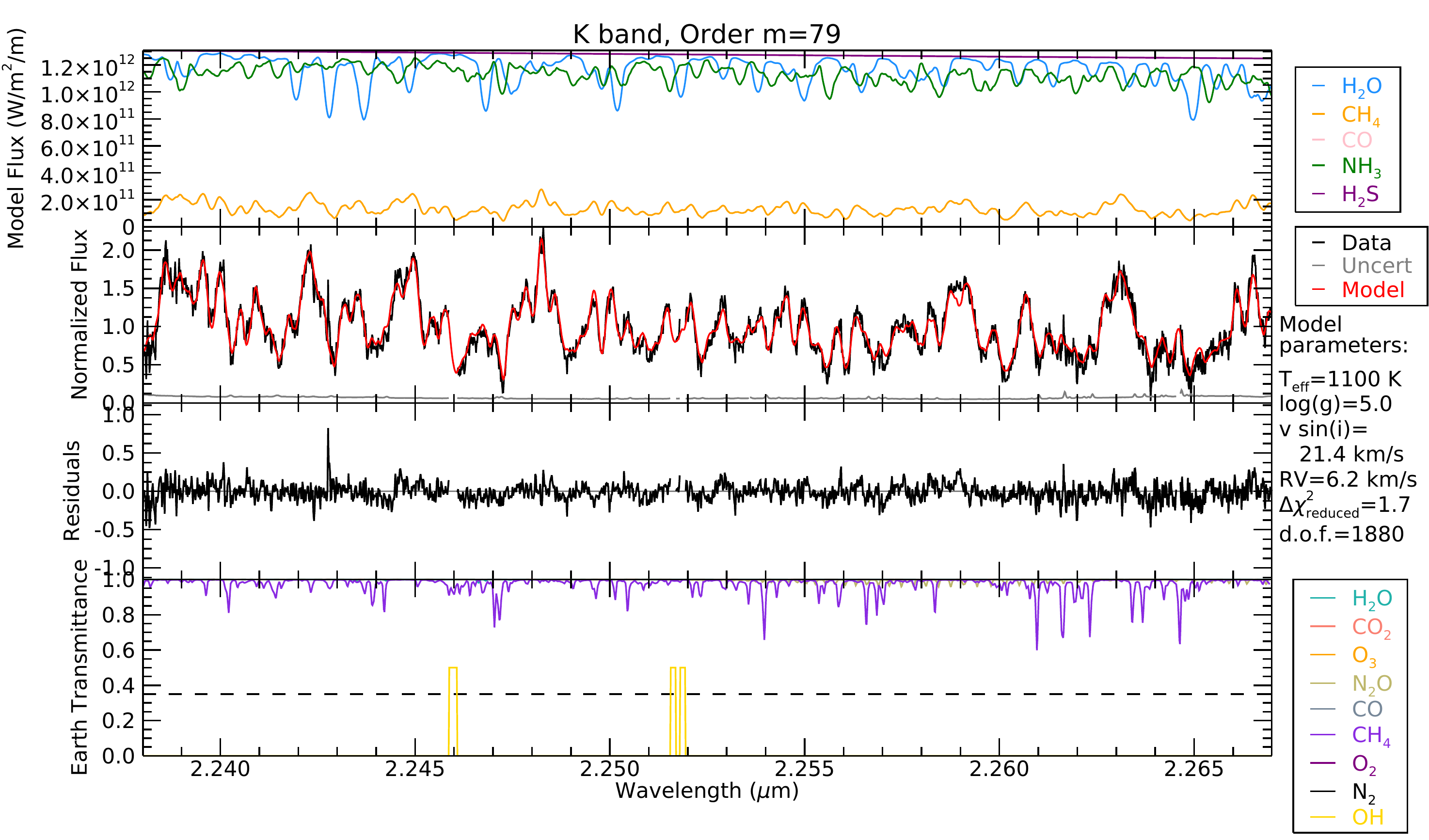} 
    \end{subfigure}
    \caption{Continued.}
\end{figure*}

\begin{figure*}
    \ContinuedFloat 
    \centering
    \begin{subfigure}{0.99\textwidth}
        \includegraphics[width=\textwidth]{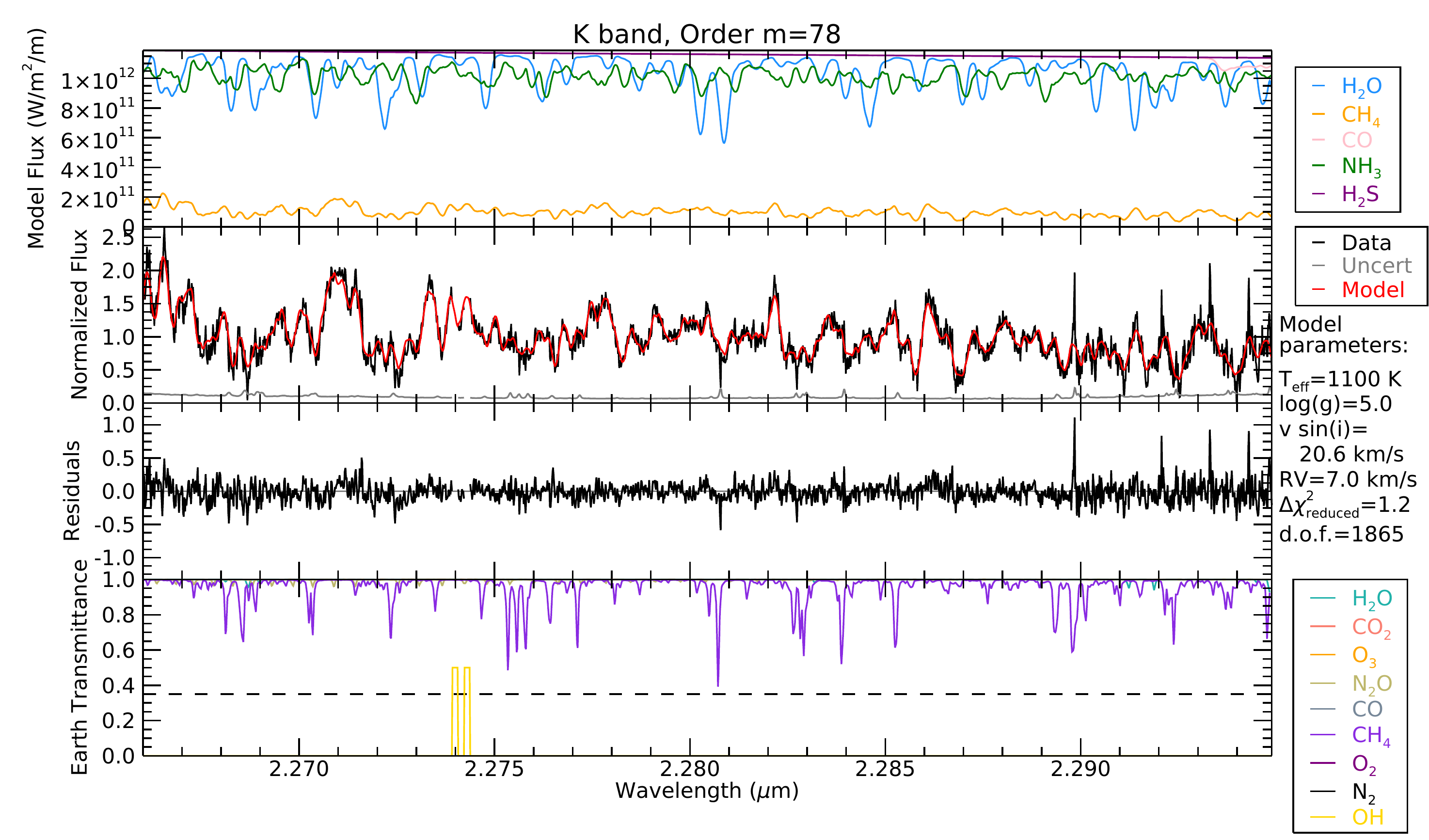}  
    \end{subfigure}
    \begin{subfigure}{0.99\textwidth}
        \includegraphics[width=\textwidth]{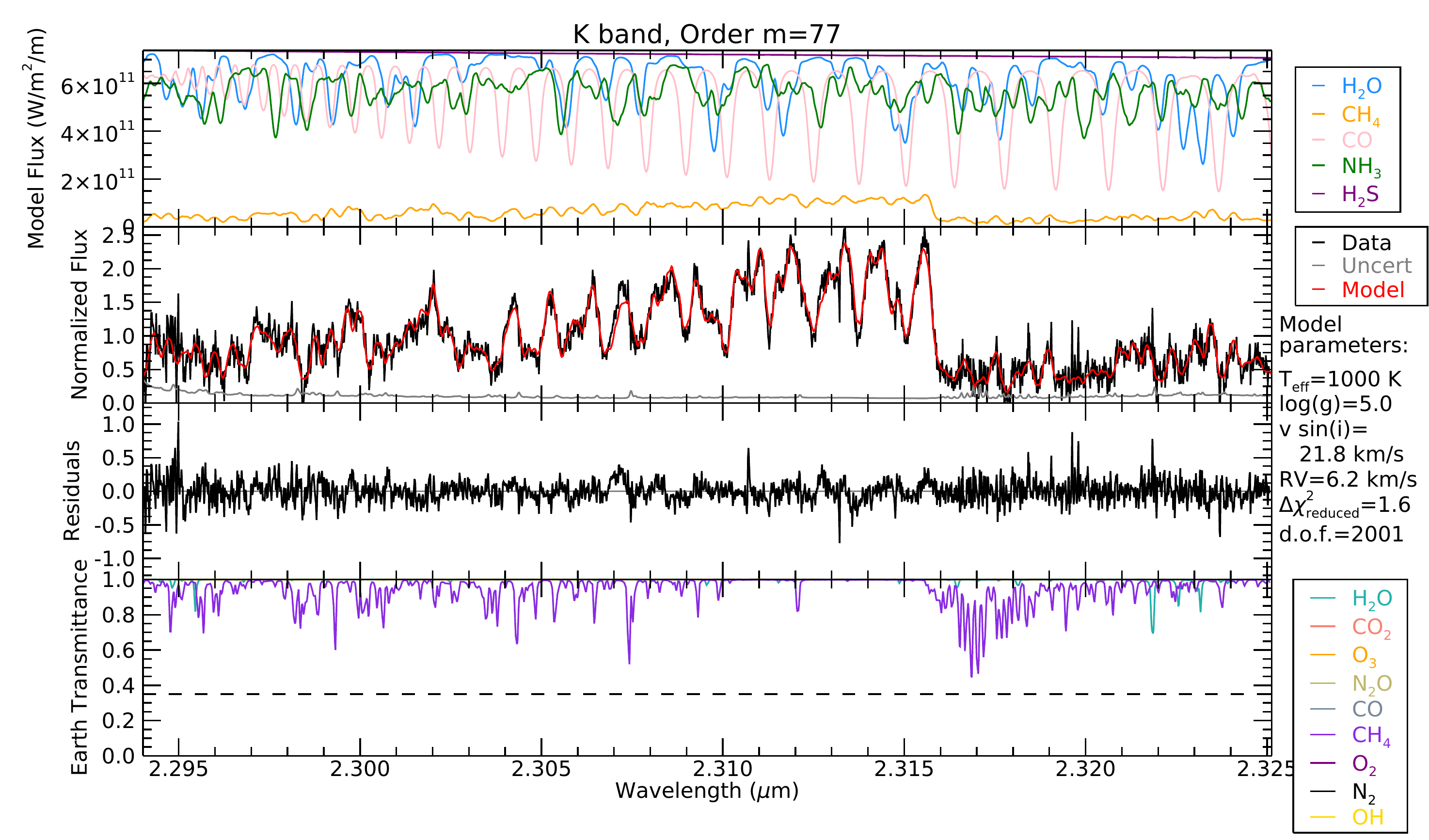}  
    \end{subfigure}
    \caption{Continued.}
\end{figure*}

\begin{figure*}
    \ContinuedFloat 
    \centering
    \begin{subfigure}{0.99\textwidth}
        \includegraphics[width=\textwidth]{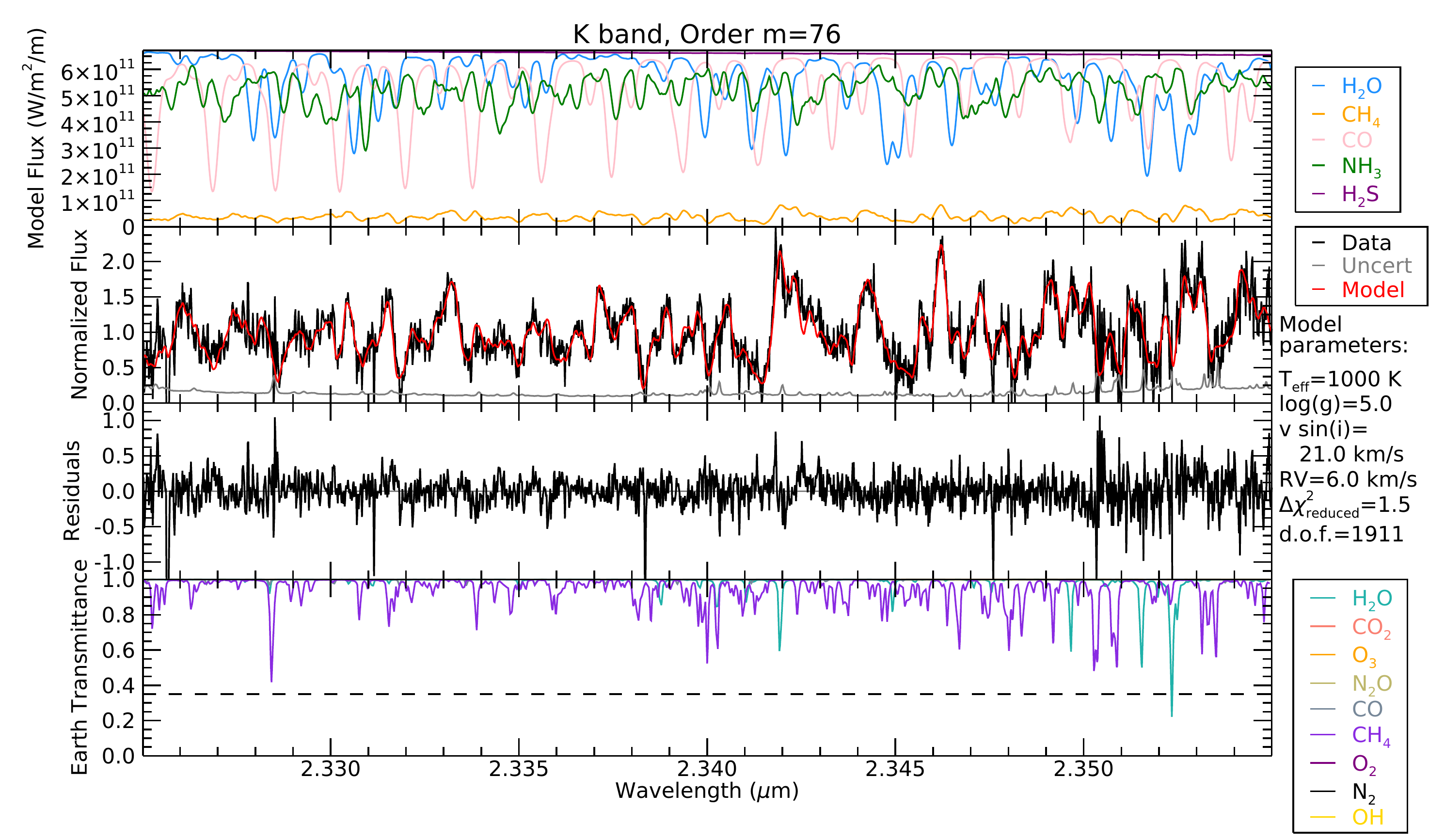} 
    \end{subfigure}
    \begin{subfigure}{0.99\textwidth}
        \includegraphics[width=\textwidth]{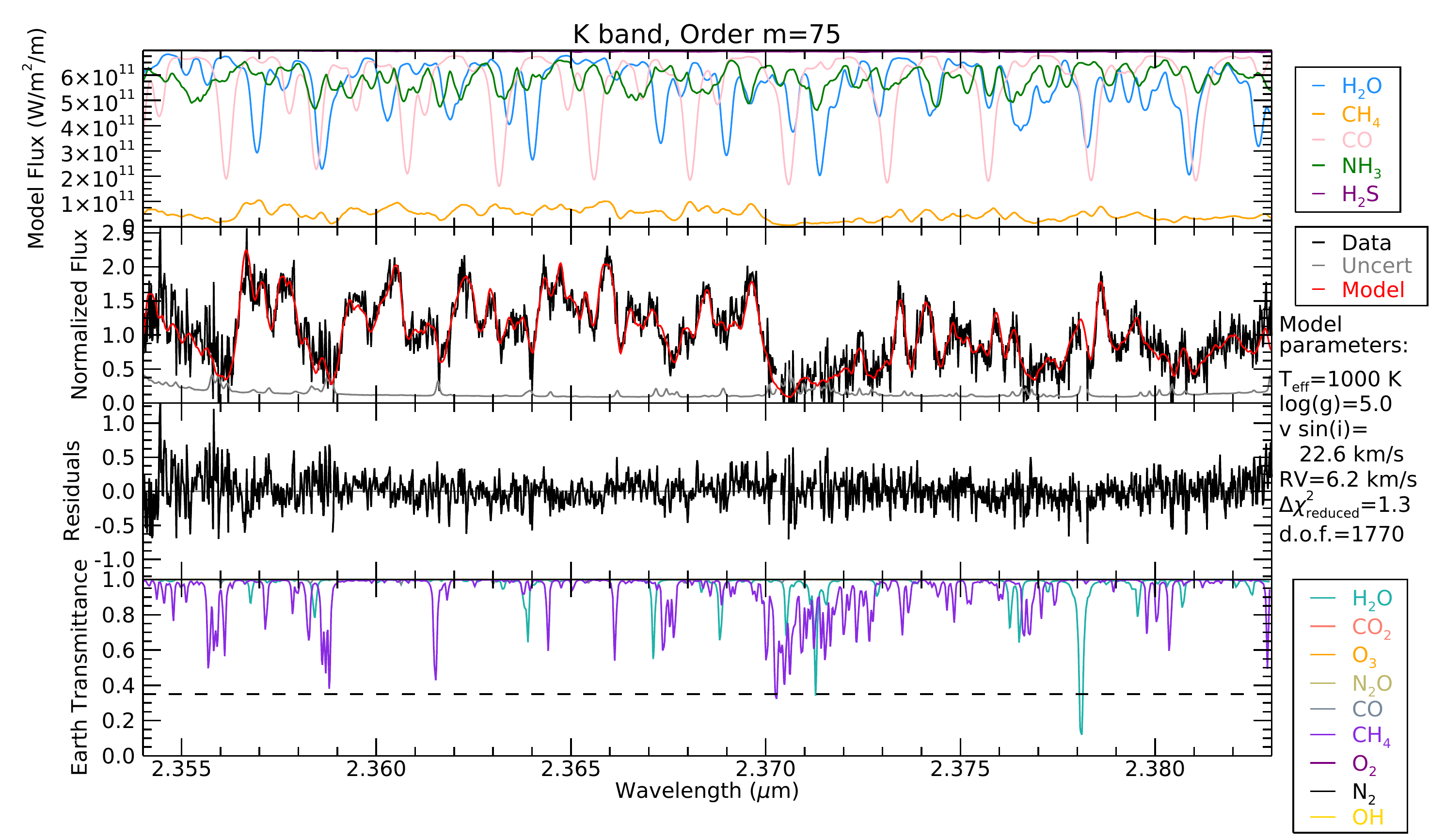}  
    \end{subfigure}
    \caption{Continued.}
\end{figure*}

\begin{figure*}
    \ContinuedFloat 
    \centering
    \begin{subfigure}{0.99\textwidth}
        \includegraphics[width=\textwidth]{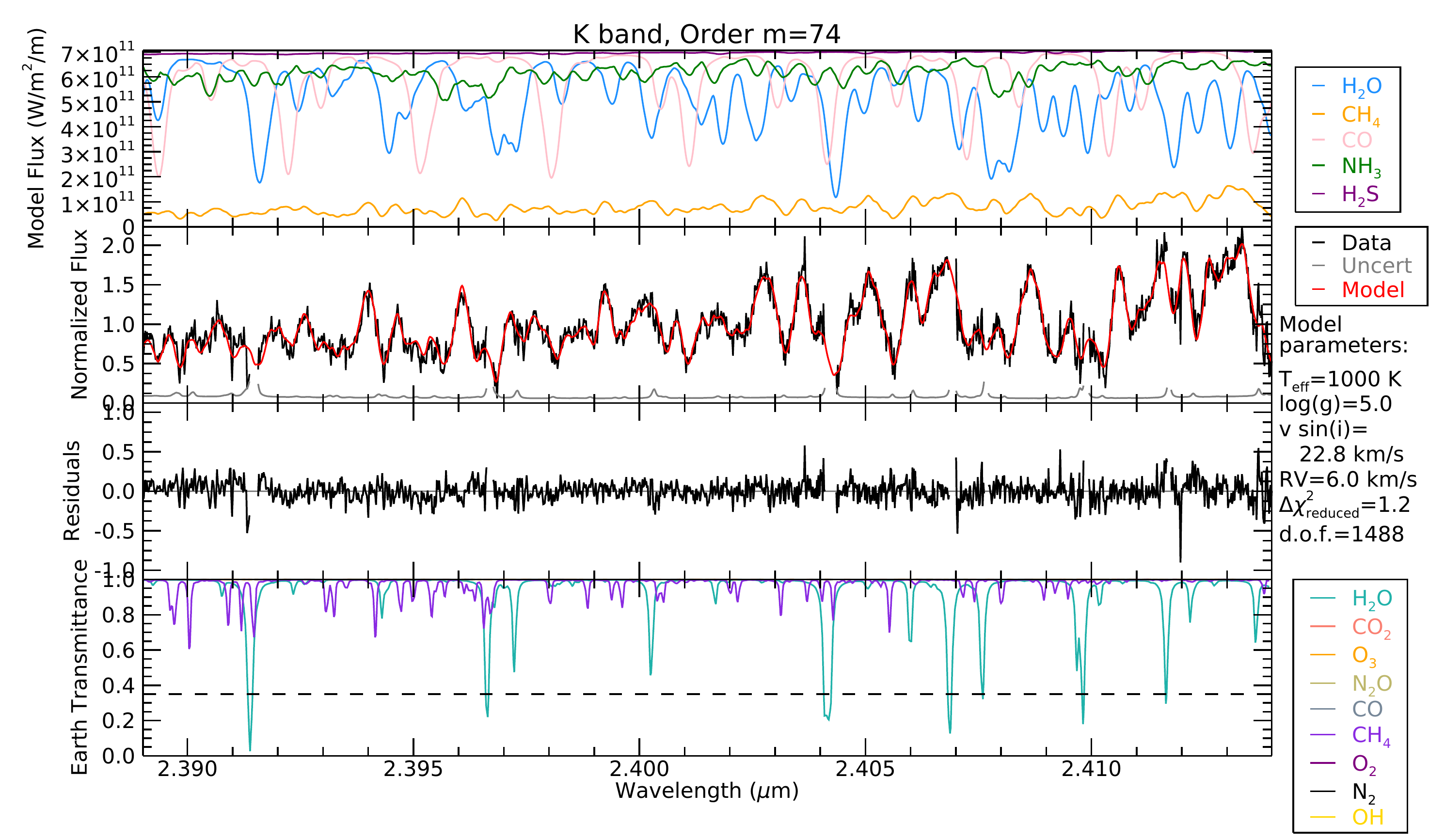} 
    \end{subfigure}
    \begin{subfigure}{0.99\textwidth}
        \includegraphics[width=\textwidth]{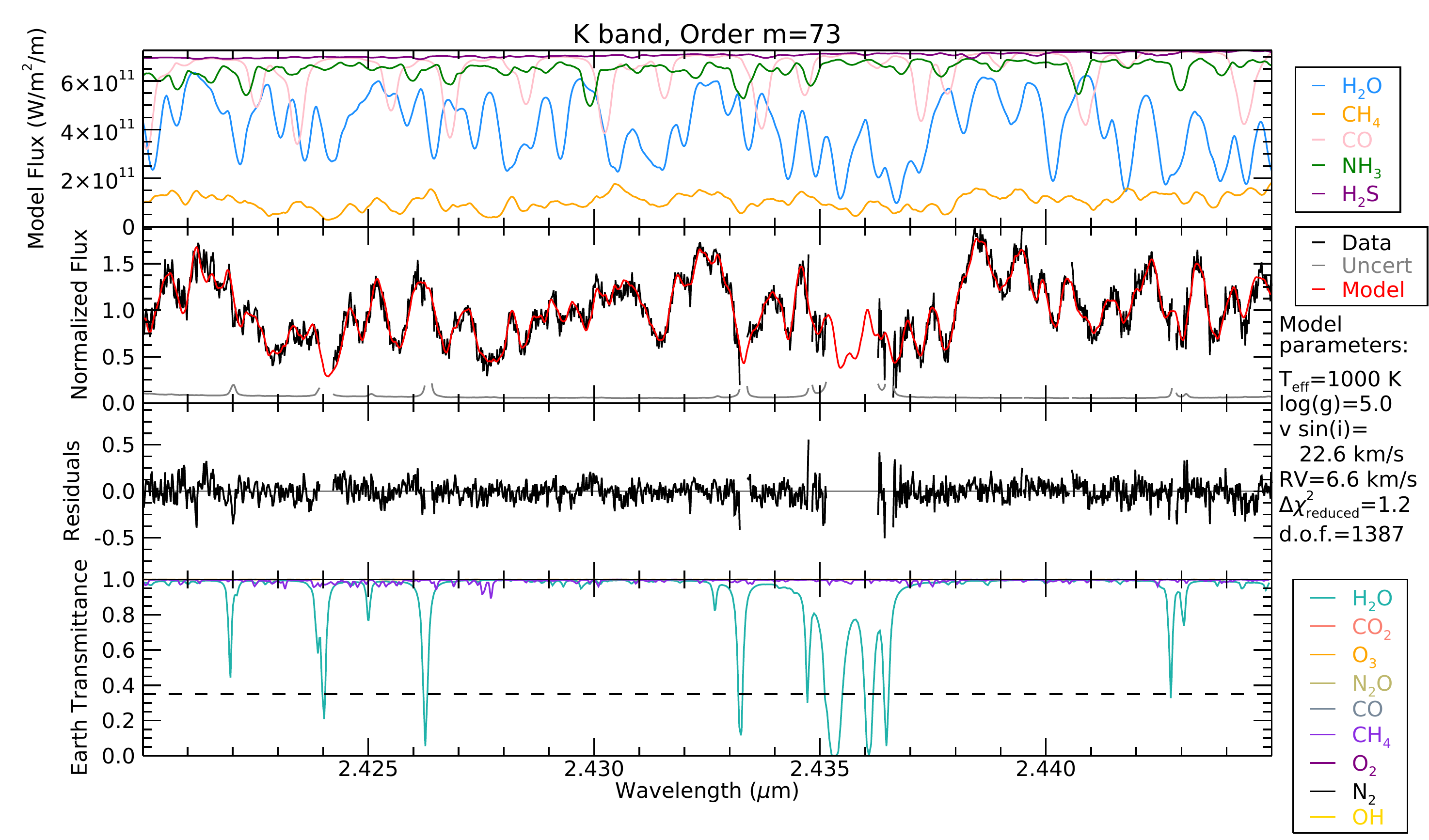} 
    \end{subfigure}
    \caption{Continued.}
\end{figure*}

\begin{figure*}
    \ContinuedFloat 
    \centering
    \begin{subfigure}{0.99\textwidth}
        \includegraphics[width=\textwidth]{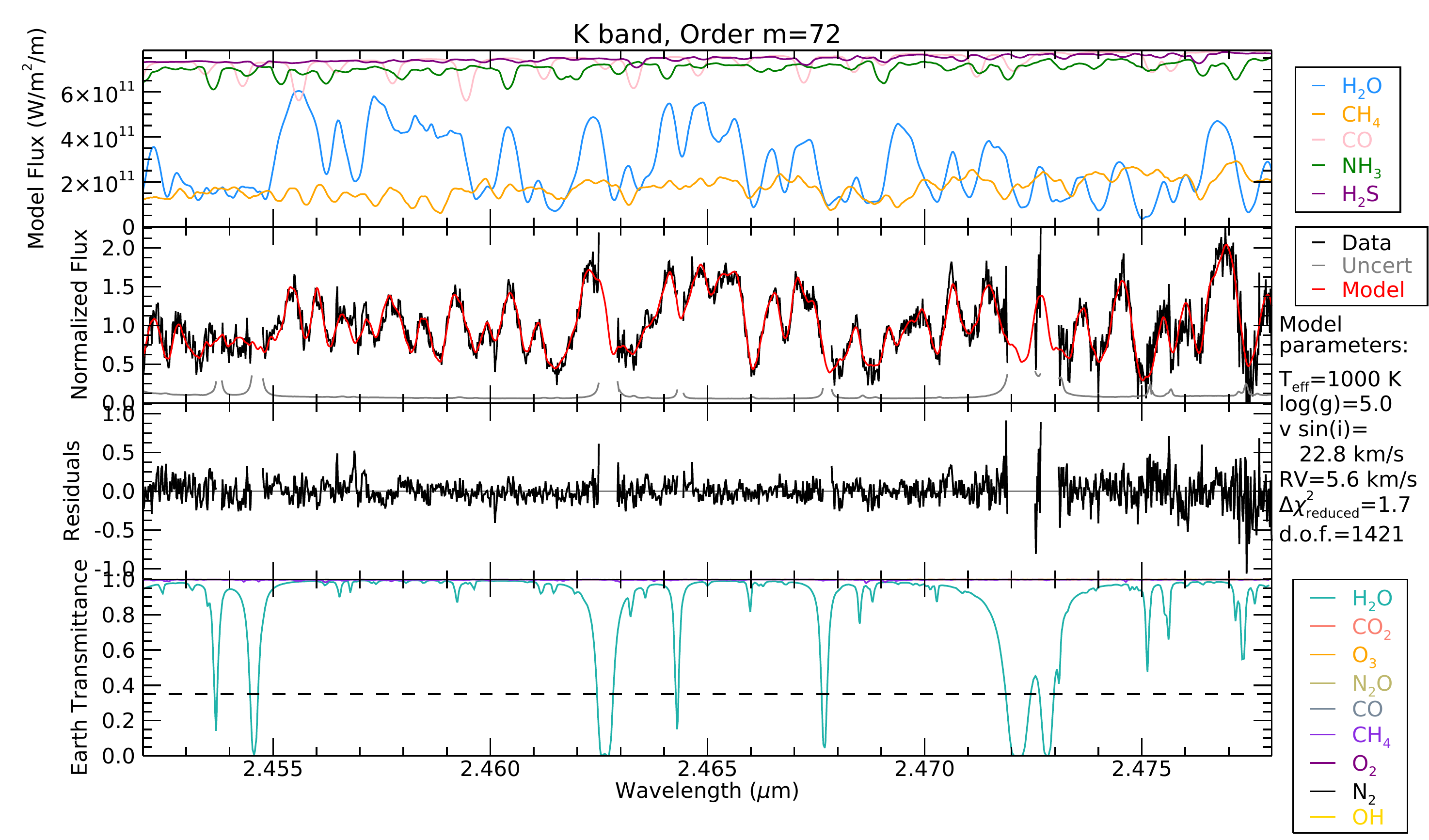} 
    \end{subfigure}
    \caption{Continued.}
\end{figure*}


\bsp	
\label{lastpage}
\end{document}